\documentclass[lettersize,journal]{IEEEtran}
\usepackage{amsmath,amsfonts}
\usepackage{algorithmic}
\usepackage{algorithm}
\usepackage{array}
\usepackage[caption=false,font=normalsize,labelfont=sf,textfont=sf]{subfig}
\usepackage{textcomp}
\usepackage{stfloats}
\usepackage{url}
\usepackage{verbatim}
\usepackage{graphicx}
\usepackage{cite}
\usepackage{multirow}
\usepackage[most]{tcolorbox}

\usepackage{tikz}
\usepackage{xcolor}
\usepackage{booktabs}
\usepackage{pifont}
\newcommand{\cmark}{\ding{51}}
\usepackage{makecell}
\usepackage{tabularx}

\begin{document}

\title{
Secure Semantic Communications via AI Defenses: Fundamentals, Solutions, and Future Directions
}

\author{
Lan Zhang,~\IEEEmembership{Senior Member,~IEEE}, 
Chengsi Liang,
Zeming Zhuang,
Yao Sun,~\IEEEmembership{Senior Member,~IEEE}, 
Fang Fang,~\IEEEmembership{Senior Member,~IEEE},
Xiaoyong Yuan,~\IEEEmembership{Member,~IEEE}, 
and Dusit Niyato,~\IEEEmembership{Fellow,~IEEE}}



\maketitle

\begin{abstract}
Semantic communication (SemCom) redefines wireless communication from reproducing symbols to transmitting task-relevant semantics. Enabled by learned encoders, decoders, and shared knowledge modules, SemCom supports efficient and robust integration with downstream and distributed AI tasks across sensing, computing, and control. However, this AI-native architecture also introduces new vulnerabilities, as semantic failures may arise from adversarial perturbations to models, corrupted training data, desynchronized priors, or misaligned inference even when lower-layer transmission reliability and cryptographic protection remain intact. This survey provides a defense-centered and system-oriented synthesis of security in SemCom via AI defense. We analyze AI-centric threat models by consolidating existing studies and organizing attack surfaces across model-level, channel-realizable, knowledge-based, and networked inference vectors. Building on this foundation, we present a structured taxonomy of defense strategies organized by where semantic integrity can be compromised in SemCom systems despite correct symbol delivery, spanning semantic encoding, wireless transmission, knowledge integrity, and coordination among multiple agents. These categories correspond to distinct security failure modes, including representation fragility, channel-realizable manipulation, semantic prior poisoning or desynchronization, and adversarial propagation through distributed inference. To bridge design and deployment, we examine security utility operating envelopes that capture tradeoffs among semantic fidelity, robustness, latency, and energy under realistic constraints, survey evaluation frameworks and representative applications, and identify open challenges in cross-layer composition and deployment-time certification. Overall, this survey offers a unified system-level perspective that enables readers to understand major threat and defense mechanisms in AI-native SemCom systems and to leverage emerging security techniques in the design and deployment of robust SemCom architectures for next-generation intelligent networks.

\end{abstract}


\section{Introduction} \label{sec:intro}
\subsection{Background}
Semantic communication (SemCom) redefines the goal of wireless communication from reproducing symbols to conveying task-relevant meaning \cite{weaver1949recent}. Rather than treating communication as a bit-level reproduction problem, SemCom systems leverage learned encoders, decoders, and shared knowledge modules to extract and convey information that is most critical for downstream tasks such as classification, decision-making, or control~\cite{xie2021deep,bourtsoulatze2019deep}. This shift enables substantial gains in bandwidth efficiency, robustness to channel impairments, and tight coupling with inference across a wide range of applications, including image and video transmission, natural language interaction, multimodal sensing, autonomous systems, and distributed AI tasks~\cite{yang2022semantic,liang2023vista, chaccour2024less,xin2024entropy,guo2024survey}.

A defining characteristic of modern SemCom systems is that they are fundamentally \emph{AI-native}. Core components of the SemCom pipeline, including semantic encoders, semantic decoders, and knowledge-access or context-inference modules, are implemented as learned models trained on data and queried at inference time~\cite{xin2024entropy,yang2022semantic,chaccour2024less,guo2024survey}. These models determine what is deemed semantically important, how it is represented, and how it is reconstructed, introducing dependencies that are absent in conventional communication architectures. Consequently, semantic fidelity depends not only on channel conditions but also on model performance, which is affected by training data quality, generalization behavior, model robustness, and context alignment. {Failures may therefore arise even when packet delivery ratios and bit error rates remain within acceptable bounds, making semantic degradation difficult to observe using conventional metrics}.

This shift toward learned, model-driven semantic processing fundamentally alters the communication security landscape. Traditional mechanisms such as encryption, authentication, and error control remain necessary, but are insufficient to protect semantic integrity \cite{yang2024secure,sagduyu2023semantic,chen2025secure}. \textit{Even when lower-layer reliability and confidentiality are preserved, adversaries can influence communication outcomes by exploiting learned model behaviors, manipulating training data, or desynchronizing shared context, without disrupting packet delivery or triggering conventional alarms.} These attacks target semantic meaning rather than symbols and therefore evade traditional communication-layer security abstractions.

\subsection{Motivation and Contribution}
While the AI security community has developed powerful techniques for improving robustness, privacy, and integrity, such as adversarial training, data poisoning defenses, membership inference protection, and secure federated learning~\cite{yuan2019adversarial,li2023trustworthy,wei2025trustworthy,wen2023survey}, they are primarily designed for standalone models and rarely account for deployment within dynamic, resource-constrained, over-the-air communication pipelines where learning, transmission, and inference are tightly coupled and interdependent. As a result, isolated model vulnerabilities can propagate across the SemCom pipeline, leading to system-level failures that are not addressed by existing AI security frameworks.

Reflecting growing awareness of these challenges, a number of recent surveys have begun to examine security issues in SemCom systems~\cite{guo2024survey,xin2024entropy,won2024resource,li2024robustsemcom,meng2025securesemcom,yang2024secure,shen2023secure,liang2025safeguarded}. As summarized in Table~\ref{tab:survey_comparison}, existing surveys span topics ranging from semantic information theory and robust transmission to network architectures and stage-wise security analysis. However, most surveys concentrate on individual components or attack surfaces, offering limited insight into how threats and defenses interact and compose across encoder and decoder models, wireless transmission, shared knowledge, and inference. As a result, a defense-centered, system-level understanding of SemCom security remains largely absent.

\begin{table*}[t]
\centering
\caption{Comparison of survey and overview papers on security in semantic communication systems.
A checkmark indicates that the corresponding security surface is explicitly discussed,
rather than system-level integration or deployment-aware analysis.
}
\label{tab:survey_comparison}
\scriptsize
\renewcommand{\arraystretch}{1.25}
\resizebox{\textwidth}{!}{
\begin{tabular}{l|p{6.0cm}|c|c|c|c|c}
\hline
\multirow{2}{*}{\textbf{Survey}} &
\multirow{2}{*}{\textbf{Primary Focus}} &
\multicolumn{5}{c}{\textbf{AI-Native Semantic Security Surfaces}} \\ \cline{3-7}
 & & \makecell{\textbf{Semantic}\\\textbf{Model}}
 & \makecell{\textbf{Wireless}\\\textbf{Channel}}
 & \makecell{\textbf{Shared}\\\textbf{Knowledge}}
 & \makecell{\textbf{Semantic}\\\textbf{Inference}}
 & \makecell{\textbf{Networked}\\\textbf{Inference}} \\
\hline
Xin et al. (2024)~\cite{xin2024entropy} 
& Theoretical foundations of SemCom, including semantic entropy,
semantic rate--distortion, and mathematical representations of knowledge resources 
&  &  & \cmark &  &  \\
\hline
Guo et al. (2024)~\cite{guo2024survey} 
& SemCom networks (SemComNet) for multi-agent systems,
emphasizing architecture, orchestration, and network-level security 
&  & \cmark &  &  & \cmark \\
\hline
Won et al. (2025)~\cite{won2024resource} 
& Resource management, scheduling, and security/privacy considerations
for SemCom systems 
&  & \cmark &  &  &  \\
\hline
Li et al. (2024)~\cite{li2024robustsemcom} 
& Robust semantic transmission under adversarial attacks and privacy risks
for learned encoders and decoders 
& \cmark & \cmark &  &  &  \\
\hline
Meng et al. (2025)~\cite{meng2025securesemcom} 
& Stage-wise analysis of security and privacy threats and countermeasures
across the SemCom life cycle 
& \cmark & \cmark & \cmark &  &  \\
\hline
Yang et al. (2024)~\cite{yang2024secure} 
& Conceptual foundations and key challenges of secure SemCom,
distinguishing semantic information security from semantic model security 
& \cmark &  & \cmark &  &  \\
\hline
Shen et al. (2023)~\cite{shen2023secure} 
& Survey of security threats and defenses across the SemCom pipeline,
including training, transmission, and knowledge base maintenance 
& \cmark & \cmark & \cmark &  &  \\

\hline
\textbf{This Survey} 
& \textbf{Defense-centered, system-level synthesis of AI-native semantic threats and defenses,
bridging AI security and SemCom, with emphasis on semantic inference integrity,
cross-layer composition, deployment constraints, and security--utility tradeoffs} 
& \cmark & \cmark & \cmark & \cmark & \cmark \\

\hline
\end{tabular}}
\end{table*}

Motivated by these gaps, \textit{this survey examines SemCom security with the focus on AI defense, integrating model-centric AI security insights with system-level SemCom design.} We advocate a system-level perspective in which robustness is treated as an explicit and central design objective rather than a post hoc add-on. Departing from prior surveys that emphasize isolated threats or individual components, we focus on cross-layer composition, deployment constraints, and runtime security utility tradeoffs. We argue that securing SemCom requires coherent alignment of threats, defenses, and performance objectives across the integrated pipeline, spanning semantic encoding and decoding, wireless transmission, shared knowledge, and networked inference and coordination in multi-agent systems.

As shown in Fig.~\ref{fig:intro}, recent advances in SemCom have rapidly expanded beyond point-to-point transmission toward model–driven, knowledge-assisted, and multi-agent systems that operate under tight resource and latency constraints \cite{chaccour2024less,xin2024entropy,guo2024survey}. At the same time, the security landscape of AI systems has evolved significantly, with demonstrated vulnerabilities in robustness, privacy, and integrity that directly impact learned semantic representations and inference pipelines~\cite{yuan2019adversarial,li2023trustworthy,wei2025trustworthy,wen2023survey}. These developments have progressed largely in parallel, creating a growing disconnect between how SemCom systems are designed and how they can be defended in practice. As SemCom technologies move closer to deployment in safety- and mission-critical applications, there is an urgent need for a unified, defense-centered perspective that bridges AI security advances with the architectural and operational realities of SemCom. This survey responds to that need by consolidating emerging threats, defenses, and deployment challenges into a coherent system-level framework.

In summary, this survey makes the following contributions.

\begin{itemize}
\item \textbf{AI-Centric Threat Model:} We synthesize and systematize an AI-centric threat model for SemCom by reviewing and consolidating existing works across SemCom security and AI security. This synthesis characterizes semantic-level adversarial objectives, attacker capabilities, and access assumptions, and organizes representative attack classes spanning model-centric threats to learned semantics, channel-realizable semantic attacks, and semantic manipulation via knowledge and context, highlighting recurring patterns and gaps in the literature. 

\item \textbf{Layered Defense Taxonomy and Synthesis:} We provide a structured, system-level synthesis of defense mechanisms across the SemCom pipeline, covering semantic encoders and decoders, wireless transmission, shared knowledge resources, and networked inference. By connecting advances in AI security with SemCom-specific architectural and operational constraints, this taxonomy clarifies how existing defenses operate, interact, and fall short when deployed in AI-native communication systems.

\item \textbf{Bridging Design and Deployment:} Drawing on insights from prior studies and practical deployment considerations, we articulate a deployment-oriented perspective centered on security–utility operating envelopes. This perspective consolidates how robustness, semantic fidelity, latency, energy, and adversarial budget tradeoffs are analyzed in the literature, and distills design principles for integrating defenses under realistic resource and runtime constraints.

\item \textbf{Evaluation and Benchmarking Gaps:} We review existing evaluation practices for SemCom security and identify critical gaps in metrics, benchmarks, and experimental settings. Based on this analysis, we outline requirements for deployment-aware evaluation, reproducible benchmarks, and adversarial testbeds that better support system-level assessment of semantic robustness and security.
\end{itemize}

\subsection{Paper Organization}
\begin{figure*}[!t]
\centering
\includegraphics[width=1.0\textwidth]{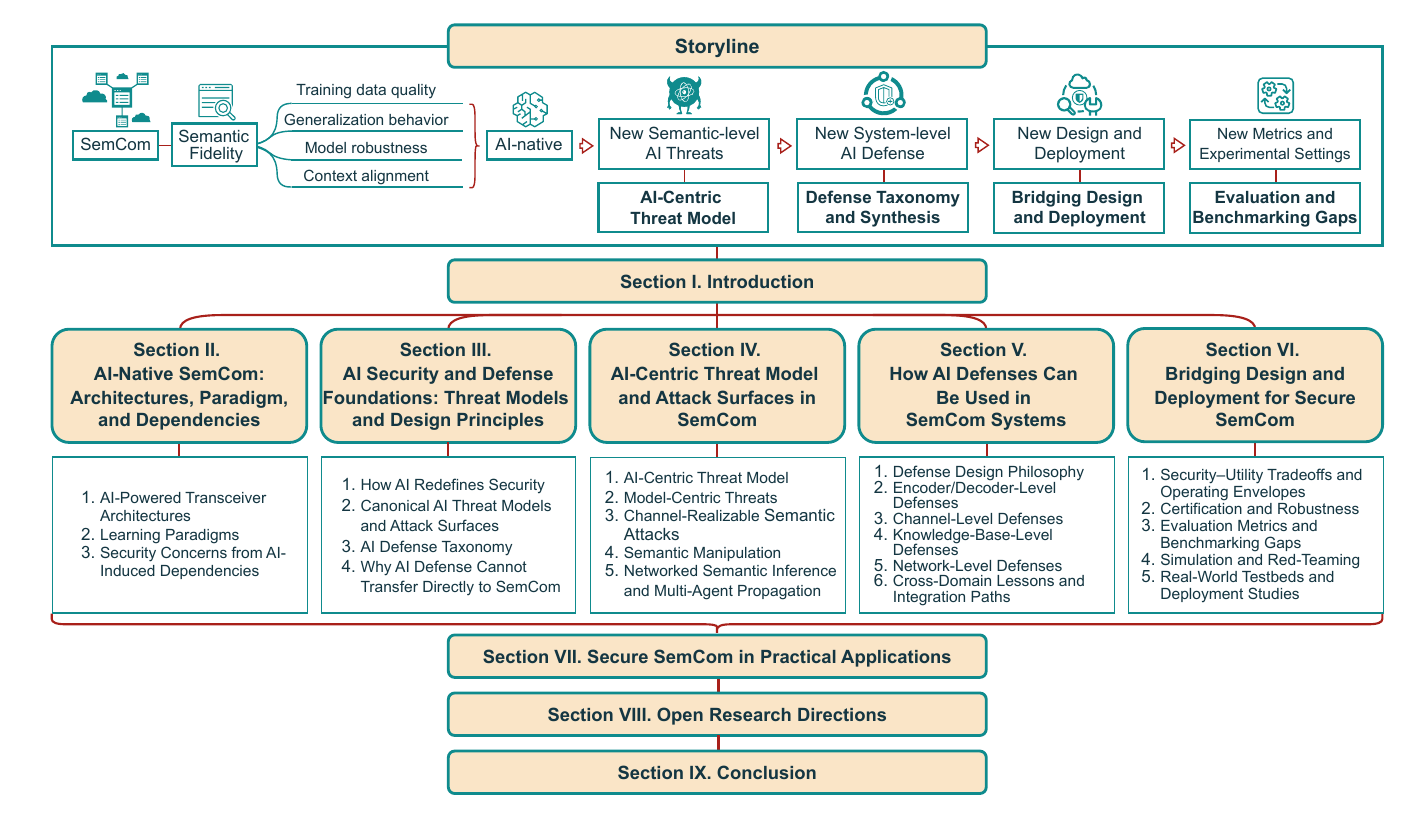}
\caption{The organization structure of this survey paper.}
\label{fig:intro}
\end{figure*}

The organization of the paper is shown in Fig.~\ref{fig:intro}. 
Section~\ref{sec:fundamentals} reviews the architectural foundations of SemCom and analyzes AI-induced dependencies that give rise to new security vulnerabilities. 
Section~\ref{sec:AI-security} formalizes an AI-centric threat model that characterizes semantic attacks targeting learned models, over-the-air transmission, and shared knowledge or contextual inference in SemCom systems.
Section~\ref{sec:4layered_defense} presents a layered defense taxonomy and synthesis, surveying security mechanisms across semantic encoding, transmission, shared knowledge, and networked inference. 
Section~\ref{sec:deployment} bridges design and deployment by examining system-level constraints and introducing security–utility operating envelopes for robust SemCom operation under real-world conditions. 
Section~\ref{sec:applications} surveys representative application domains of secure SemCom across sensing, control, and distributed AI systems.
Finally, Section~\ref{sec:open} outlines open research challenges and future directions toward secure and trustworthy SemCom systems.

\section{AI-Native Semantic Communication: Architectures, Learning Paradigm, and Dependencies} \label{sec:fundamentals}


SemCom departs from classic communication architectures by shifting the design objective from bit-level fidelity to task-level semantic correctness\cite{xie2021deep}. This shift is fundamentally enabled by AI, which learns task-relevant representations and governs how information is extracted, compressed, transmitted, and interpreted under uncertainty. A central architectural abstraction is the AI-powered semantic transceiver~\cite{yang2022semantic,chaccour2024less,guo2024survey}, which consists of three tightly coupled components: a semantic encoder, a semantic decoder, and a shared knowledge base. The semantic encoder maps raw inputs into compact latent representations that preserve task utility, while the semantic decoder reconstructs the intended outcome at the receiver by leveraging shared context or knowledge.

In this section, we focus on AI-based semantic transceivers and knowledge-assisted architectures that govern how meaning is encoded, transmitted, and reconstructed, rather than access network- or network-level architectures, which are beyond the scope of this section. We introduce the architectural foundations and learning paradigms of modern SemCom systems, highlighting the unique dependencies and design affordances that arise from embedding intelligence throughout the communication pipeline. These foundations set the stage for a deeper security analysis in Section~\ref{sec:AI-security}.

\subsection{{AI-Powered Semantic Transceiver Architectures}}\label{subsec:ai-arch}

This architectural abstraction builds on the end-to-end learning paradigm with an autoencoder that represents a complete communication system by jointly learning modulation and demodulation without explicitly separating source coding, channel coding, and modulation stages \cite{oshea2017intro}. Extending this idea beyond symbol transmission, architectures such as DeepSC for text \cite{xie2021deep} and DeepJSCC for images \cite{bourtsoulatze2019deep} demonstrate that learned semantic encoders and decoders can transmit task-relevant representations directly over the channel and recover meaning even under severe channel impairments. In these systems, semantic correctness is determined by the alignment between the encoder–decoder mapping and the task objective, rather than by symbol-level fidelity.

As SemCom architectures mature, the encoder–decoder pipeline is increasingly augmented with explicit semantic knowledge. Knowledge-assisted SemCom incorporates shared priors, such as ontologies, knowledge graphs, or structured semantic memories, directly into the encoding and decoding process. A line of work proposed mapping sentences into knowledge-graph triplets and selectively transmitting semantically important units, enabling receivers to reconstruct meaning even under limited communication resources \cite{jiang2022semantic,wang2023knowledge}. More recent work integrates structured knowledge with neural encoders using graph neural networks (GNNs) \cite{hello2024semantic} and Transformer-based models \cite{liang2025knowledge}, allowing semantic representations to be grounded in entities and relations rather than surface-level symbols . In such architectures, semantic correctness depends not only on the transmitted signal but also on the consistency, integrity, and synchronization of shared knowledge.

Building on these foundations, modern SemCom systems increasingly adopt advanced neural backbones and adaptive structures to cope with dynamic environments and diverse task requirements. Transformer-based joint source–channel coding (JSCC) improves robustness to contextual and syntactic variations in text, while adaptive semantic transceivers dynamically adjust coding strategies to meet latency and resource constraints \cite{liu2023transformer,xu2023deep,liang2025knowledge,zhang2023prompt}. At the frontier, foundation models and generative architectures are being embedded directly into SemCom pipelines, where large language model (LLM) based tokenizers are developed to transmit compact semantic cues and reconstruct rich outputs \cite{wang2025llmsc,salehi2025llm,cheng2024wireless,cheng2024wireless,xia2025generative,liang2024generative}. These generative designs enable extremely high compression ratios and flexible semantic reconstruction, but they also expand the trust boundary of SemCom systems toward learned models and shared semantic priors.

Overall, modern SemCom transceiver architectures are AI-native systems composed of learned semantic-level encoders and decoders, often augmented with shared knowledge or generative models. While these architectures provide substantial gains in efficiency and flexibility, they also introduce new dependencies on learning algorithms, training data, adaptation mechanisms, and external knowledge. These dependencies fundamentally shape system behavior and directly influence robustness and security. These evolving architectures raise new questions about how meaning is learned, updated, and preserved. These questions will be addressed next by examining the learning paradigms that underpin SemCom systems.

\subsection{Learning Paradigms in Semantic Communication}
SemCom systems are fundamentally learning-driven: semantic encoders, decoders, and knowledge modules are realized through data-driven training and adaptation rather than fixed signal mappings. Consequently, the learning paradigm directly determines how meaning is defined, represented, shared, and updated, and thus shapes the dominant security dependencies of a SemCom system.

Most early SemCom designs rely on supervised learning (SL), where encoders and decoders are trained in an end-to-end  manner using labeled data and task-oriented objectives \cite{xie2021deep,bourtsoulatze2019deep}. In this setting, semantic correctness is implicitly tied to the integrity, coverage, and bias of the training data, making learned representations sensitive to data manipulation and distributional mismatch between training and deployment.

Self-supervised learning (SSL) is increasingly adopted to reduce reliance on labeled data and improve generalization. By exploiting intrinsic data structure through contrastive, masked, or multimodal objectives, SSL enables transferable semantic abstractions that can be reused across tasks and environments. At the same time, semantics learned via SSL depend strongly on pretraining corpora and alignment assumptions, introducing new representation-level dependencies~\cite{zhao2025multi,tang2024contrastive,zou2025self}.

Reinforcement learning (RL) arises naturally in SemCom systems operating in dynamic or closed-loop settings, where semantic transmission decisions are optimized with respect to long-term task rewards rather than explicit reconstruction losses~\cite{yan2025review,lu2021rl_semcom,zhao2023joint}. While RL enables adaptive semantic behavior under time-varying conditions, it tightly couples semantics to reward design, feedback fidelity, and exploration dynamics.

Learning paradigms also differ in how training and adaptation are distributed. Centralized training enables global optimization but scales poorly under data locality and privacy constraints, motivating distributed and federated learning approaches \cite{xu2024federated,nguyen2024efficient}. These methods allow collaborative learning of semantic models and shared knowledge without raw data exchange, at the cost of additional dependencies on synchronization, aggregation, and trust.

Finally, many SemCom systems~\cite{si2024post,zhang2025o2sc,zhang2023prompt,liu2024ofdm} rely on online adaptation or continual learning to remain effective after deployment, as channel conditions, tasks, and context evolve over time. While such mechanisms support long-lived operation in non-stationary environments, they also extend the temporal window over which semantic behavior can be influenced.

Together, these learning paradigms define how semantic meaning is formed and maintained in SemCom systems, and explain why vulnerabilities can arise even when lower-layer communication remains reliable. We build on this foundation in the next section by analyzing the resulting threat surfaces from an AI defense perspective.

\begin{figure*}[!tb]
\centering
\includegraphics[width=0.85\textwidth]{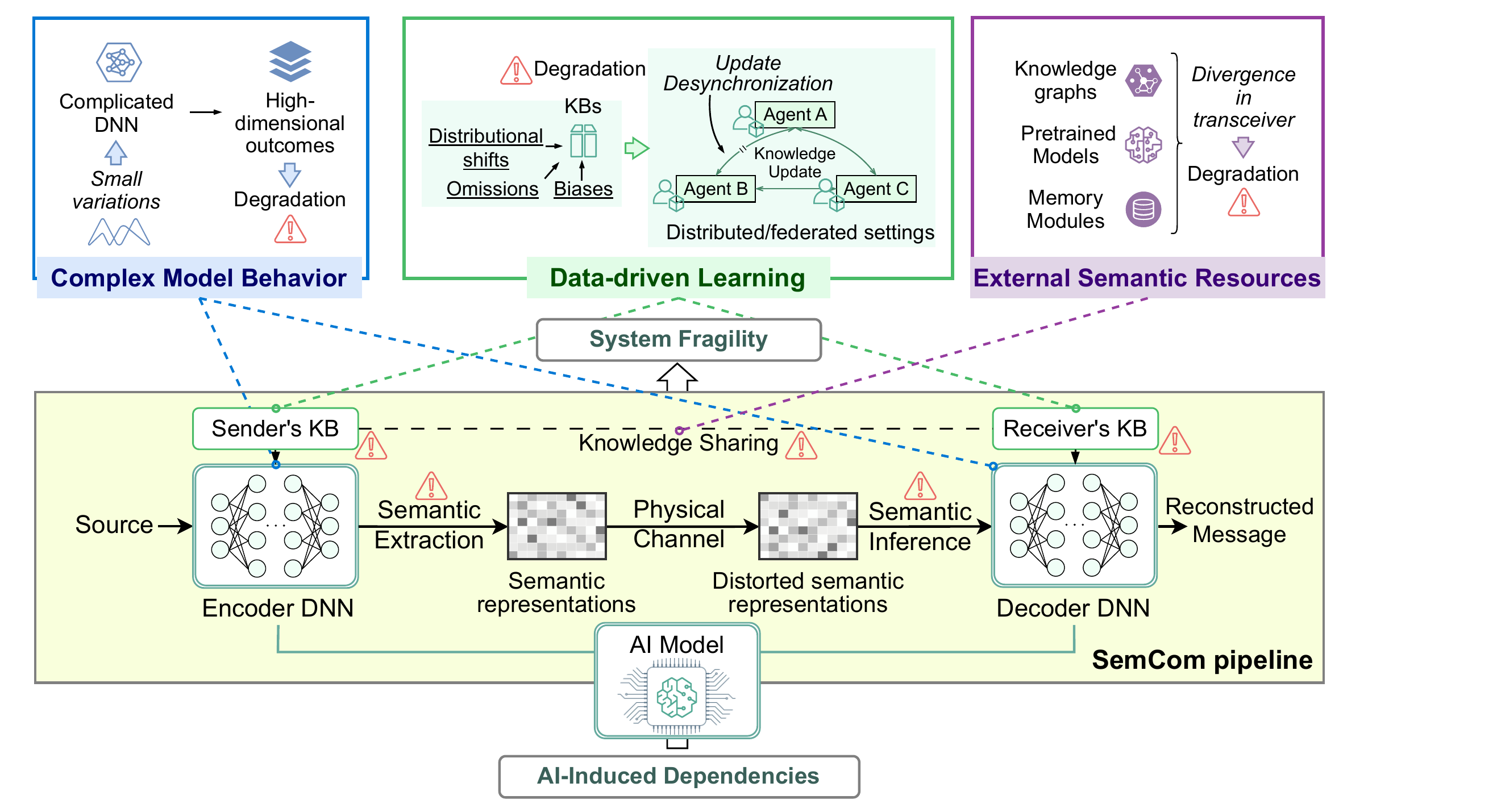}
\caption{System fragility across the AI-native semantic communication inference pipeline: 1) the reliance on complex model behavior, 2) the training corpora to build semantic behavior, and 3) the use of external semantic resources.}
\label{fig:2sec}
\end{figure*}

\subsection{Lessons Learned: Security Concerns in SemCom from AI-Induced Dependencies}
The integration of AI into SemCom  introduces new forms of adaptability and efficiency, but also gives rise to \textbf{security concerns rooted in how meaning is defined, learned, and maintained}. As illustrated in Fig. \ref{fig:2sec}, unlike traditional communication pipelines where information fidelity is grounded in analytically specified encoders and decoders, SemCom systems depend on model-driven inference pipelines, where meaning is shaped by the behavior of trained representations under varying conditions. Three main security concerns in SemCom are summarized in Table \ref{tab:semantic_fragility}.

The first concern is from dependence on complex learned models. Semantic encoders and decoders are typically implemented as deep neural networks trained on high-dimensional data, making their behavior sensitive to subtle variations in input distributions, latent structure, or contextual cues \cite{xie2021deep,bourtsoulatze2019deep}. As a result, even when lower-layer transmission succeeds and syntactic correctness is preserved, small discrepancies between training and deployment conditions can lead to degraded or inconsistent semantic outcomes\cite{liu2025knowledge}. For example, by crafting inputs that exploit vulnerabilities in the semantic encoder, an adversary can induce misleading latent semantic representations even though the channel encoder and physical transmission operate normally, resulting in incorrect semantic reconstruction at the decoder.

Data-driven learning further binds semantic behavior to the properties of the training corpus. Biases, omissions, or distributional skew in training data can manifest as systematic performance degradation at deployment, particularly in dynamic environments. This effect is amplified in systems that incorporate continual learning or online adaptation, where accumulated updates may gradually shift semantic representations away from their original operating regime. In distributed or federated settings, such drift can occur unevenly across agents, leading to desynchronization of shared semantic models. For example, by poisoning the data used to train or update the semantic encoder/decoder pair, an adversary can gradually shift the mapping between raw data and semantic symbols, causing semantic drift that persists even when the channel encoder preserves bit-level fidelity.

Additional concerns comes from reliance on external semantic resources, such as pretrained foundation models, shared knowledge bases, or memory modules. In these architectures, effective semantic reconstruction depends not only on local inference, but also on the consistency and alignment of shared priors across communicating endpoints. Divergence in these priors, even in the absence of explicit transmission failures, can distort interpretation or render received content ambiguous or misleading. For example, tampering with foundation models, shared embeddings, or external memory modules that serve as semantic priors can create mismatches between endpoints, causing ambiguous or misleading semantic decoding even though the channel decoder returns error-free symbols.

Crucially, these AI-induced dependencies do not represent failures or vulnerabilities in isolation. Fidelity is no longer determined solely by symbol-level integrity, but by the coherence of the end-to-end semantic pipeline, encompassing learned model behavior, shared knowledge, and adaptation dynamics. In the next section, we adopt a security-centered perspective that examines how adversarial objectives can interact with these semantic dependencies, and how attacks emerge from the same learned mechanisms.

\begin{table*}[t]
\centering
\caption{Comparison of AI-Induced Security Concerns in SemCom}
\begin{tabular}{|p{0.15\linewidth}|p{0.16\linewidth}|p{0.20\linewidth}|p{0.20\linewidth}|p{0.17\linewidth}|}
\hline
\textbf{Security Concern} & \textbf{Underlying Dependency} & \textbf{Failure Mode in SemCom} & \textbf{Representative Attack Vector} & \textbf{Security Implication} \\ \hline
Dependence on Complex Learned Models & Semantic encoder/decoder implemented as deep learned models & Semantic meaning shifts under distributional or contextual perturbation despite error-free channel delivery & Adversarial input manipulation targeting semantic encoders (e.g., semantic misclassification) & Latent meaning becomes attackable without violating channel-level integrity \\ \hline
Dependence on Data-Driven Learning Corpora & Training data, continual learning updates, federated semantic alignment & Drift or mismatch in semantic mappings, especially across agents or over time & Data poisoning or federated desynchronization attacks & Semantic boundaries move over deployment, silently altering meaning \\ \hline
Dependence on External Semantic Resources & Shared embeddings, knowledge bases, foundation models, memory priors & Inconsistent semantic reconstruction arising from mismatched or tampered shared priors & Ontology manipulation or priors mismatch injection & Meaning reconstruction relies on integrity of shared external semantics \\ \hline
\end{tabular}
\label{tab:semantic_fragility}
\end{table*}

\section{AI Security and Defense Foundations: Threat Models and Design Principles}
\label{sec:AI_security_new}
The observations in Section \ref{sec:fundamentals} highlight a fundamental shift in how correctness and failure arise in SemCom systems. When meaning is defined and reconstructed through learned models, shared knowledge, and adaptive inference pipelines, semantic degradation can occur without explicit transmission errors or protocol violations. These failure modes are not incidental artifacts of implementation, but direct consequences of adopting AI-native designs.

This change raises security questions that differ from classical channel- and protocol-centric settings. If semantic correctness depends on learned behavior, data distributions, and contextual alignment, then adversaries need not tamper with signals or protocols to induce failure. Instead, they may exploit the same mechanisms that enable semantic efficiency and adaptability, including representation learning, training dynamics, and inference-time sensitivity \cite{goodfellow2014explaining,biggio2018wild,yuan2019adversarial}. Reasoning about secure SemCom therefore requires a security framework that goes beyond reliability and explicitly addresses robustness, integrity, and privacy of learned behavior.

To establish this foundation, we step back from SemCom-specific mechanisms and summarize how security is modeled and defended in AI systems more broadly. Importantly, we do so with an explicit view toward communication-centric instantiations of these concepts. For example, threat models that are classically defined over input perturbations or training data in AI naturally map, in SemCom systems, to perturbations of learned joint source–channel representations, poisoning of end-to-end JSCC training pipelines, or manipulation of shared semantic priors between transmitters and receivers. This section therefore introduces core concepts in AI security and defense not as abstract machine learning constructs, but as building blocks that directly inform the threat surfaces and defense strategies of semantic communication systems. These concepts provide the necessary vocabulary and analytical lens for translating AI-induced dependencies into concrete, communication-relevant attack models and defense principles for SemCom.

\subsection{From Reliability to Robustness: How AI Redefines Security}
Traditional communication security is rooted in a reliability-centric abstraction, where correctness is defined by faithful symbol reproduction under noise, interference, and adversarial disruption. Within this paradigm, failures are explicit and observable, manifesting as decoding errors, packet loss, or violated cryptographic checks. Protecting the channel and enforcing protocol-level integrity are therefore sufficient to ensure correct system behavior.

AI-driven systems depart fundamentally from this model. Learned components do not implement analytically specified transformations with predictable error behavior. Instead, they infer high-dimensional decision functions from data, making system behavior sensitive to subtle variations in inputs, context, or operating conditions. As a result, failures may arise even when inputs are syntactically valid, transmission is reliable, and cryptographic protections remain intact \cite{goodfellow2014explaining,biggio2018wild,yuan2019adversarial}. This shift reframes security from a question of reliability to one of robustness.

In AI systems, robustness characterizes the stability of a learned model under admissible perturbations to its inputs. 
Let $(x,y)\sim\mathcal{D}$ denote a data distribution over inputs $x\in\mathcal{X}$ and task labels (or targets) $y\in\mathcal{Y}$. 
Let $f_\theta:\mathcal{X}\rightarrow\mathcal{Y}$ be a learned predictor parameterized by $\theta$, and let 
$\ell:\mathcal{Y}\times\mathcal{Y}\rightarrow\mathbb{R}_{\ge 0}$ denote a task loss.
Given a threat model specified by an admissible perturbation set $\mathcal{S}(x)$, robustness is commonly formalized as the worst-case loss in a neighborhood of $x$:
\[
\mathcal{L}_{\mathrm{rob}}(f_\theta;x,y)
~=~
\sup_{\delta \in \mathcal{S}(x)}
\ell\!\left(f_\theta(x+\delta),\,y\right),
\label{eq:robust_loss}
\]
and the corresponding robust risk is
$R_{\mathrm{rob}}(f_\theta)= \mathbb{E}_{(x,y)\sim\mathcal{D}}
\left[
\mathcal{L}_{\mathrm{rob}}(f_\theta;x,y)
\right].$ Here, the perturbation set $\mathcal{S}(x)$ encodes the assumed adversary model, and robustness is evaluated with respect to the worst-case degradation in task performance rather than symbol-level error.
Robust learning then seeks model parameters that minimize this worst-case risk $R_{\mathrm{rob}}(f_\theta)$, which recovers the standard min--max formulation underlying adversarial training and related robustness methods \cite{madry2018towards}.

This distinction fundamentally alters security objectives. Whereas reliability-centric systems protect symbols and protocols, AI-centric systems must protect learned behavior itself. Robustness seeks to limit worst-case semantic or task degradation under an explicit threat model \cite{madry2018towards,zhang2019theoretically}. Privacy aims to prevent sensitive information about inputs or training data from being inferred through model outputs or representations. Integrity focuses on ensuring that learned models behave consistently with intended objectives despite malicious data manipulation or adversarial interaction.

Crucially, robustness in AI is not an intrinsic or universal property of a model. Instead, it is defined relative to explicit threat models, perturbation budgets, and operational constraints, and it often trades off against accuracy, efficiency, or adaptability \cite{zhang2019theoretically}. Complementary notions such as distributional robustness further emphasize robustness under deployment shift rather than pointwise perturbations \cite{sinha2018certifying}. This perspective aligns directly with SemCom, where semantic fidelity, latency, energy, and robustness must be jointly managed. Understanding this shift from reliability to robustness is therefore essential before instantiating threat models and defenses for SemCom systems \cite{cohen2019certified}.

\subsection{Canonical AI Threat Models and Attack Surfaces}
Security analysis in AI systems begins with an explicit threat model that characterizes what the adversary can access, manipulate, and observe, as well as what constitutes a successful attack. Unlike traditional communication systems, where threat models are often defined at the signal or protocol level \cite{zou2016survey,nichols2001wireless}, AI threat models are centered on learned representations, data-driven behavior, and inference dynamics. This shift leads to attack surfaces that are semantic, statistical, and often indirect.

A first organizing dimension is the \emph{stage of interaction} at which the adversary operates. Training-time threats target the data and procedures used to construct or update models. In supervised or self-supervised learning pipelines, adversaries may inject poisoned samples to bias learned representations or implant backdoors that activate under specific conditions \cite{biggio2012poisoning,gu2017badnets}. In distributed and federated learning settings, which are increasingly common in networked and edge deployments, attackers can further exploit aggregation and synchronization mechanisms, introducing subtle but persistent deviations that are difficult to detect through local validation alone \cite{bhagoji2019analyzing,wang2020attack}. These attacks are particularly concerning because their effects may remain dormant until deployment, at which point mitigation options are limited.

Inference-time threats arise when adversaries interact with a deployed model through its inputs or query interface. Adversarial examples are the most well-known manifestation, where carefully crafted inputs induce incorrect predictions while remaining perceptually or syntactically valid \cite{goodfellow2014explaining}. Beyond worst-case perturbations, inference-time attacks also include semantic manipulations such as paraphrasing, reordering, or context injection that exploit inductive biases of learned representations \cite{ilyas2019adversarial}. Because inference-time attacks do not require access to training data or model internals, they are especially relevant in open or service-based deployments.

A second dimension concerns \emph{information exposure and leakage}. Learned models often encode rich information about their training inputs in internal representations and outputs. Adversaries can exploit this through membership inference, model inversion, or representation probing to extract sensitive attributes or reconstruct training samples \cite{shokri2017membership,fredrikson2015model}. Even when models are queried through restricted interfaces, repeated interaction may enable model extraction attacks that approximate decision boundaries or replicate functionality \cite{tramer2016stealing}.

From a security objectives perspective, these dimensions can be interpreted—rather than replaced—through the classical confidentiality–integrity–availability (CIA) triad \cite{nist80012}. Information exposure and leakage directly relate to confidentiality, training- and inference-time manipulation primarily threaten semantic integrity, and adaptation- or feedback-driven attacks can undermine availability by inducing cascading or persistent semantic failure \cite{huang2011adversarial,barreno2010security}. However, unlike traditional systems, these objectives are often violated in SemCom without explicit service disruption or syntactic corruption \cite{papernot2016towards}, motivating a threat taxonomy that emphasizes where and how learned behavior is exploited rather than categorizing attacks solely by security objectives.

A third dimension captures \emph{adaptation and feedback-driven threats}. Many AI systems operate in closed-loop settings, where model outputs influence future inputs, data collection, or learning updates. RL, online adaptation, and continual learning introduce feedback paths that adversaries can exploit by shaping observations, rewards, or environment dynamics \cite{huang2017adversarial}. Such attacks need not cause immediate failure. Instead, they may induce gradual policy drift, degraded exploration, or biased adaptation over time, making them difficult to attribute to malicious behavior.

Across these dimensions, AI threat models are commonly classified by the adversary’s access assumptions. White-box adversaries possess full knowledge of model architecture and parameters, gray-box adversaries observe partial internal signals, and black-box adversaries interact only through inputs and outputs. In SemCom systems, these access assumptions correspond to different degrees of visibility into the semantic pipeline rather than direct access to symbols or protocols. For instance, adversarial access may range from knowledge of learned semantic mappings, to observation of intermediate semantic signals or adaptation feedback, to interaction only through end-to-end semantic inputs and task-level outputs. Importantly, many effective AI attacks operate under gray- or black-box assumptions, exploiting transferability, query access, or statistical side channels rather than explicit parameter access \cite{papernot2017practical}. We defer a detailed SemCom-specific instantiation of these access models to Section~\ref{sec:AI-security}, where they are embedded into an AI-centric threat model tailored to semantic communication pipelines.

Taken together, these threat models operationalize the departure from classical security abstractions by exposing how learned behavior becomes an attack surface. Attacks on AI systems do not necessarily violate syntactic correctness, protocol compliance, or signal integrity. Instead, they exploit the statistical and semantic structure of learned behavior, often leveraging the same mechanisms that enable generalization and adaptability, and that underpin semantic efficiency in communication pipelines.

\subsection{AI Defense Taxonomy: Preventive, Reactive, and Certifiable Defenses}\label{sec:IIIC}
Defenses for AI systems are designed in response to these threat models and are typically organized by when and how they intervene in the learning and inference pipeline. Unlike traditional security mechanisms that enforce correctness through fixed rules or protocol guarantees, AI defenses aim to shape, monitor, or constrain learned behavior under uncertainty. Defense effectiveness is therefore tied to explicit assumptions about adversarial capabilities, perturbation sets, and system constraints, with the learned model serving as the primary unit of protection.

A first category consists of \emph{preventive defenses}, which are applied prior to deployment to reduce vulnerability to anticipated attacks. Representative examples include adversarial training and its variants \cite{madry2018towards,zhang2019theoretically}, robust augmentation \cite{hendrycks2020augmix,zhang2018mixup,yun2019cutmix,cubuk2019autoaugment}, and representation-level regularization that enforces stability through Jacobian or Lipschitz control \cite{rifai2011contractive,ross2018improving,yoshida2017spectral,cisse2017parseval,tsuzuku2018lmt}. Distributionally robust objectives further improve resilience under deployment drift rather than pointwise perturbations \cite{sinha2018certifying}. Preventive defenses are attractive because they incur no runtime overhead, but their effectiveness depends critically on how well the assumed threat model matches the deployment conditions of the model.

A second category comprises \emph{reactive defenses}, which operate during inference or system execution. These defenses monitor model behavior and intervene when anomalous or suspicious conditions are detected, using detection tests, reconstruction checks, or input transformations \cite{ma2018lid,meng2017magnet,xu2018featuresqueezing,guo2018countering,xie2018randomization}. However, reactive defenses are often brittle under adaptive attackers that explicitly optimize against the detector or the transformation, making careful evaluation essential \cite{carlini2017detecting,athalye2018obfuscated}. In addition, reactive defenses introduce latency and computation overhead, which can be prohibitive in resource- and latency-constrained inference settings.

A third category focuses on \emph{certifiable and verifiable defenses}, which provide formal guarantees on model behavior under explicitly defined perturbation sets. Certified robustness via randomized smoothing yields probabilistic guarantees under additive noise models \cite{salman2020denoised,cohen2019certified}. Verification and bound-propagation methods provide deterministic certificates for certain architectures by propagating relaxations through the network \cite{gowal2018ibp,wong2018provable,zhang2019crownibp}. While such guarantees are attractive in safety- and mission-critical settings, they often trade coverage and tightness for rigor and scalability.

These defense categories are not mutually exclusive. Preventive defenses can reduce the frequency and severity of failures encountered at runtime, reactive defenses can handle residual or unmodeled threats, and certifiable components can serve as safety envelopes when failure is unacceptable. This compositional view is especially important in systems that operate under dynamic conditions, where adversarial strategies, data distributions, and system objectives may evolve over time. These categories originate from model-centric AI security and treat the learned model as the primary unit of protection; in Section \ref{sec:4layered_defense}, we reinterpret this taxonomy under the system-level constraints, physical channels, and cross-layer dependencies unique to SemCom.

\subsection{Lessons Learned: Why AI Defense Cannot Transfer Directly to Semantic Communication}
Despite the maturity of AI security research, defense techniques developed for standalone learning systems cannot be applied directly to SemCom systems. This gap arises not from missing mechanisms, but from a fundamental mismatch in system assumptions. Classical AI defense frameworks typically consider a single model operating on well-defined inputs and producing outputs that can be evaluated in isolation. SemCom systems, by contrast, embed learned models within a tightly coupled pipeline that spans physical transmission, shared knowledge, and distributed inference, where failures emerge from cross-layer interactions rather than from any single component.

A primary challenge stems from \textit{the presence of the physical communication channel}. Most AI defenses assume direct access to model inputs and outputs, with perturbations defined in abstract feature or input spaces. In SemCom, however, semantic representations are transmitted over noisy and bandwidth-constrained wireless channels, where perturbations are shaped by propagation effects, interference, and protocol dynamics. As a result, admissible attacks are neither arbitrary nor norm-bounded in the conventional sense. Instead of unconstrained $\ell_p$ perturbations applied directly to model inputs, semantic perturbations in SemCom must be realized through physical channels and protocol operations, and are therefore constrained by noise statistics, bandwidth, modulation, and coding \cite{bourtsoulatze2019deep,oshea2017intro}. Defenses that rely on carefully constructed adversarial examples or certified input neighborhoods may thus fail to capture the structure of channel-realizable semantic distortions, limiting their effectiveness when deployed over the air.

A second challenge arises from \textit{the use of shared knowledge and contextual priors}. Many AI defense techniques implicitly assume a fixed and trusted context under which inference is performed. SemCom systems violate this assumption by design. Transmitters and receivers may rely on external knowledge bases, pretrained models, or contextual memories that evolve over time and may not remain perfectly synchronized. Attacks that manipulate, poison, or desynchronize this shared context can induce semantic failure without directly affecting the learned encoder or decoder \cite{li2025synchronizing,hoang2025adversarial}. Such failure modes fall outside the scope of most existing AI defenses, which focus on protecting individual models rather than the integrity and alignment of distributed semantic priors.

\textit{Resource and latency constraints} further complicate defense deployment. AI defenses are often evaluated under offline or compute-rich settings, where increased model complexity, ensemble inference, or repeated sampling is acceptable. SemCom systems, particularly those operating at the edge or in real-time control loops, must satisfy stringent constraints on latency, energy, and bandwidth \cite{liang2025knowledge,liu2023transformer,xu2023deep}. Defensive mechanisms that significantly increase computation, communication overhead, or decision delay may negate the efficiency gains that motivate SemCom in the first place. This tension fundamentally alters what constitutes a feasible and deployable defense in SemCom systems.

Crucially, SemCom systems exhibit emergent failure modes driven by feedback and interaction. Many SemCom deployments involve closed-loop operation, multi-agent coordination, or adaptive rate and representation selection based on inferred semantic confidence. In such settings, small semantic distortions introduced at one point in the pipeline can be amplified through feedback, propagated across agents, or reinforced through adaptation. Traditional AI defenses, which are largely evaluated on single-shot inference tasks, are ill-equipped to reason about these cascading, temporal, and collective effects.

Taken together, these challenges underscore that securing SemCom is not a matter of directly importing AI defense techniques, but of reinterpreting them within a system-level framework. Preventive, reactive, and certifiable defenses must be redesigned to account for physical channels, shared knowledge, resource constraints, and distributed inference. This observation motivates the need for SemCom-native threat models and defense taxonomies, which we develop in the next section to explicitly capture where semantic failures arise and how robustness can be enforced across the entire communication pipeline.


\begin{figure*}[!t]
\centering
\includegraphics[width=1.02\textwidth]{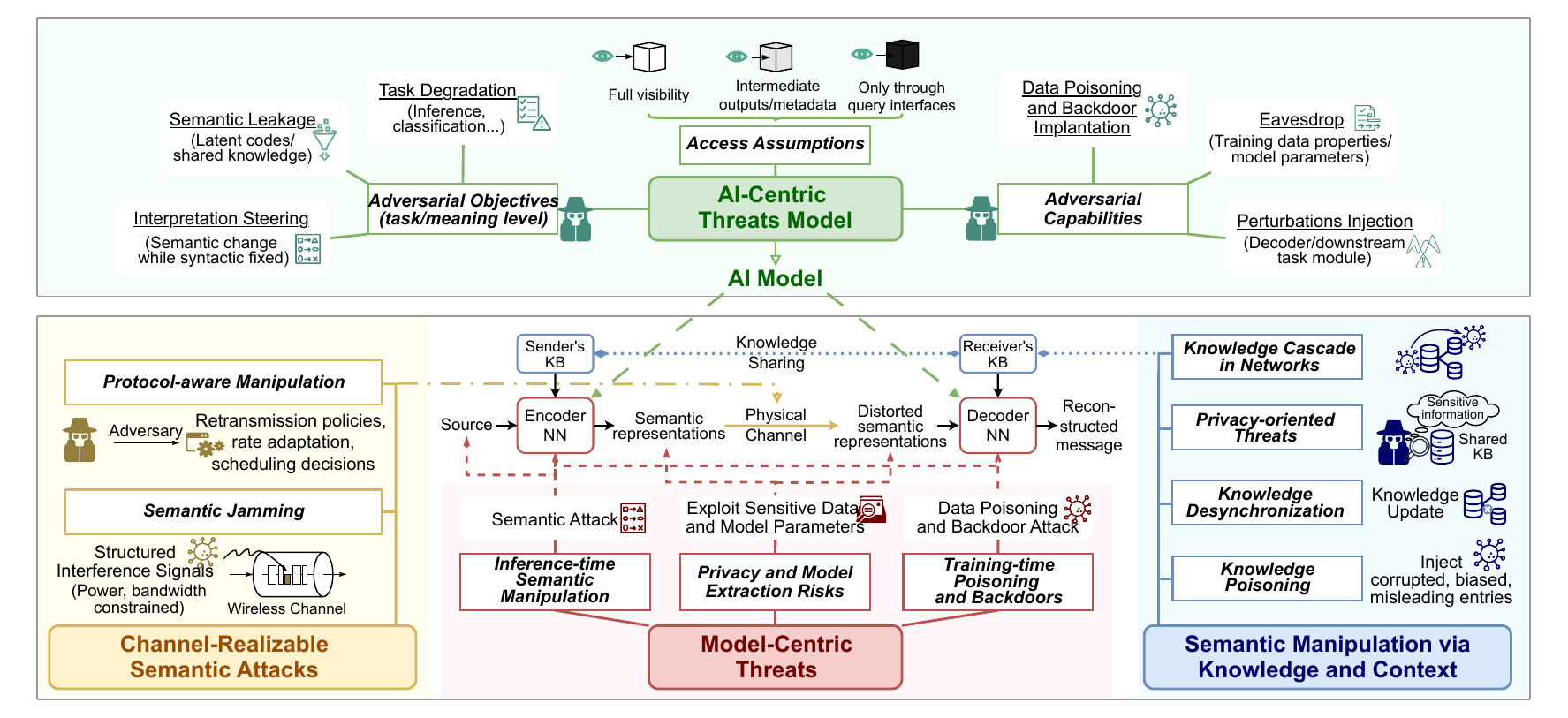}
\caption{Threat model and attack surfaces in semantic communication. \textit{Top}: an AI-centric threat model over adversarial objectives, adversarial capabilities, and access assumptions; \textit{Bottom}: how threats manifest differently across model, channel, and knowledge layers.}
\label{fig:sec3}
\end{figure*}

\section{AI-Centric Threat Model and Attack Surfaces in Semantic Communication}
\label{sec:AI-security}
Building on the AI-centric security foundations introduced in Section \ref{sec:AI_security_new}, we now formalize a threat model tailored to SemCom systems. 
Rather than focusing on symbol corruption or protocol violations, this model characterizes adversarial actions that target learned semantic representations, shared contextual priors, and adaptive inference behavior across the communication pipeline as illustrated in Fig.~\ref{fig:sec3}. We begin by identifying adversarial assumptions specific to SemCom pipelines, including attacker capabilities related to model training, knowledge sharing, inference behavior, and feedback loops. We then introduce a taxonomy of semantic attack surfaces, organized by the system layers or dependencies they target, ranging from encoder–decoder representations and context-dependent adaptation mechanisms to the integrity of shared semantic priors. 


\subsection{AI-Centric Threat Model}
We adopt an AI-centric threat model that reflects the unique vulnerabilities introduced by learning-based semantic communication. Unlike conventional threat models that target symbol-level integrity or channel confidentiality, the emphasis here is on compromising the interpretation of meaning itself. In SemCom systems, adversaries aim to induce failures at the semantic level such as shaping, revealing, or corrupting the conveyed meaning, while potentially leaving lower-layer communication metrics such as packet delivery and bit error rates unaffected \cite{yang2024secure,meng2025securesemcom}.

\textbf{Adversarial objectives.} The adversary's goals are defined at the task or meaning level, rather than at the physical or syntactic levels. Three primary objectives arise: (1) \textit{task performance degradation}\footnote[1]{Throughout this paper, the term \emph{task} is used in a broad sense and includes traditional data reconstruction objectives (e.g., minimizing reconstruction error or bit error rate) in conventional communication systems as special cases.}, inducing incorrect semantic inference, misclassification, or failure in downstream decision-making~\cite{sagduyu2023semantic,dreossi2018semantic}; (2) \textit{semantic leakage}, extracting sensitive information from latent codes or shared knowledge \cite{chen2023model}; and (3) \textit{interpretation steering}, shaping the receiver’s understanding toward attacker-chosen semantics without altering the syntactic structure~\cite{li2025synchronizing,hoang2025adversarial}. These objectives may manifest as explicit task failures or silent behavioral shifts and are especially dangerous in applications where decisions are made autonomously based on semantic understanding.

\textbf{Adversarial capabilities.} To achieve these goals, adversaries may exploit vulnerabilities across different phases of a SemCom pipeline. At inference time, they may manipulate semantic inputs using adversarial perturbations that preserve syntactic validity while inducing misinterpretation by the decoder or downstream task module \cite{sagduyu2023semantic}. These attacks often bypass integrity checks, highlighting a mismatch between semantic and bit-level robustness \cite{biggio2018wild}. During model development or online adaptation, adversaries may poison training data \cite{peng2024adversarial}, implant backdoors, or compromise the integrity of federated or continual learning processes used to refine semantic models and shared knowledge bases \cite{sagduyu2023vulnerabilities,guo2025persistent}. Other attacks exploit model observability: by issuing structured queries or monitoring outputs, adversaries may reconstruct latent representations, infer training data properties, or extract model parameters.


\textbf{Access assumptions.} In SemCom systems, adversarial access is shaped by how semantic encoders/decoders, channels, and shared knowledge are exposed and coordinated \cite{yang2024secure,meng2025securesemcom}. In \textit{white-box} SemCom settings, an adversary may have access to semantic encoder or decoder architectures, pretrained weights, or update mechanisms used for semantic adaptation or continual learning, consistent with canonical AI threat models \cite{yuan2019adversarial}. \textit{Gray-box} access commonly arises when the adversary can observe intermediate semantic representations, confidence scores, channel-aware adaptation signals, or metadata exchanged for coordination, rate control, or semantic alignment \cite{papernot2017practical}. In \textit{black-box} SemCom scenarios, the adversary interacts only through semantic inputs and outputs, such as transmitted semantic features, decoded task outputs, or query-based access to semantic services, where attacks exploit transferability and query-based probing rather than parameter access \cite{papernot2017practical}.

Importantly, SemCom attacks need not target isolated components. Because meaning is reconstructed through the joint operation of learned models, channels, and contextual priors, partial access at one interface can exploit cross-module dependencies, such as encoder–decoder mismatch, context desynchronization, or biased adaptation. 

Throughout this survey, we focus on semantic threat scenarios that persist even when conventional link- and physical-layer protections such as encryption, authentication, or error correction, function correctly. This framing underscores that semantic attacks exploit vulnerabilities in meaning-making rather than signal delivery, and motivates the need for semantic-aware defenses beyond traditional communication security.

\subsection{Model-Centric Threats to Learned Semantics}
At the core of SemCom systems lie learned models that encode, decode, and interpret semantic information. Because these models determine how meaning is abstracted, compressed, and reconstructed, they form a primary attack surface in semantic communication pipelines \cite{yuan2019adversarial,li2023trustworthy,wei2025trustworthy,wen2023survey}. Model-centric threats arise from the fact that semantic representations are learned from data and optimized for task objectives, rather than being analytically specified or symbolically verified.

\textit{Inference-time semantic manipulation.}
At inference time, adversaries may craft inputs that are syntactically valid and perceptually plausible, yet induce incorrect task-level semantics at the decoder or downstream inference module. Unlike classical adversarial examples that focus on pixel- or waveform-level perturbations, semantic manipulation often exploits higher-level structure such as paraphrasing, reordering, or contextual ambiguity to steer meaning while preserving surface validity \cite{sagduyu2023semantic,dreossi2018semantic}. Even without access to model internals, repeated queries can enable black-box probing of fragile semantic regions, exposing sensitivity in learned representations.

\textit{Training-time poisoning and backdoors.}
When semantic encoders and decoders are trained or adapted from data, adversaries may inject poisoned samples that bias the learned semantic space \cite{sagduyu2023vulnerabilities,zhou2024backdoor}. Clean-label poisoning can gradually distort semantic boundaries without degrading nominal accuracy, while backdoor attacks implant hidden triggers that activate attacker-chosen semantic behavior under specific conditions. Because SemCom models are often trained end to end and reused across tasks or deployments, such training-time compromises can propagate through the communication pipeline and affect multiple downstream functions.

\textit{Privacy leakage and model extraction.}
Semantic representations typically compress rich, high-level information, which creates risks of unintended semantic leakage. Latent codes may encode sensitive attributes of inputs or training data, enabling membership inference, model inversion, or representation probing attacks \cite{chen2023model,tang2025towards,yuan2022membership}. In deployed SemCom services, repeated semantic queries can further facilitate model extraction, allowing adversaries to approximate semantic encoders or decoders and replicate or manipulate their behavior. 

In SemCom, learned models act as intermediate semantic interfaces within an end-to-end communication pipeline, so privacy leakage, extraction, and manipulation can have system-level consequences beyond a single inference task.
As a result, model-level attacks need not cause immediate misclassification to be effective. Small distortions in semantic representations can propagate through channel transmission, knowledge-assisted reconstruction, and downstream inference, leading to amplified or delayed semantic failures even when lower-layer communication remains reliable. Moreover, because semantic encoders and decoders are jointly optimized with communication objectives and often reused across tasks, model-centric attacks in SemCom can simultaneously affect compression efficiency, robustness to channel impairments, and task-level correctness. This coupling motivates examining additional threat surfaces that arise when semantic models interact with channels, shared knowledge, and networked inference, which we discuss next.


\subsection{Channel-Realizable Semantic Attacks}\label{sec:2c_channelattack}
Beyond direct manipulation of learned models, SemCom systems are vulnerable to attacks that exploit the physical communication channel while remaining within realistic signal, power, and protocol constraints. Unlike classical jamming or interference, which aim to disrupt symbol decoding or synchronization, semantic attacks over the channel target the learned semantic inference process at the receiver.

\textit{Channel-constrained semantic perturbations.}
In SemCom, channel noise and interference are not merely impairments to be corrected; they directly interact with the learned semantic decoders that map received signals to semantic representations \cite{liu2024manipulating,rong2025semantic,fallahreyhani2024countering}. As a result, adversaries can craft channel-realizable perturbations that leave conventional metrics such as bit error rate, frame error rate, or CRC checks largely unaffected, yet cause substantial degradation in task-level semantic outcomes.

A prominent class of threats is semantic jamming \cite{tang2023gan_jamming,zhou2025rome,chen2025coding_jamming}, in which the adversary injects carefully structured interference signals constrained by power, bandwidth, and spectral masks. Rather than maximizing energy or randomness, the adversary optimizes interference to distort specific semantic features extracted by the decoder. For example, small waveform perturbations may selectively alter latent features corresponding to key entities, labels, or commands, leading to misclassification, mistranslation, or incorrect control actions even when packets are successfully decoded.

\textit{Protocol-aware semantic manipulation.}
Another attack vector arises from protocol-aware manipulation \cite{zhou2022adaptive,gong2023adaptive}, where the adversary exploits retransmission policies, rate adaptation, or scheduling decisions that are coupled with semantic confidence or task feedback. By subtly degrading semantic quality without triggering retransmissions or fallback modes, an attacker can induce persistent semantic drift while avoiding detection by lower-layer reliability mechanisms.

Importantly, channel-realizable semantic attacks expose a fundamental mismatch between syntactic reliability and semantic integrity. Because semantic decoders are trained under assumed channel models, adversarial perturbations that exploit modeling gaps or worst-case channel realizations can push received signals into regions of the semantic decision space that were rarely observed during training. This highlights that robustness to random noise does not imply robustness to adversarial, structure-aware channel manipulation.

\subsection{Semantic Manipulation via Shared Knowledge and Context} \label{sec:IIID_knowlegdeThreat} 
A defining feature of modern SemCom architectures is the use of shared knowledge, contextual priors, and coordinated inference across multiple agents to guide semantic encoding and decoding \cite{zhou2022adaptive,liu2025knowledge}. Knowledge bases, pretrained models, ontologies, and external memory modules enable aggressive semantic compression and robust reconstruction at the link level, while networked deployment allows semantic information to be reused, aggregated, and refined across nodes. At the same time, these capabilities introduce a broad and underexplored attack surface that extends beyond individual encoders or channels.

\textit{Knowledge poisoning and semantic prior manipulation.}
At the knowledge and context level, adversaries may target the integrity, availability, or alignment of semantic priors used by transmitters and receivers \cite{yang2022semantic,chaccour2024less}. Unlike attacks on encoders or physical channels, such manipulations may not modify transmitted signals. Instead, they alter how meaning is grounded and inferred. Knowledge poisoning attacks inject corrupted, biased, or misleading entries into a semantic knowledge base, causing systematic misinterpretation of transmitted representations \cite{peng2024adversarial,zhou2024stealthy,hoang2025adversarial}. 
Because semantic encoders rely on knowledge to decide what information to transmit and semantic decoders rely on the same priors to reconstruct missing details, even small perturbations to shared knowledge can silently bias both sides of the communication.

\textit{Context desynchronization and semantic divergence.}
SemCom also implicitly assumes sufficient alignment between the contextual priors held by the transmitter and receiver. Adversaries may exploit update delays, partial synchronization, or version mismatches to induce semantic divergence, where identical transmitted representations are interpreted differently across endpoints~\cite{li2025synchronizing,hoang2025adversarial,ren2024knowledge}. Such desynchronization attacks are particularly subtle in distributed systems, as similar effects may arise from benign context drift, making malicious manipulation difficult to distinguish from natural system evolution.

\textit{Inference-time semantic leakage.}
Privacy-oriented threats further arise when semantic reconstruction depends on rich contextual knowledge. Adversaries who gain access to shared knowledge, or who can issue queries to semantic models conditioned on that knowledge, may infer sensitive information that was never explicitly transmitted~\cite{han2025manipulating,tang2025towards}. In these cases, semantic leakage occurs through inference rather than communication, blurring the boundary between data exposure and semantic reasoning.

\subsection{Networked Semantic Inference and Multi-Agent Propagation}
\label{sec:IIIE_networkThreat}
Beyond individual links, SemCom systems are frequently deployed in multi-agent and networked settings, where semantic information is relayed, aggregated, or jointly processed across nodes~\cite{he2025invisible,yang2024inviins,li2023catfl}. In such architectures, semantic inference is no longer localized to a single transmitter–receiver pair, but emerges from collective processing, shared models, and feedback-driven coordination among multiple agents \cite{guo2024survey}. This networked inference paradigm enables scalability and robustness, but also introduces a distinct and underexplored attack surface.

\textit{Propagation and amplification of semantic corruption.}
In networked SemCom systems, attacks need not compromise a single component to be effective. Corrupted semantic representations, biased contextual updates, or manipulated inference outputs introduced at one node may propagate through shared models, cooperative inference, or adaptive feedback loops \cite{shi2021semantic,uysal2022semantic}. Since downstream agents may treat received semantic information as reliable context, even small perturbations can be amplified through collective decision processes \cite{he2019model,ding2024patrol}, leading to correlated errors, cascading misinterpretations, or emergent misbehavior at the system level.

\textit{Coordination-level manipulation and collective misalignment.}
Adversaries may further exploit coordination protocols, consensus mechanisms, or task-sharing strategies that govern how agents fuse semantic information \cite{shen2021coordinated,zhao2020shielding}. By selectively influencing a subset of agents or disrupting coordination signals, an attacker can induce semantic disagreement, delayed convergence, or biased collective inference without necessarily attacking individual semantic encoders or channels. Such effects are particularly difficult to diagnose, as failures manifest only at the global system level and may resemble benign coordination noise or nonstationary environments.

Networked semantic inference threats highlight that security in SemCom cannot be assessed solely at the level of individual models or links. 
When meaning is inferred through distributed, cooperative processes, semantic integrity becomes a collective property of the system, requiring defenses that explicitly account for propagation dynamics, inter-agent trust, and coordination robustness.

\begin{figure*}[!t]
\centering
\includegraphics[width=0.85\textwidth]{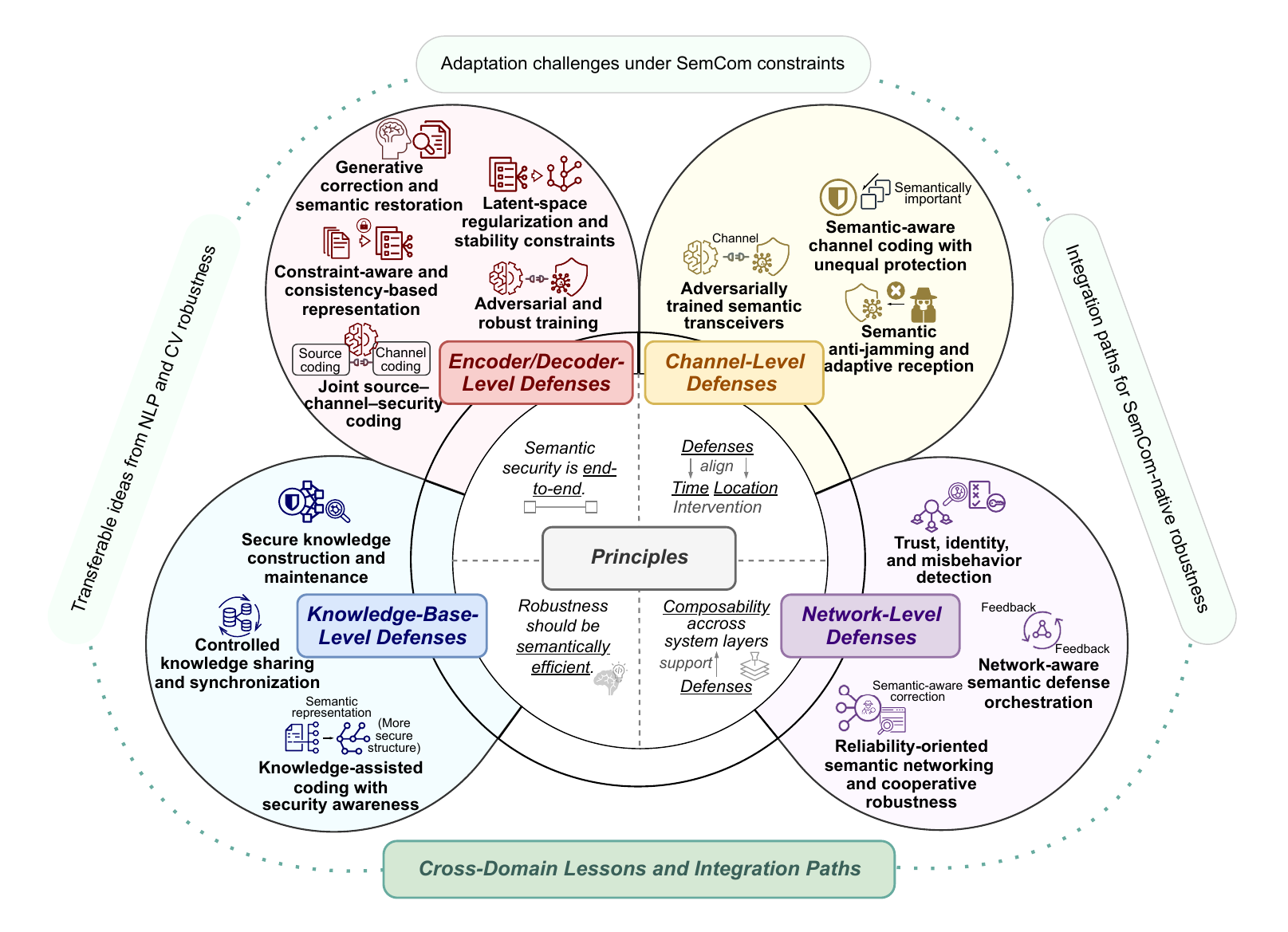}
\vspace{-1em}
\caption{Layered defenses for semantic communication security. According to how and where they intervene in the communication pipeline, defenses are organized into five categories: encoder/decoder level, channel level, knowledge-base level, network level, and cross-domain integration. The layered structure highlights how AI-centric defenses must be instantiated and composed across multiple SemCom subsystems to provide end-to-end semantic security.}
\label{fig:sec4}
\end{figure*}

\subsection{Lessons Learned}
The threats surveyed in this section reveal that vulnerabilities in semantic communication systems arise fundamentally from their AI-native design rather than from unreliable channels or protocol violations. By targeting learned representations, training and adaptation dynamics, shared semantic priors, and networked inference processes, adversaries can induce semantic failures while preserving syntactic correctness and lower-layer integrity. This decoupling between semantic integrity and conventional communication metrics challenges traditional security abstractions based on bit-level reliability, encryption, and authentication.

A central lesson is that semantic failures rarely originate from isolated components. Instead, they emerge from interactions among models, channels, knowledge resources, and coordination mechanisms, where seemingly localized or benign perturbations can propagate and amplify through cross-module dependencies. As a result, assessing SemCom security requires reasoning about semantic inference as a coupled, end-to-end process rather than as a collection of independently secured blocks.

Another key insight is that robustness, security, and privacy in SemCom are tightly intertwined. Design choices that improve semantic efficiency through aggressive compression, context reuse, or cooperative inference often increase exposure to poisoning, leakage, and misalignment risks. This tension underscores the need to analyze security as a first-class design objective, shaping architectural choices and operational trade-offs.

Finally, these observations indicate that defending semantic communication systems requires mechanisms that are aware of meaning, context, and coordination, rather than solely signal fidelity or protocol correctness. \emph{We defer a systematic discussion of defenses and mitigation strategies to the following sections.}

\section{How AI Defenses Can Be Used in Semantic Communication Systems} \label{sec:4layered_defense}

As discussed in Section~\ref{sec:AI-security}, SemCom enables attacks that target meaning, leakage, or misalignment while leaving syntactic integrity and lower-layer metrics largely intact. Defending SemCom therefore requires system-level protections that span semantic transceivers, channels, shared knowledge, and networked coordination. Following Fig.~\ref{fig:sec4}, we organize defenses by intervention interface and feasibility under rate, latency, energy, and deployment constraints, and summarize design principles in Table~\ref{tab:design_principle} and the taxonomy in Table~\ref{tab:defense_taxonomy}.

\begin{table*}[t]
\caption{Design Principles for AI Defense in Semantic Communication Systems}
\label{tab:design_principle}
\centering
\renewcommand{\arraystretch}{1.15}
\footnotesize

\begin{tabular}{%
p{2.5cm}   
p{2.6cm}   
p{5.0cm}   
p{5.8cm}   
}
\hline
\textbf{Principle} &
\textbf{Defense Focus} &
\textbf{Key Observation} &
\textbf{Design Implication} \\
\hline
\textbf{P1: End-to-end semantic security} &
System-level correctness &
Semantic correctness arises from interactions among learned and non-learned components across the entire SemCom pipeline. &
Defenses must address vulnerabilities at semantic encoders/decoders, channel effects, shared knowledge, and multi-agent coordination, rather than protecting individual models in isolation. \\
\hline
\textbf{P2: Time--location aligned defense} &
Intervention alignment &
Defenses differ in \emph{when} they act (training-time, inference-time, certified) and \emph{where} they intervene within the pipeline. &
Effective protection requires mapping defense mechanisms to concrete intervention points at the model, channel, knowledge-base, and network levels. \\
\hline
\textbf{P3: Semantically efficient robustness} &
Cost-aware robustness &
SemCom operates under strict constraints on rate, latency, energy, and computation. &
Defense mechanisms should maximize semantic fidelity or task robustness per unit cost, avoiding excessive redundancy or model complexity that negates SemCom efficiency gains. \\
\hline
\textbf{P4: Layered and composable defenses} &
Cross-layer protection &
No single defense is sufficient to secure system-level semantics in SemCom. &
Layered defenses across pipeline stages enable complementary protection, adaptive responses, and safety envelopes for critical components. \\
\hline
\end{tabular}
\end{table*}

\subsection{Defense Design Philosophy for Semantic Communication} \label{sec:principle}
Traditional AI defenses are typically developed in a model-centric setting, where the objective is to protect a single learned function against adversarial manipulation, data poisoning, or information leakage. 
In contrast, SemCom systems embed learned components within a broader pipeline that couples semantic encoding and decoding with wireless transmission, shared knowledge, and distributed decision-making. 
As a result, defenses that are effective for standalone models do not directly translate to SemCom without reinterpretation at the system level. 
In SemCom, the security objective is not to preserve intermediate representations, but to preserve task-relevant meaning after the entire pipeline has acted on the message.

Designing defenses for SemCom requires reasoning along two fundamental dimensions. The first concerns \emph{when} a defense intervenes in the pipeline. Defenses may be preventive, such as training-time hardening of semantic representations; reactive, such as inference-time detection and mitigation; or certifiable, providing formal guarantees under explicit threat models, as discussed in Section~\ref{sec:AI_security_new}. The second concerns \emph{where} the defense intervenes. Because semantics is created, transformed, and consumed across multiple interfaces, natural intervention points arise at the encoder/decoder, channel, knowledge-base, and network levels. Effective protection therefore requires mapping AI defense mechanisms to these concrete system interfaces, rather than treating security as an add-on to a single model.

Beyond this time–location alignment, SemCom introduces domain-specific constraints that fundamentally shape defense feasibility. Transmission is inherently lossy: noise, interference, and bandwidth limits can irreversibly distort semantic representations. Many SemCom designs rely on shared or external knowledge, introducing risks of poisoning, desynchronization, and context leakage. Moreover, SemCom systems often operate under strict rate, latency, energy, and computation budgets, making heavyweight redundancy or expensive inference impractical. Finally, semantic failures can be subtle: outputs may remain fluent or syntactically valid while becoming task incorrect or unsafe.

These constraints motivate four SemCom-specific principles in Table~\ref{tab:design_principle}: end-to-end semantic security (P1), time–location aligned defense (P2), semantically efficient robustness (P3), and layered composability (P4). 
Guided by these principles, we organize AI defenses for SemCom along the same pipeline through which semantic information is generated, transmitted, contextualized, and propagated. Rather than grouping defenses by algorithmic family, we group them by their points of intervention within the SemCom system. This organization highlights how different defense mechanisms address distinct semantic failure modes and how protections at one layer interact with vulnerabilities at others. 
Table~\ref{tab:defense_taxonomy} further refines each interface by stating the dominant threat objective and the corresponding AI defense primitives, making explicit the translation from model-centric defense mechanisms to SemCom intervention points. 
The remainder of this section follows this structure, examining encoder/decoder-level, channel-level, knowledge-base-level, and network-level defenses in turn.

\begin{table*}[t]
\centering
\caption{Layered defense taxonomy for semantic communication systems, organized by intervention interface, threat objective, and AI defense primitives.}
\label{tab:defense_taxonomy}
\renewcommand{\arraystretch}{1.18}
\footnotesize
\begin{tabular}{|p{0.09\linewidth}|p{0.18\linewidth}|p{0.20\linewidth}|p{0.38\linewidth}|}
\hline
\centering\textbf{Defense Layer} &
\centering\textbf{Defense Category} &
\centering\textbf{Threat / Defense Goal} &
\centering\arraybackslash\textbf{Defense primitives and representative works} \\ \hline

\multirow{4}{0.09\linewidth}{\centering\textbf{Encoder/\\Decoder-Level}}
& Adversarial and Robust Training
& Semantic evasion; task degradation under meaning-preserving perturbations
& \textbf{Robust optimization} over semantic neighborhoods using adversarial or minimax training objectives \cite{hu2022robust,wei2025robust};
\textbf{robust architecture} via constrained bottlenecks (e.g., masked VAE) \cite{hu2023robust}. \\ \cline{2-4}

& Representation Shaping
& Representation instability; enlarged semantic attack surface
& \textbf{Stability regularization} through Jacobian, spectral, or Lipschitz constraints \cite{cisse2017parseval,yoshida2017spectral};
\textbf{latent compression and quantization} to reduce adversarial degrees of freedom \cite{chen2024lightweight};
\textbf{semantic consistency regularization} across meaning-preserving views \cite{peng2022robust,peng2024robust_text}. \\ \cline{2-4}

& Generative Correction and Semantic Restoration
& Test-time corruption; adversarial or channel-induced semantic distortion
& \textbf{Purification and projection} of corrupted representations using GANs or diffusion models \cite{song2018pixeldefend,weng2025generative};
\textbf{generative denoising} for semantic manifold recovery \cite{weng2025generative,cai2025robust}.\\ \cline{2-4}

& Joint Source--Channel--Security Coding
& Semantic leakage; eavesdropping under compressed representations
& \textbf{Secrecy-aware representation learning} that embeds robustness and confidentiality into the encoder \cite{tung2023deepjscec};
\textbf{security-by-design JSCC} without higher-layer cryptography \cite{tung2023deepjscec,kalkhoran2023securedeepjscc}.
\\ \hline

\multirow{3}{0.09\linewidth}{\centering\textbf{Channel-Level}}
& Adversarial Channel Modeling
& Over-the-air semantic degradation under physically realizable interference
& \textbf{Robust training under structured channel noise} combining stochastic fading and adversarial perturbations \cite{nan2023physical,zhao2025secdiff};
\textbf{worst-case utility optimization} under power-bounded perturbation models \cite{nan2023physical,zhao2025secdiff}.
\\ \cline{2-4}

& Semantic-Aware Channel Coding
& Unequal semantic importance; disproportionate task impact
& \textbf{Importance-weighted protection} and unequal semantic error control in Deep JSCC \cite{bourtsoulatze2019deep,peng2024robust_image};
\textbf{cost-sensitive robustness} \cite{peng2024robust_image}. \\ \cline{2-4}

& Semantic Anti-Jamming and Adaptive Reception
& Adaptive jamming targeting task-level meaning
& \textbf{Detection and co-training} against learned semantic jammers \cite{tang2023gan_jamming};
\textbf{adaptive reception} via task-driven decoding and interference-aware adaptation \cite{tang2023gan_jamming,chen2025coding_jamming}.
\\ \hline

\multirow{3}{0.09\linewidth}{\centering\textbf{Knowledge-Base-Level}}
& Secure Knowledge Construction and Maintenance
& Knowledge poisoning; integrity and provenance loss
& \textbf{Provenance tracking, auditing, and verification} for KB ingestion and updates \cite{shen2023secure,yang2024secure};
\textbf{rollback and isolation} of untrusted sources \cite{shen2023secure}. \\ \cline{2-4}

& Controlled Knowledge Sharing and Synchronization
& Knowledge desynchronization; context leakage across agents
& \textbf{Access control and authenticated synchronization} of shared semantic knowledge \cite{lu2024efficient,liu2025knowledge};
\textbf{separation of global and private knowledge} \cite{liu2025knowledge}. \\ \cline{2-4}

& Secure Knowledge-Assisted Semantic Coding
& Malicious grounding; semantic inference by unauthorized receivers
& \textbf{Verified grounding and cross-source consistency checks} \cite{hu2022knowledge,liu2024semantic};
\textbf{private-knowledge separation} as an implicit semantic security key \cite{liu2025knowledge}.
\\ \hline

\multirow{3}{0.09\linewidth}{\centering\textbf{Network-Level}}
& Reliability-Oriented Semantic Networking
& Cascading semantic drift in multi-hop or multi-agent propagation
& \textbf{Network-level semantic correction} using generative reconstruction and relay refinement \cite{erdemir2023generative,chen2023commin};
\textbf{cooperative robustness} \cite{luo2022relay,ma2023intelligentrelay}.
\\ \cline{2-4}

& Trust, Identity, and Semantic Misbehavior Detection
& Impersonation; malicious semantic injection
& \textbf{Semantic-aware authentication and trust evaluation} using environment or content-level features \cite{gao2024esanet,tan2024sae};
\textbf{semantic anomaly detection} \cite{gao2024esanet}.
\\ \cline{2-4}

& Feedback-Aware Semantic Defense Coordination
& Adaptive, multi-agent semantic attacks over time
& \textbf{Closed-loop defense orchestration} via feedback signals and reinforcement learning \cite{shokrnezhad2025arc};
\textbf{graceful degradation under partial compromise} as a system-level resilience objective.
\\ \hline
\end{tabular}
\end{table*}

\subsection{Encoder/Decoder-Level Defenses}\label{subsec:encdec}
Guided by the design philosophy outlined above, we begin by examining defenses at the semantic encoder/decoder level. This level is the most direct interface between AI security and SemCom: it is where meaning is first abstracted into learned representations and where semantic errors most immediately translate into task failure. {As a result, many vulnerabilities and defenses studied in model-centric AI security reappear here as direct effects on semantic representations and task performance.}

At this first stage of the pipeline, defenses focus on shaping how meaning is encoded and recovered before transmission. 
At this layer, the dominant threats are semantic evasion and task degradation under meaning-preserving perturbations, so defenses primarily instantiate robust optimization, stability regularization, purification, and secrecy-aware coding primitives.
Early work adapts adversarial training to semantic representations, while subsequent approaches increasingly emphasize representation geometry, correction at inference time, and finally architectural integration of robustness and confidentiality. 
Together, these efforts reflect a progression from reactive hardening toward robustness by construction.

\begin{enumerate}
    \item \emph{Adversarial and robust training of semantic encoders and decoders.} 
This category extends adversarial training from pixel- or waveform-level perturbations to \emph{semantic neighborhoods} defined by meaning-preserving transformations. Rather than norm-bounded noise, adversaries are constructed through paraphrasing, masking, reordering, or content-preserving edits that preserve syntactic validity while inducing task-level errors, enabling encoders and decoders to be trained directly against worst-case semantic failures~\cite{hu2022robust}. Robustness can be further strengthened through architectural choices such as masked Variational Autoencoder (VAE) bottlenecks that promote discrete, semantically meaningful latent codes~\cite{hu2023robust}, as well as task-aligned objectives that enforce semantic agreement across multiple views of the same content~\cite{peng2022robust,peng2024robust_text}. At a more principled level, robustness has been formalized as a minimax semantic rate--distortion problem, where training explicitly optimizes worst-case semantic distortion rather than average-case performance~\cite{wei2025robust}. Compared with standard adversarial training targeting norm-bounded perturbations~\cite{madry2018towards}, these approaches align robustness objectives with task utility and semantic correctness.

\item \emph{Representation shaping via regularization and semantic consistency.} 
Rather than explicitly generating adversarial examples, this class improves robustness by shaping the geometry and invariances of learned semantic representations. Latent-space regularization constrains encoder sensitivity by limiting Jacobian norms, spectral radii, or latent dispersion, reducing vulnerability to small but semantically harmful perturbations under noise and interference. Masked or quantized codebooks further compress semantics into compact latent spaces, shrinking the adversary’s effective action range while improving stability and efficiency~\cite{hu2023robust,chen2024lightweight}. Complementary semantic consistency constraints encode task- or knowledge-derived invariances directly into the learning objective, encouraging paraphrases or cross-lingual variants to map to nearby representations with aligned task outputs~\cite{peng2022robust,peng2024robust_text}. These techniques parallel classical stability and Lipschitz regularization in AI security~\cite{cisse2017parseval,yoshida2017spectral}, but are adapted to semantic fidelity objectives and communication constraints, making them attractive for resource-constrained SemCom deployments.

\item \emph{Generative correction and semantic restoration.} 
Instead of hardening the encoder or decoder alone, this family introduces generative modules that actively correct corrupted semantic representations at inference time by projecting noisy signals or latent codes back onto a manifold of semantically plausible content. Generative Adversarial Networks (GAN) and diffusion-based pre- and post-processors have been proposed to filter adversarial artifacts before decoding in speech and multimodal SemCom settings~\cite{weng2024ross,weng2025generative}. Generative models can also serve a dual role by acting as adaptive attackers during training and as purification modules at inference~\cite{cai2025robust}. These approaches parallel generative purification in AI defense~\cite{song2018pixeldefend}, but must remain lightweight and interruptible to respect latency, computation, and energy constraints in communication systems.

\item \emph{Joint source--channel--security coding at the encoder/decoder.} 
A distinct line of work embeds robustness and confidentiality directly into the semantic transceiver architecture. Rather than treating security as an external layer or post-hoc correction, neural encoders are trained so that latent representations simultaneously enable reliable semantic recovery at the intended receiver while remaining statistically uninformative to eavesdroppers~\cite{kalkhoran2023securedeepjscc}. Representative designs such as deep joint source--channel and encryption coding unify semantic compression, channel coding, and secrecy within a single learned mapping~\cite{tung2023deepjscec}. In contrast to adversarial training or generative purification, these approaches enforce robustness and confidentiality \emph{by construction}, making them attractive when higher-layer cryptographic support is unavailable or incompatible with semantic objectives.

\end{enumerate}
    
Overall, encoder/decoder-level defenses in SemCom closely mirror the AI security toolbox, including adversarial training~\cite{madry2018towards}, stability regularization~\cite{cisse2017parseval,yoshida2017spectral}, consistency-based learning, and generative purification~\cite{song2018pixeldefend}. However, they differ fundamentally in evaluation criteria and operating constraints. Robustness in SemCom is evaluated not by $\ell_p$ distortion or bit error rate, but by task-level semantic utility, often assessed via BLEU score~\cite{papineni2002bleu}, semantic similarity~\cite{cer2017semeval}, or downstream accuracy under rate, latency, and energy budget constraints in wireless systems.

Despite promising advances, several challenges remain. Most existing defenses assume static or bounded semantic perturbations and are developed for single-shot inference, lacking support for multi-turn or interactive SemCom workflows where feedback loops and adaptation play key roles. Furthermore, robustness is rarely co-optimized with rate–distortion and latency objectives, leaving open how much semantic robustness can be achieved without negating the efficiency gains of SemCom. 
These gaps underscore the need for SemCom-native robustness evaluation protocols that integrate task utility, communication constraints, and system feedback. Progress will require tighter co-design across encoder training, JSCC architecture, and physical-layer dynamics, where robustness is not bolted on, but emerges from end-to-end semantic fidelity under adversarial and constrained conditions.

\subsection{Channel-Level Defenses under Semantic Objectives}\label{subsec:channel}
Even when semantic representations are robustly encoded, transmission over a wireless channel introduces distortions that cannot be controlled at the model level alone. 
This shifts the defense objective from representation robustness to resilience under physically constrained interference. 
Wireless channels convey encoded semantic representations over noisy, interference-limited links, making them a primary locus where adversaries can induce semantic degradation under strict physical constraints. 
Unlike conventional attacks that aim to disrupt symbol recovery, semantic attackers may inject carefully crafted over-the-air perturbations that preserve syntactic reliability while inducing task-level semantic distortion, as discussed in Section~\ref{sec:2c_channelattack}. 
Channel-level defenses in SemCom must therefore operate within spectrum, power, and protocol constraints while explicitly optimizing semantic utility rather than bit-level fidelity.

Channel-level defenses therefore evolve along three complementary directions, corresponding to how much control the system assumes over the channel during training, coding, and runtime adaptation.

\begin{enumerate}
    \item \emph{Adversarial channel modeling for robust SemCom.} 
    A direct defense strategy hardens semantic transceivers by explicitly modeling the wireless channel as a composition of stochastic fading and physically realizable adversarial perturbations constrained by power and bandwidth budgets. Semantic encoders and decoders are jointly trained under this mixed channel to optimize worst-case task utility rather than bit-level fidelity~\cite{nan2023physical}. Extensions of this approach incorporate diffusion-based secure DeepJSCC to improve robustness against structured adversarial interference~\cite{zhao2025secdiff}, task- and context-adaptive objectives to handle non-stationary environments~\cite{wijesinghe2025taco}, and sparse semantic coding schemes that enhance resilience under dynamic channel and task variations~\cite{zhan2025sparse}. Collectively, these methods emphasize training-time exposure to realistic worst-case channels as a foundation for semantic robustness.

    \item \emph{Semantic-aware channel coding with unequal protection.} 
    A complementary line of work focuses on allocating channel resources asymmetrically according to semantic importance, ensuring that task-critical information receives stronger protection. Deep JSCC inherently exhibits unequal error protection by prioritizing salient semantic features during end-to-end training~\cite{bourtsoulatze2019deep}. Building on this property, explicit semantic importance maps have been introduced to guide adaptive power allocation, redundancy, and diversity across semantic representations~\cite{peng2024robust_image,peng2025deepscri}. These approaches resemble cost-sensitive robust optimization in AI, but must additionally respect physical-layer constraints such as spectral efficiency, latency, and transmit power.

    \item \emph{Semantic anti-jamming and adaptive reception.} 
    Semantic jamming targets task-level meaning rather than maximizing bit errors, allowing attacks to remain effective even when signal-level metrics appear benign. Recent defenses therefore operate at the semantic level, either by co-training receivers against learned semantic jammers or by embedding coding-aware interference strategies that selectively degrade an eavesdropper while preserving legitimate semantic recovery. Representative examples include GAN-inspired adversarial frameworks that jointly train a semantic jammer and a robust receiver~\cite{tang2023gan_jamming}, as well as coding-enhanced jamming schemes that strengthen secrecy through encoder-side design under rate and reliability constraints~\cite{chen2025coding_jamming}. These approaches highlight both the effectiveness and the adaptivity of semantic jamming, underscoring the need for dynamic, task-aware reception strategies.
\end{enumerate}

Taken together, these defenses reflect three complementary design philosophies: hardening semantic transceivers through channel-realizable adversarial training, reshaping physical and link-layer resources around semantic importance, and incorporating runtime mechanisms that adapt transmission or decoding in response to semantic-aware interference. While differing in mechanism and assumptions, all three shift the protection objective from symbol fidelity to task-level semantic utility under physically realizable attacks.

From an AI defense perspective, channel-level defenses can be viewed as robustness under structured measurement noise, where adversarial perturbations are constrained by wireless physics rather than arbitrary $\ell_p$ bounds. The defining challenge in SemCom is that robustness must be engineered jointly with spectral efficiency, latency, and regulatory constraints. Defenses cannot arbitrarily increase redundancy or computation and must interoperate with standardized physical- and link-layer protocols.

Despite growing interest, channel-level defenses remain limited by simplified channel models and narrow adversary assumptions. Most evaluations rely on AWGN or single-antenna fading channels and do not capture dense, multi-user, or multi-RAT environments where attackers may exploit protocol features and scheduling dynamics. Interactions with cryptographic mechanisms such as PHY-layer authentication and link encryption are also underexplored, as semantic robustness and confidentiality are often treated as separate objectives. A key takeaway is that channel-level defenses should be evaluated using SemCom-native metrics that couple semantic utility with adversarial, spectrum- and power-constrained interference, and that robust transceiver design should be co-optimized with adaptive, attack-aware resource management.



\subsection{Knowledge-Base-Level Defenses}\label{subsec:kb}
While channel-level defenses protect semantic signals during transmission, they implicitly assume a stable and trusted semantic context at the receiver. In traditional systems with centralized knowledge bases, this assumption is often reasonable: the KB is treated as a backend resource whose integrity and access are enforced through perimeter security, access control, and centralized governance. In SemCom, however, encoding and decoding explicitly depend on shared or external semantic knowledge that is distributed, replicated, and continuously updated across communicating endpoints \cite{yang2022semantic,shen2023secure}. As a result, semantic context itself becomes part of the communication process rather than a passive background resource.

Knowledge bases (KBs), pretrained foundation models, ontologies, and structured memories serve as semantic priors that enable aggressive compression and robust reconstruction under limited communication resources \cite{yang2022semantic,liang2025knowledge}. Unlike traditional centralized KBs that primarily support query answering or decision support, semantic knowledge in SemCom directly shapes what information is transmitted and how missing content is inferred at the receiver. \emph{Despite this central role, the security implications of semantic knowledge in communication pipelines have received comparatively limited attention relative to encoder- and channel-level defenses}.

As discussed in Section~\ref{sec:IIID_knowlegdeThreat}, this tighter coupling between knowledge and communication introduces a unique attack surface: semantic failures may arise from manipulation, inconsistency, or overexposure of knowledge even when learned encoders, decoders, and wireless channels behave nominally \cite{yang2024secure}. Because semantic knowledge simultaneously influences encoding decisions and decoding interpretation, attacks on the knowledge layer can silently induce semantic misalignment without triggering signal-level anomalies or reliability failures \cite{li2025synchronizing,goodfellow2014explaining}. As semantic knowledge becomes an active participant in encoding and decoding, defenses naturally evolve along three complementary directions that address integrity, synchronization, and secure semantic grounding.

\begin{enumerate}
    \item \emph{Secure knowledge construction and maintenance.} 
    The first line of defense treats the semantic KB as a critical system asset whose ingestion, update, and evolution must be explicitly secured. Threats to semantic KBs have been systematized, with poisoning, tampering, and stale or inconsistent updates identified as primary risk factors~\cite{shen2023secure}. Provenance-aware KB pipelines tag each update with source identity, temporal metadata, and verification status, enabling suspicious or conflicting updates to be quarantined or rolled back. Secure maintenance workflows can further incorporate access control, logging, and auditability to bound the blast radius of compromised contributors and support post-hoc forensic analysis~\cite{yang2024secure}. These mechanisms parallel data lineage and governance practices in secure ML pipelines, but are uniquely critical in SemCom because KB corruption affects both semantic compression and reconstruction.

    \item \emph{Controlled knowledge sharing and synchronization.} 
    The second category focuses on regulating how semantic knowledge is shared and synchronized across devices, agents, and administrative domains. SemCom systems often implicitly assume sufficiently aligned knowledge at the transmitter and receiver; adversaries can exploit partial synchronization, delayed updates, or version mismatches to induce semantic divergence without disrupting signal delivery. Recent works therefore distinguish between global knowledge and private or personalized knowledge, and enforce authenticated synchronization policies \cite{lu2024efficient,tian2025model}. A dedicated knowledge management layer can coordinate access control, update authentication, and selective synchronization, preventing adversaries from reconstructing the exact semantic context used by legitimate endpoints~\cite{liu2025knowledge}. From a defense standpoint, such mechanisms act as semantic access control, limiting both semantic leakage and desynchronization-induced misinterpretation.

    \item \emph{Secure knowledge-assisted SemCom pipeline.} 
    Rather than treating knowledge solely as a background resource, this category integrates structured knowledge directly into semantic encoding and decoding in a security-aware manner. Knowledge-assisted semantic coding schemes inject graph-based or symbolic priors into representation learning so that encoders compress messages in ways consistent with trusted knowledge, while decoders use knowledge-guided reasoning to recover missing semantics under noise. Incorporating verified knowledge sources and cross-source consistency checks can improve robustness against both channel noise and semantic ambiguity~\cite{hu2022knowledge,yang2024secure,liu2024semantic}. More recent designs emphasize separating public and private knowledge during encoding, such that accurate semantic reconstruction requires access to the intended private knowledge context~\cite{liu2025knowledge}. This effectively turns private knowledge into an implicit security key: even if an adversary intercepts semantic representations and possesses strong global priors, the absence of synchronized private knowledge prevents correct interpretation of sensitive semantics.
\end{enumerate}

From an AI defense viewpoint, KB-level defenses share commonalities with robust data management, trustworthy knowledge graphs, and secure retrieval-augmented generation. Provenance tracking and access control mirror secure data pipelines in ML, while robust graph learning limits the influence of corrupted nodes or edges during message passing. The key distinction in SemCom is that KB integrity and availability directly influence \emph{both} encoding and decoding behavior. Small perturbations in knowledge can therefore induce large semantic shifts without triggering any signal-level anomaly.

Despite growing recognition of their importance, KB-level defenses remain relatively underdeveloped in SemCom. There is no standardized benchmark for evaluating robustness or privacy of semantic knowledge bases under dynamic updates and partial trust assumptions. Many proposed defenses rely on strong provenance or trust signals that may be unavailable in open or cross-domain deployments. Moreover, restricting knowledge access to improve security can complicate verification, auditing, and synchronization, creating new trade-offs between robustness, privacy, and system usability.

A key takeaway is that semantic knowledge should be studied as a first-class security asset rather than a benign background resource. Its construction, synchronization, exposure, and integration into semantic models should be co-designed with explicit robustness and privacy objectives, and evaluated under adversarial conditions that reflect realistic SemCom deployments.

\subsection{Network-Level Defenses}\label{subsec:network}
Unlike encoder-, channel-, or knowledge-level defenses, network-level defenses address semantic failures that emerge from interaction, propagation, and coordination among multiple terminals over time, rather than from any single compromised component. In networked SemCom systems, meaning is not only decoded locally but also propagated, aggregated, and acted upon across agents, making semantic security an inherently collective property.

From an AI defense perspective, network-level semantic defenses parallel robust multi-agent learning, distributed anomaly detection, and trust-aware decision-making under adversarial conditions. Generative correction mechanisms resemble robust generative modeling under distribution shift, while trust evaluation and authentication mirror reputation systems in adversarial multi-agent environments. The key distinction in SemCom is that the protected object is not packet delivery or throughput, but the consistency and integrity of shared meaning as it propagates through the network.

Existing network-level defenses for SemCom manifest in three recurring defense patterns.

\begin{enumerate}
    \item \emph{Reliability-oriented semantic networking and cooperative robustness.} 
    The first category addresses semantic failures that emerge from multi-hop transmission and multi-agent propagation, where small local distortions can accumulate into global task failure. Generative and semantic-aware correction mechanisms can be deployed at intermediate nodes, relays, or edge servers to stabilize semantics before forwarding. Representative examples include InverseJSCC and GenerativeJSCC, which use generative models to denoise or reconstruct semantically plausible outputs from heavily corrupted DeepJSCC transmissions~\cite{erdemir2023generative}. Semantic reconstruction has also been formulated as an inverse problem, combining invertible neural networks with diffusion models to recover high-quality semantics from degraded intermediate representations~\cite{chen2023commin}. 
    Beyond pure correction, intelligent semantic relaying allows intermediate nodes to partially decode, refine, or re-encode semantic information using shared or local knowledge, reducing semantic drift across hops~\cite{luo2022relay,ma2023intelligentrelay}. Collectively, these approaches act as network-level semantic filters that prevent localized corruption from cascading through the system.

    \item \emph{Trust, identity, and semantic misbehavior detection.} 
    Even when links are reliable, adversaries may impersonate legitimate agents, inject malicious semantic content, or manipulate coordination protocols. Network-level defenses therefore extend beyond packet-level authentication to reasoning about semantic consistency and task impact. Trust establishment and authentication can leverage semantic and environmental features that are difficult to forge at scale, such as environment-level semantics extracted from massive MIMO channels for physical-layer authentication~\cite{gao2024esanet}. Authentication mechanisms have also been co-designed with semantic objectives through metrics that explicitly balance authentication reliability and throughput~\cite{tan2024sae}. 
    More broadly, misbehavior detection in SemCom networks must identify agents whose semantic outputs systematically distort shared meaning or degrade task performance, even when syntactic validity is preserved.

    \item \emph{Feedback-aware semantic defense coordination.} 
    A third category treats semantic defense as a coordinated, adaptive process across the network rather than a static mechanism at individual nodes. In this view, detectors, trust scores, and semantic performance metrics serve as feedback signals that inform routing, scheduling, model selection, or fallback strategies. Detection of semantic jamming or anomalous behavior at one node can trigger route changes, decoder switching, or semantic rate adaptation elsewhere in the network~\cite{gao2025agenticsemcom}. Although still in an early stage, such orchestration aligns closely with AI-based control and reinforcement learning frameworks for autonomous networks~\cite{shokrnezhad2025arc}. 
    From an AI defense perspective, these mechanisms resemble robust multi-agent learning and trust-aware decision-making, with the key distinction that SemCom networks protect \emph{shared meaning and semantic coherence}, rather than packet delivery or throughput alone.
\end{enumerate}

Despite their importance, many existing schemes assume benign cooperation or random impairments and do not explicitly model adaptive adversaries that exploit network dynamics, feedback loops, or generative priors. Scalability and latency pose additional challenges: trust management and generative correction may introduce communication, computation, and coordination overhead that is incompatible with real-time semantic tasks. Moreover, trust is often defined at the agent level rather than at the level of semantic content or knowledge items, leaving open how to quantify, propagate, and act on trust in meaning itself. More broadly, these mechanisms aim for \emph{graceful degradation} under partial compromise, where semantic utility degrades predictably rather than failing catastrophically.

A key takeaway is that semantic security is an emergent, system-level property. Ensuring robustness in networked SemCom requires defenses that jointly reason over agents, links, knowledge, and tasks, and that enable graceful degradation under partial compromise rather than catastrophic failure.

\subsection{Cross-Domain Lessons and Integration Paths}
The layered defenses discussed above demonstrate that securing SemCom requires coordinated protection across representation learning, physical transmission, shared knowledge, and networked interaction. At each layer, many proposed mechanisms draw inspiration implicitly or explicitly from robustness techniques developed in mature AI domains, particularly NLP and CV. To place these defenses in a broader context and to identify paths toward principled integration, it is instructive to examine how related challenges have been addressed in NLP and CV.

Both NLP and CV confront adversaries that manipulate learned representations to alter meaning without obvious syntactic corruption, a threat model that closely mirrors semantic attacks in SemCom. However, while robustness techniques in these domains are typically developed for standalone inference pipelines, SemCom embeds learned models within communication-constrained, distributed systems where robustness must coexist with rate, latency, energy, and coordination constraints. This subsection distills transferable lessons from NLP and CV robustness, clarifies where direct adoption breaks down, and outlines integration paths for developing SemCom-native defenses that respect system-level constraints.

\paragraph{Transferable Ideas from NLP and CV Robustness}
In NLP, consistency-based learning and semantic-preserving perturbations, such as paraphrasing or synonym substitution, have improved robustness against ambiguity and adversarial rewriting~\cite{wang2022measure,goyal2023survey}. Similarly, in CV, defenses have evolved from norm-bounded perturbation resistance to include semantic shifts and natural distributional corruptions. Notable techniques include representation smoothing, input transformations, and ensemble-based strategies~\cite{drenkow2021systematic,li2023trade}. These advances have inspired emerging SemCom defenses, such as semantic consistency objectives~\cite{peng2022robust}, latent-space regularization~\cite{chen2024lightweight}, and generative purification~\cite{weng2025generative}, which aim to align encoded representations with underlying meaning under constrained conditions.

\paragraph{Adaptation Challenges Under SemCom Constraints}
Despite these parallels, SemCom introduces domain-specific challenges that limit direct transfer. As highlighted in the “Distinct Constraints” of Section \ref{sec:principle}, SemCom pipelines must meet strict rate, power, and latency budgets, operate over noisy and lossy physical channels, and maintain coherence with shared or external knowledge modules. Defenses must therefore be communication-aware and lightweight, often requiring co-design with encoder–decoder architectures or joint source–channel coding schemes. Standard robustness techniques from NLP or CV, such as computationally expensive adversarial training or always-on ensembles, cannot be directly applied without violating these deployment constraints.

Evaluation practices from NLP and CV also offer useful reference points. Benchmark datasets like ImageNet-C~\cite{hendrycks2019benchmarking} and adversarial GLUE~\cite{nie2020adversarial} have catalyzed standardized robustness evaluation. A SemCom analog should incorporate over-the-air channel traces, semantic misalignment detection, and knowledge desynchronization scenarios that reflect practical operation. These additions would encourage evaluation under constraints that match the semantic and physical realities of deployment.

\paragraph{Integration Paths for SemCom-Native Robustness}
To meaningfully integrate these lessons, robustness in SemCom should be treated as a system-level resource allocation problem. Rather than uniformly deploying defenses, protection budgets, such as redundancy, abstention, or adaptive filtering, should be allocated based on marginal utility under task priorities and system constraints. Lightweight mechanisms such as real-time anomaly detection, fallback control strategies, or contract-based semantic monitoring may offer better tradeoffs than heavy, static defenses.

Future SemCom pipelines will benefit from semantically aware, composable defenses that are compatible with constrained environments and adversarial uncertainty. These defenses should be designed with explicit assumptions, verified through contextual testing, and reported using deployment-aware robustness profiles rather than isolated metrics. Lessons from NLP and CV robustness will be instrumental in shaping these systems, but success will require adapting them to the unique architectural and operational demands of SemCom.
\section{Bridging Design and Deployment for Secure Semantic Communications} \label{sec:deployment}

\begin{figure*}[!t]
\centering
\includegraphics[width=1.0\textwidth]{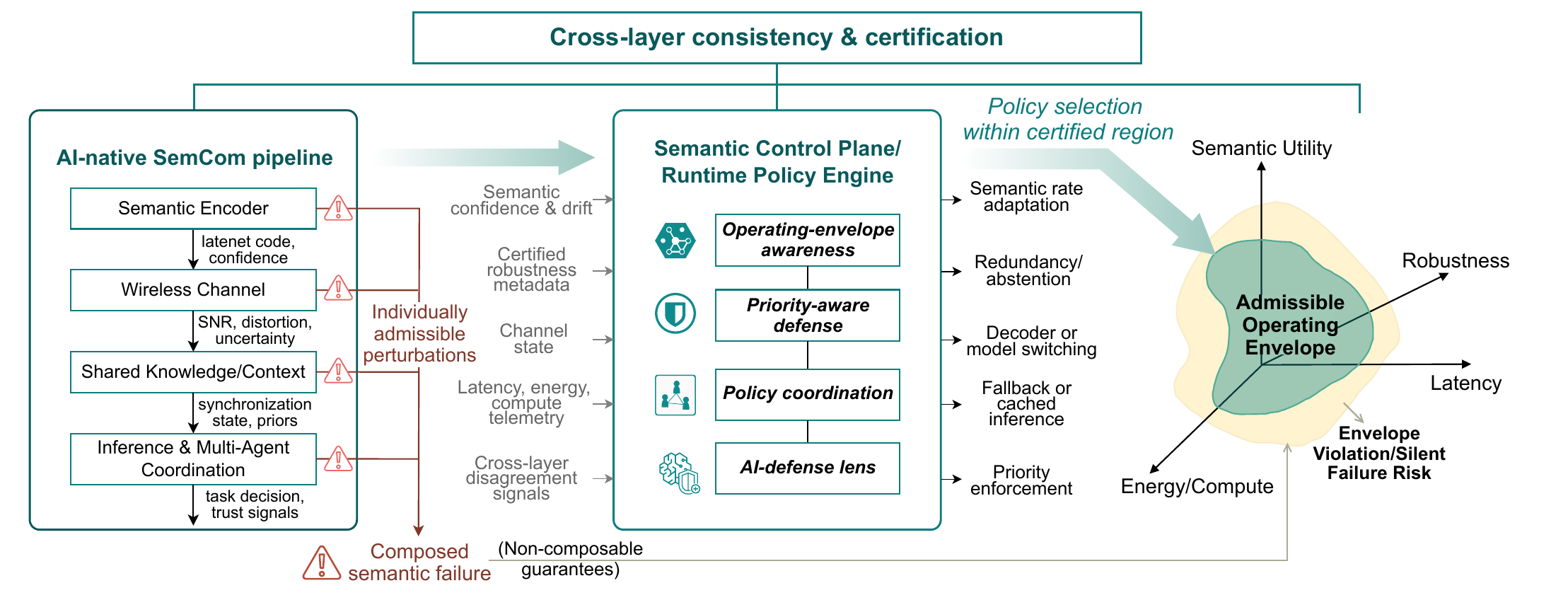}
\caption{Bridging design and deployment through operating envelopes and runtime control in secure semantic communication. Robustness is enforced as a managed system property via a semantic control plane that integrates operating-envelope awareness, resource and priority telemetry, policy coordination, and cross-layer consistency signals from the AI-defense lens.}
\label{fig:sec5}
\end{figure*}

Despite advances in semantic encoding, adversarial robustness, and threat detection, many SemCom defenses remain demonstrated only in isolated models or simplified simulations rather than integrated into deployable systems. Bridging this gap requires system architectures that support runtime introspection, policy enforcement, and adaptive behavior under stringent rate, latency, energy, and compute constraints. This section highlights three deployment challenges: (i) managing security--utility tradeoffs within constrained operating envelopes, (ii) enforcing robustness without violating resource or timing budgets, and (iii) enabling verifiable behavior through lightweight, system-facing assurance signals.
We emphasize that enforcing robustness under resource and timing budgets is primarily realized through runtime policy enforcement, adaptive fallback, and envelope-aware monitoring, which recur across the certification, red-teaming, and testbed discussions below.

\subsection{Security–Utility Tradeoffs and Operating Envelopes}
Security mechanisms applied in SemCom often conflict with task utility, latency, and energy constraints. While these defenses improve robustness under attack or drift, they impose computational or communication overhead that can degrade system throughput or violate timing budgets. This creates a core tension: securing SemCom systems cannot be done in isolation from their performance envelopes.

To reason about these tradeoffs, we adopt the notion of \textit{operating envelopes}: 
regions in the multidimensional space of semantic utility, latency, robustness, and system cost that represent acceptable operating points under bounded resources and threat models. This concept, widely used in real-time and networked systems \cite{aastrom2021feedback,liu2000real}, has recently found traction in robust AI and adaptive communication design, where runtime decisions are governed by policies that trade off robustness and performance under explicit budgets \cite{madry2018towards,zappone2019model}.
Within this framework, robustness mechanisms are viewed as policy choices constrained by explicit system budgets. For instance, increasing abstention thresholds may improve resilience to semantic drift, but at the cost of higher decision latency or degraded task completion rates.

In practice, most current SemCom pipelines are tuned offline using fixed thresholds and evaluated under average-case conditions. Runtime enforcement of semantic risk budgets is largely absent. 
Moreover, the lack of standard metrics for semantic degradation under resource constraints makes it difficult to compare or certify defenses across systems. 
In addition, we also need to consider the resource limitations of SemCom systems. 
Thus, to transition from design to deployment, future SemCom systems should support \textbf{envelope-aware introspection}: mechanisms that continuously monitor semantic fidelity, latency, energy usage, and channel state to dynamically adjust semantic security policies under explicit resource budgets. This requires coordination with lower-layer scheduling and system monitors to ensure defenses remain within operating envelopes.

\subsection{Certification and Verifiable Robustness}
Certification represents a promising bridge between principled robustness analysis and deployable SemCom systems. In the SemCom context, certification is not merely about proving worst-case guarantees for isolated models, but about enabling system-level assurances that support safe and predictable behavior under real-world uncertainty. This includes certifying task-level properties, such as intent preservation or safe control actions, under physically realizable perturbations and dynamic operating conditions.

Unlike classic AI settings, SemCom systems operate under stochastic fading, time-varying bandwidth, feedback-driven protocols, and adaptive control mechanisms. As a result, robustness guarantees are inherently multi-layered and context-dependent. While techniques such as randomized smoothing, Lipschitz bounding, and convex relaxations provide useful foundations in static learning pipelines \cite{li2023sok,zhang2022rethinking,cohen2019certified}, their assumptions rarely hold end-to-end in wireless systems. Over-the-air attackers are constrained by physical limits such as transmit power, spectral locality, and temporal coherence rather than abstract $\ell_p$ norms, and semantic utility itself may depend on protocol state, decoding slack, or application feedback. Adaptive behaviors such as HARQ, link adaptation, and semantic fallback further complicate static certification.

To address these challenges, certification in SemCom should be treated as a \emph{runtime system artifact} rather than an offline proof obligation. Certificates, such as confidence radii, abstention indicators, or semantic bounds, should be attached to semantic outputs as lightweight metadata. 
Such metadata can be derived from calibrated uncertainty estimates, decoder disagreement, or bounded sensitivity of semantically stable representations.
Additionally, deployable certification requires threat models aligned with physical-layer realities. Certifiable perturbation sets should be defined in terms of SNR ranges, error vector magnitude (EVM) bounds, or semantic edit distances that reflect realistic wireless degradation. Certification efforts should prioritize semantically stable representations, such as latent embeddings or task-relevant decision statistics, where guarantees are more meaningful and resilient to transmission variability.

Rather than producing static worst-case bounds, certified robustness should be embedded within the \emph{operating envelope} of the system. 
Operationally, these regions can be implemented as lookup tables or lightweight predictors that map observed channel and workload telemetry to whether a certificate is valid and what fallback action is required.
Future research directions include composing certificates across encoder–channel–decoder pipelines, caching and reusing certified inference regions, and integrating certification signals into adaptive abstention or fallback policies. 
Ultimately, certification in SemCom must evolve into a dynamic, observable, and composable mechanism that supports system-level decision making under real-world conditions.

\subsection{Evaluation Metrics and Benchmarking Gaps}
\begin{figure}[t!]
\centering
\includegraphics[width=0.70\linewidth] {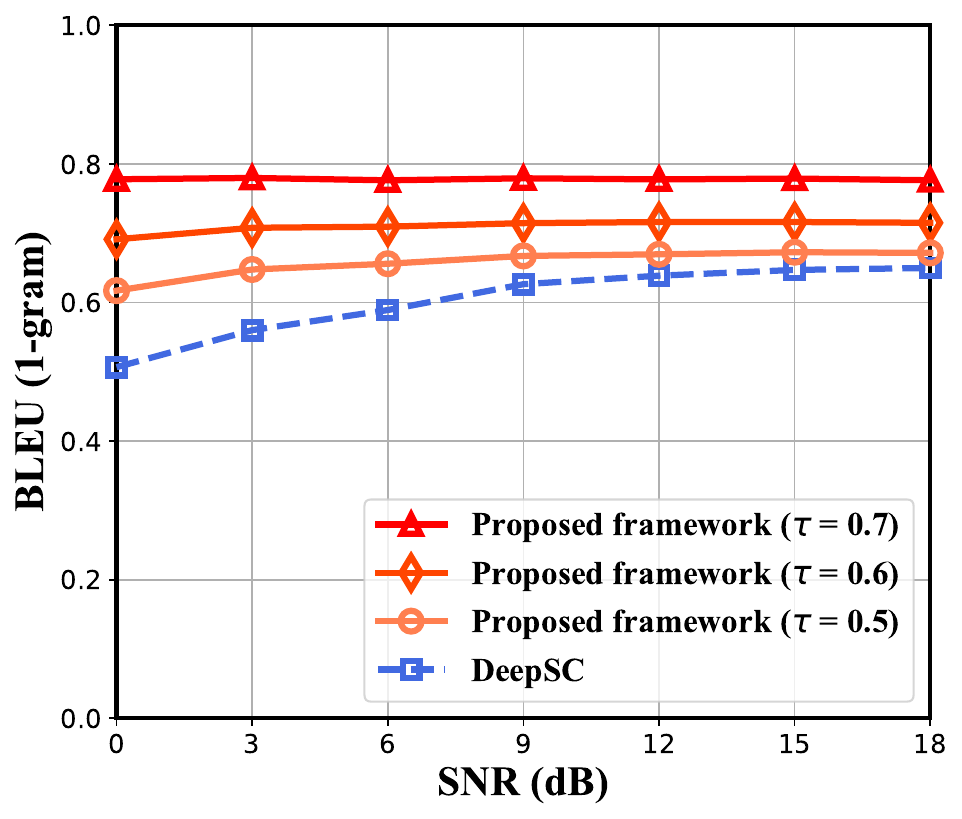}
\caption{BLEU (1-gram) scores for the proposed secure SemCom framework with varying thresholds ($\tau =0.5, 0.6, 0.7$) vs. DeepSC \cite{liang2025safeguarded}. The parameter $\tau$ is the safeguarded threshold guaranteed in the proposed framework. DeepSC is a traditional SemCom framework with no consideration of security.}
\label{fig:result-safeguraded}	
\vspace{-.5em}
\end{figure}

The evaluation of security in SemCom systems requires a fundamental rethinking of what constitutes robustness, utility, and performance. Traditional metrics such as bit error rate (BER), SNR, or downstream task accuracy offer limited visibility into the security posture of SemCom pipelines, especially under adversarial conditions. Unlike classical norm-bounded robustness metrics in machine learning \cite{carlini2017towards}, adversarial budgets in SemCom reflect semantic corruption, protocol behavior, and physical-layer constraints. This calls for a shift toward \textbf{semantic robustness}, an evaluation axis that measures degradation in meaning preservation, task utility, or reconstruction fidelity under worst-case semantic perturbations or cross-layer interference. For instance, in visual SemCom, degradation can be quantified via mean intersection over union (mIoU) or classification accuracy under semantic perturbation, while in language tasks, BLEU or semantic similarity scores may better reflect task failure. 
Fig.\ref{fig:result-safeguraded} reports BLEU (1-gram) under varying channel conditions as an example of task-level reliability evaluation in secure SemCom \cite{liang2025safeguarded}.

Semantic robustness should also account for context dependence and knowledge misalignment \cite{chaccour2024less}, where identical transmitted representations may yield divergent interpretations due to corrupted priors or desynchronized knowledge bases. For example, attacks may not directly perturb the transmitted signal but instead induce semantic failure by poisoning the receiver’s grounding assumptions. 
Benchmarks should reflect this fragility by modeling both data-space divergence and knowledge-space divergence within the pipeline. 
All evaluation pipelines should support dual reporting under benign and adversarial conditions to expose hidden robustness–utility tradeoffs that would otherwise remain latent under average-case testing.

Beyond robustness, security evaluations should quantify \textbf{semantic leakage}, which reflects the extent to which transmitted representations inadvertently reveal private or structured information about the source input \cite{nasr2019comprehensive,liu2021machine,elliott2022ai}. Leakage can be measured via mutual information, inversion success rates, or recoverability of sensitive attributes from latent codes. For systems employing modular or cross-task semantic encoding, leakage may also span unrelated downstream applications, requiring broader auditing.

\begin{figure}[t!]
\centering
\includegraphics[width=0.70\linewidth] {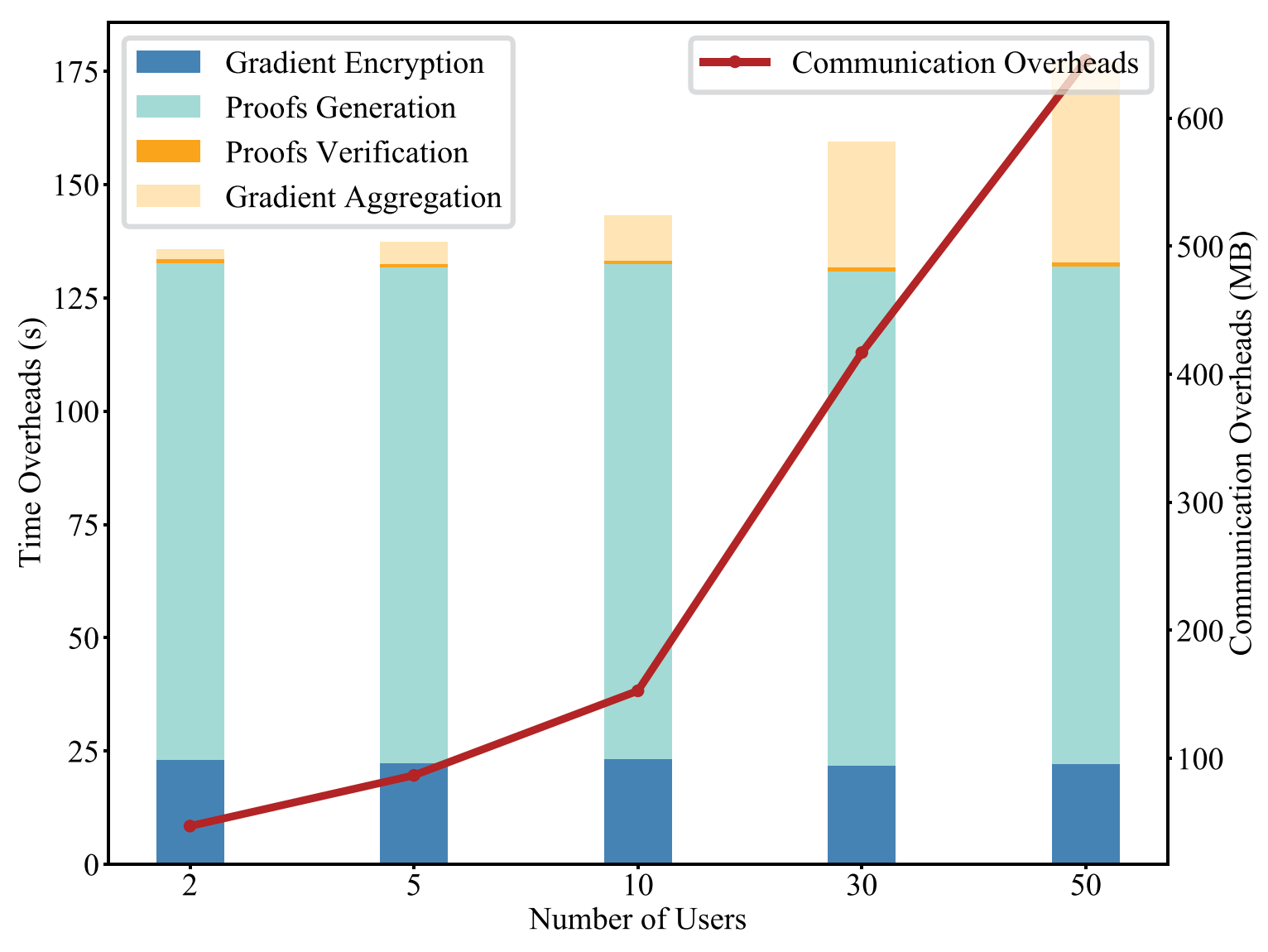}
\caption{Time overheads and communication overheads of reliable training method for secure SemCom\cite{shen2023secure}.}
\label{fig:result-overhead}	
\vspace{-1em}
\end{figure}

In deployment scenarios, \textbf{operational tradeoffs} connect robustness with system viability. These include latency–robustness curves, semantic fidelity versus throughput, and energy per bit of meaning \cite{saad2019vision,zhu2020toward,zhang2025resource,hua2025bandwidth,ma2025power,liu2025joint}. Semantic envelope violation rates, \textit{i.e.}, instances where performance falls below acceptable thresholds under attack, are especially useful for quantifying resilience limits in constrained environments such as edge devices or time-critical inference pipelines. 
Similarly, latent trust drift, where decoder predictions diverge gradually due to upstream semantic desynchronization or poisoned priors, can be monitored to trigger adaptive defenses or trust recalibration protocols. For example, Fig.\ref{fig:result-overhead} examines the time overhead and communication overhead of training reliable SemCom coding models \cite{shen2023secure}. 

Despite these needs, current SemCom evaluations remain fragmented and ad hoc \cite{getu2025semantic}. Most reuse task-oriented metrics from AI or wireless domains without addressing SemCom-specific risks. The field urgently requires reproducible, open-source benchmark suites that simulate realistic attacker models, environmental variations, and cross-layer semantics. Only with such foundations can future work offer rigorous guarantees and practical resilience at scale. To capture SemCom-specific risks under realistic stress, static metrics should be complemented with adaptive adversarial evaluation, as elaborated in the next subsection.

\subsection{Threat Simulation and Red-Teaming}
While static benchmarks and norm-bounded attacks provide valuable baselines, they fail to capture the adaptive, layered nature of real-world adversaries. Red-teaming fills this gap by simulating intelligent, feedback-driven agents that dynamically probe, manipulate, and stress-test SemCom  
pipelines under realistic conditions \cite{morris2020textattack,zhao2023evaluating,brundage2020toward}.

\textbf{Red-teaming agents} adapt their strategies based on observed system responses, such as retransmission patterns, decoding confidence, or semantic drift. For instance, a semantic-aware jammer may evolve its interference by monitoring task failure rates, while a gray-box adversary may exploit repeated queries to reconstruct latent embeddings or trigger meaning misalignment. These agents can operate across layers, such as coordinating encoder poisoning, channel perturbations, and knowledge-base desynchronization, to induce emergent failures that span semantic, protocol, and physical layers—failures that remain latent in isolated testing. Effective implementations may rely on adaptive optimization and co-training to co-evolve with system defenses \cite{brundage2020toward,zhao2023evaluating}.

\textbf{Threat roles and surfaces} vary in observability and access: white-box agents emulate insider threats with full model access; gray-box attackers observe intermediate features or metadata; 
black-box adversaries interact only with system inputs and outputs \cite{papernot2017practical}.
Attack surfaces span semantic-aware jamming, backdoor insertion, knowledge poisoning, privacy leakage, and coordinated misinformation in multi-agent settings. A modular red-teaming framework should support diverse attacker profiles and allow composition of complex, multi-phase campaigns that reflect real-world threat evolution.

\textbf{Integrated evaluation} embeds red-teaming agents into training, adaptation, or runtime inference to reveal resilience under dynamic pressure. Defenses such as decoder switching, abstention, or knowledge re-synchronization can be stress-tested using runtime triggers like semantic mismatch, latent trust drift, or envelope violations \cite{meng2025securesemcom}. Beyond degrading performance, red-teaming enables \textit{semantic risk auditing}—a diagnostic process that identifies brittle input modalities, fragile concept classes, and cascading failure paths. These insights support the development of semantic risk budgets that quantify acceptable degradation bounds under adversarial stress, particularly for safety-critical applications such as autonomous driving, remote surgery, or multi-agent coordination.

Ultimately, red-teaming is not a replacement for theoretical analysis or benchmark testing but a necessary complement. By exposing emergent, cross-layer vulnerabilities that elude static analysis, red-teaming provides a pathway toward more trustworthy, resilient SemCom systems. Future work should prioritize open-source red-teaming toolkits tailored to SemCom, incorporating domain-specific artifacts such as latent codes, shared priors, and task-conditioned decoders. However, even the most sophisticated simulations should ultimately be validated under physical constraints, motivating real-world testbed deployments.

\subsection{Real-World Testbeds and Deployment Studies}
Simulation and red-teaming offer powerful abstractions for evaluating SemCom security. However, validating resilience under real-world constraints requires deployment-centric testbeds that expose the unpredictability of physical hardware, asynchronous control loops, and imperfect communication channels.

\begin{figure}[t!]
\centering
\includegraphics[width=0.80\linewidth] {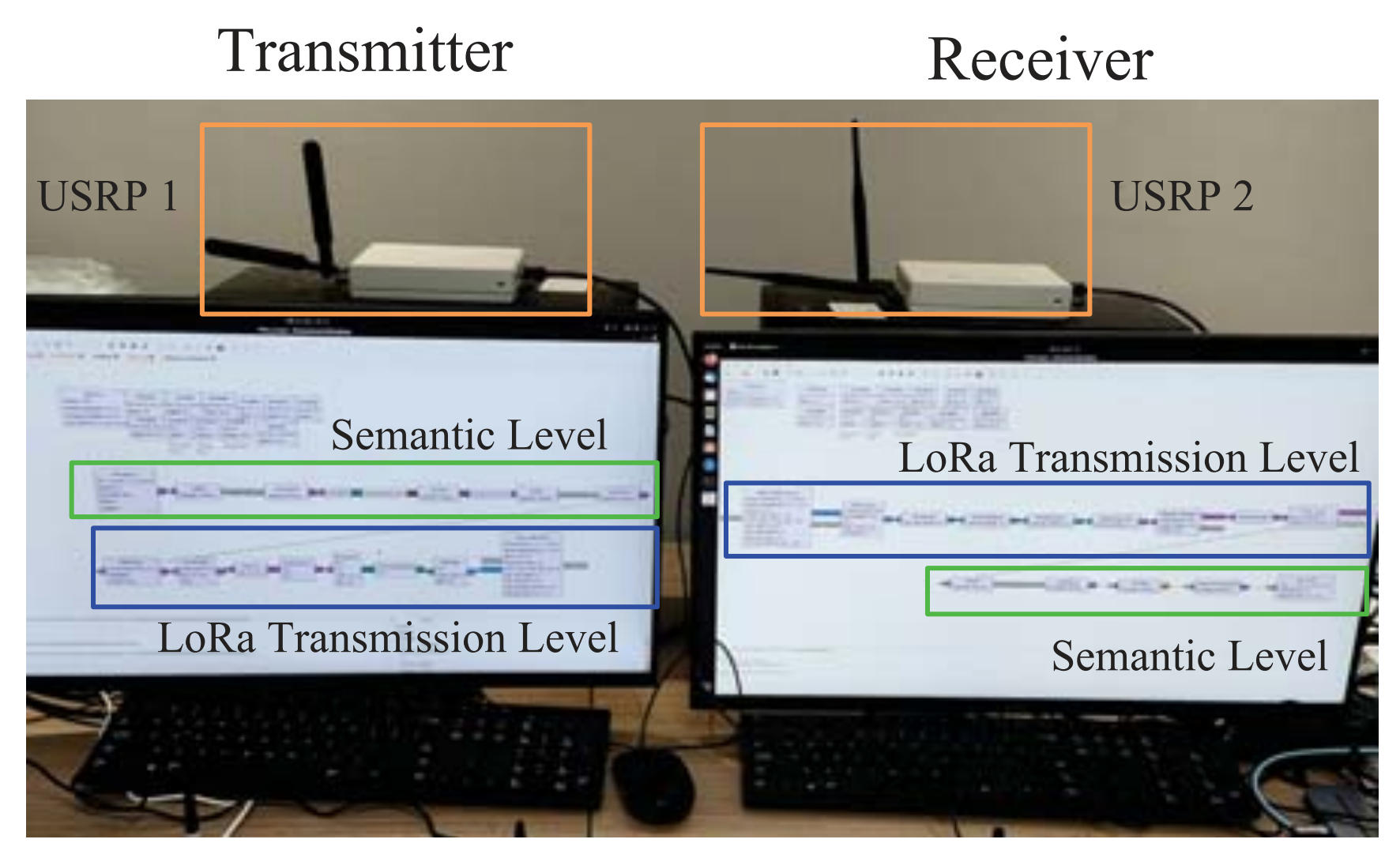}
\caption{Semantic communication on SDR-based LoRa communication platform \cite{ma2024implementation}.}
\label{fig:result-implenmentation}	
\end{figure}

\textbf{Semantic observability} is central to effective deployment studies. Unlike conventional wireless testbeds that focus on throughput, delay, or error rates, SemCom evaluation requires introspection into latent code distributions, semantic drift, decoder disagreement, and knowledge desynchronization. Emerging 6G/O-RAN platforms \cite{bonati2021intelligence,polese2023understanding}, augmented with software defined radios (SDRs), xApps, and edge AI accelerators, provide partial support for this layered introspection. These platforms should be extended to monitor high-level behaviors such as task-level degradation under semantic mismatch and adversarial drift. For example, Fig.\ref{fig:result-implenmentation} shows a simple SemCom platform built on SDR-based LoRa system, which includes two computers and USRPs to perform practical semantic coding and data transmission over a wireless link \cite{ma2024implementation}. 

Testbeds expose \textbf{practical system constraints} that often go unnoticed in simulation \cite{ma2024implementation,shen2023secure}. Knowledge re-synchronization may fail under lossy control channels or when agents are asynchronously updated \cite{lu2024efficient,liu2025knowledge}. Redundancy-based defenses, such as ensemble decoding or semantic voting, may conflict with stringent real-time latency constraints typical of edge environments \cite{erdemir2023generative,chen2023commin}. Practical issues such as hardware timing violations, adaptation delays, and resource contention often emerge as performance bottlenecks during deployment and must be explicitly profiled and reported as part of security-utility operating envelopes \cite{luo2022relay}.

Beyond synthetic benchmarks, testbeds allow \textbf{cross-domain robustness evaluation} across realistic and heterogeneous settings, ranging from remote robotic control and telehealth to rural IoT deployments \cite{qiu2018can}. These domains introduce asymmetric vulnerabilities: low-power encoders at the edge may rely on powerful, semantically informed decoders in the cloud, resulting in one-sided failure modes when semantic codes are distorted or desynchronized. Edge–cloud heterogeneity, link asymmetry, and mobility further stress semantic security under real-world operating envelopes.

To consolidate these deployment-centric evaluation needs, Table~\ref{tab:evaluation-framework} summarizes core security objectives, evaluation methods, and representative failure modes that characterize deployment-aware SemCom security studies.

\begin{table*}[!t]
\centering
\caption{Deployment-Aware Evaluation Dimensions for Semantic Communication Security.}
\label{tab:evaluation-framework}
\begin{tabular}{|p{1.7cm}|p{4.7cm}|p{5.3cm}|p{4.5cm}|}
\hline
\textbf{Evaluation Goal} & \textbf{Security Objective} & \textbf{Evaluation Method} & \textbf{Representative Threats and Failure Modes} \\
\hline
\textbf{Semantic Robustness} & Meaning preservation and task utility under adversarial semantic perturbations & Semantic fidelity metrics, adversarial perturbation tests, semantic envelope violation rates & Knowledge misalignment, decoder drift, semantic corruption \\
\hline
\textbf{Semantic Leakage} & Privacy exposure from latent representations or shared knowledge & Mutual information analysis, inversion success rate, sensitive attribute recovery & Latent inversion, encoder leakage, cross-task information leakage \\
\hline
\textbf{Operational Resilience} & Tradeoffs between robustness, latency, and energy under adversarial and runtime constraints & Latency–robustness curves, energy per bit of meaning, graceful degradation tests & Time-critical failures, resource exhaustion, edge deployment stress \\
\hline
\textbf{Threat Adaptability} & System behavior under adaptive, multi-phase adversaries & Adaptive red-teaming agents, multi-stage and feedback-driven stress testing & Gray-box jammers, encoder poisoning, semantic desynchronization \\
\hline
\textbf{Deployment Robustness} & Resilience under real-world hardware, protocol, and update constraints & SDR-based testbeds, fault injection, runtime observability and monitoring & Semantic-aware jamming, asynchronous updates, decoder disagreement \\
\hline
\end{tabular}
\end{table*}

\textbf{Controlled fault injection} provides additional levers for testing adversarial resilience. SDR-based waveform injection can simulate semantic jamming at the physical layer. Compromised agents can be used to poison models or propagate misinformation through shared knowledge graphs. These setups further support fallback evaluations, including abstention, human-in-the-loop verification, and automated triggers for semantic anomalies.
To enable reproducible and modular evaluation, future efforts should prioritize open-source testbeds that integrate semantic communication pipelines, red-teaming agents, and runtime observability. Platforms such as Powder-RENEW \cite{powderrenew}, B5G Playground \cite{b5gplayground}, OpenAirInterface \cite{oai2023}, ARA \cite{ara2020testbed}, and RISE-6G \cite{rise6g2023} provide promising foundations for SemCom-aware security experimentation and system-level validation.

\subsection{Lessons Learned}
In summary, bridging design and deployment for secure semantic communication requires a system-aware approach that goes beyond isolated model hardening to address interdependencies among semantic fidelity, robustness, latency, and resource constraints. Practical deployment motivates security--utility operating envelopes for managing tradeoffs under real-world conditions. Table~\ref{tab:evaluation-framework} summarizes the evaluation dimensions in this section, highlighting metrics and methodologies for assessing SemCom security under adversarial and deployment-realistic conditions.

This section reveals that securing SemCom is fundamentally a co-design challenge, requiring tight integration of AI defenses with communication protocols, knowledge management, and network coordination. Success depends not only on advancing robustness techniques but also on developing adaptive, lightweight assurance mechanisms that can operate within strict operational envelopes, supported by continuous adversarial evaluation and deployment-focused validation. In this way, SemCom systems can achieve the resilience and trustworthiness required for real-world, safety-critical applications.

\section{Secure Semantic Communications in Practical Applications} \label{sec:applications}
SemCom has been gaining lots of attention in several application domains where distributed intelligence, perception, and coordinated decision-making play a central role. While SemCom offers considerable advantages in reducing communication load and aligning transmitted information with downstream tasks, its deployment also exposes security concerns that arise from the manipulation, degradation, or misinterpretation of meaning rather than raw syntactic data. This section surveys 4 typical applications and highlights the corresponding security challenges and threat surfaces of SemCom for each application.

\subsection{Cooperative Perception}
Cooperative perception allows multiple terminals (e.g., autonomous vehicles, robotic platforms, and sensor nodes) to share processed interpretations of their local observations to achieve a more complete, accurate, and robust understanding of the environment. Traditional cooperative perception methods often rely on exchanging high‑level features or compressed raw data \cite{chen2020performance, yang2023spatio, dao2024practical}. However, these traditional methods may suffer from information loss, limited adaptability to dynamic environments, and reduced perception accuracy under limited communication and computing resources. 
Considering the promises of SemCom, several works start to exploit SemCom to enhance the performance of cooperative perception by aligning transmitted information with downstream fusion tasks. 
For example, the authors in \cite{sheng2024semantic} proposed an importance-aware semantic encoder that prioritizes perceptually salient regions for cooperative automotive perception, reducing communication load while preserving safety-critical information. 
More recently, the authors in \cite{lu2025cross} designed a cross-modal SemCom framework to support heterogeneous perception modalities (e.g., camera and LiDAR) during collaborative perception. In general, these works indicate the significant potential of SemCom in cooperative perception.

However, incorporating SemCom into multi-agent cooperative perception introduces new security risks, as adversarial agents may inject false detections, misleading trajectories, or fabricated hazards, and semantic desynchronization can lead to inconsistent world models. Conventional channel security mechanisms, such as authentication and encryption, do not address semantic integrity violations, where packets are syntactically correct but adversarial at the meaning level. This gap motivates secure SemCom mechanisms that ensure shared semantics remain trustworthy and resilient under adversarial conditions.

Secure cooperative perception can be realized through the four layers of semantic defense outlined in Section~\ref{sec:4layered_defense}. Encoder- and decoder-level defenses improve robustness against semantic poisoning through constrained and adversarially robust representations. Channel-level mechanisms, including semantic-aware coding and unequal protection, preserve safety-critical semantics under interference and jamming. Knowledge-base defenses enforce semantic provenance and consistency across shared world models, mitigating desynchronization. Network-level defenses further support trust, misbehavior detection, and semantic consensus to ensure reliable collective perception in adversarial environments.

\subsection{Remote Robotic Systems}
Remote robotic systems, referred to as telerobotics or teleoperation—enable human operators to control robotic platforms at a distance by coupling human cognition with machine embodiment in remote or hazardous environments \cite{Sheridan1992Telerobotics, Niemeyer2008telerobotics}. Leveraging SemCom for remote robotics has recently emerged as a means to reduce communication overhead and improve task alignment/completion by transmitting task-level semantics instead of raw sensor streams. In \cite{Talli2023SemanticEffective}, the authors introduced SemCom strategies for remote robotic manipulation, showing that task-aware feature selection leads to reduced bandwidth consumption without degrading control performance. Knowledge-based SemCom frameworks for robotic edge intelligence have further been proposed to match transmitted task semantics with stored knowledge graphs for low-latency robotic exploration and control assistance \cite{zeng2024knowledge}. 

While SemCom improves communication efficiency in remote robotic systems, it also introduces semantic security risks. Adversaries may manipulate affordance descriptors, inject false constraints, or corrupt task objectives, leading to unsafe behavior or task failure. Semantic ambiguity between operator intent and robot execution can further be exploited at the meaning level, creating mismatches between perceived and actual system states that are not addressed by traditional control-channel security mechanisms.

Secure SemCom for remote robotics can be realized through the four-layer defense framework in Section~\ref{sec:4layered_defense}. Encoder- and decoder-level defenses enforce robust and constrained semantic representations, channel-level mechanisms protect safety-critical semantics under interference, knowledge-base defenses preserve consistency between task models and environment semantics, and network-level mechanisms enable trust, misbehavior detection, and semantic consensus among distributed robots. Together, these layers extend security from channel integrity to semantic integrity, enabling safe and reliable remote robotic operation.

\subsection{Agentic AI Systems}

Agentic AI systems refer to autonomous or semi-autonomous agents that perceive environments, plan actions, and interact with digital or physical systems to achieve specific goals.
These systems integrate perception, decision-making, and actuation loops to operate with minimal human intervention \cite{acharya2025agentic, sapkota2025ai}. Recent research has explored the use of SemCom to support agentic AI coordination by transmitting task-oriented semantics instead of raw data of state observations or control messages. For example, in \cite{li2024semantic}, a SemCom framework for multi-agent coordination was proposed to compress and align exchanged semantics for cooperative decision-making. Knowledge-enhanced SemCom has also been leveraged to support semantic alignment among autonomous agents via shared or partially shared knowledge bases, enabling more robust inference and coordination \cite{zeng2024knowledge}. Moreover, cognitive and goal-oriented SemCom frameworks have demonstrated improved communication efficiency in distributed planning and navigation tasks, where semantics are ranked according to their value for downstream decision modules \cite{jiang2025large}. 

However, SemCom-enabled agentic AI systems introduce new semantic security and trust challenges \cite{tang2026rethinking}. Malicious agents may manipulate goals, beliefs, or task outcomes to influence collective planning, while semantic misalignment can lead to inconsistent task interpretations. Conventional cryptographic and access-control mechanisms prevent unauthorized participation but do not guarantee semantic correctness or consistency across distributed decision-making processes, making semantic integrity a critical concern in agentic AI systems.

Secure SemCom for agentic AI can be realized through the four-layer defense framework in Section~\ref{sec:4layered_defense}. Encoder-and decoder-level defenses enforce robust and verifiable semantic representations, channel-level mechanisms protect critical task semantics under interference, knowledge-base defenses preserve consistency of shared world models through provenance and synchronization, and network-level mechanisms enable misbehavior detection and semantic consensus. Together, these layers extend security from communication integrity to semantic integrity, enabling trustworthy coordination among distributed agentic AI systems.

\subsection{Smart Manufacturing}

With the development of Industry 4.0/5.0, smart manufacturing has been envisioning highly automated, interconnected, and data-driven production systems that integrate sensing, computation, and control across cyber-physical infrastructures, factory equipment, and supervisory systems \cite{zuehlke2010smartfactory, lee2015cyber}. 
Recently, SemCom has been exploited to enhance smart manufacturing by transmitting task-oriented or context-aware semantics rather than raw sensor streams. For example, the work of \cite{luo2022semantic} discusses how semantic communications enable intelligent, goal-centric machine interactions in smart factories, improving operational efficiency by transmitting only the semantic intent of monitoring and control information. Besides, the authors in \cite{pokhrel2022learning} develop a SemCom framework with continuous federated reinforcement learning capabilities for smart factories and industry IoT scenarios. Moreover, digital twin-driven semantic communication has been investigated for synchronizing simulation models with factory-floor equipment, exploiting semantic compression to reduce update overhead between physical and virtual entities \cite{jagatheesaperumal2023semantic, thomas2023causal}. 

Integrating SemCom into smart manufacturing introduces semantic-level vulnerabilities, as adversarial nodes may manipulate fault semantics, scheduling intent, or digital-twin updates to distort factory-state representations. 
For example, in a robotic assembly line where sensors transmit semantic fault descriptors rather than raw data, an adversary can subtly alter these semantics to misclassify a critical fault as benign, delaying maintenance and causing cumulative equipment damage. 
Unlike traditional industrial cybersecurity, which focuses on protocol-level attacks or unauthorized actuation, semantic attacks target meaning and task context, potentially causing performance degradation or safety hazards without violating syntactic integrity.

Secure SemCom for smart manufacturing can be achieved through the four-layer defense framework in Section~\ref{sec:4layered_defense}. Encoder- and decoder-level defenses enforce physically plausible semantic representations, channel-level mechanisms protect critical semantics under interference, knowledge-base defenses ensure secure digital-twin synchronization and semantic provenance, and network-level mechanisms enable trust-aware semantic fusion and misbehavior detection. Together, these layers protect semantic integrity in industrial cyber-physical systems, enabling safe and reliable manufacturing operations.

\subsection{Lessons Learned}
Across domains, SemCom is promising to reduce communication overhead, align transmitted information with tasks, and support distributed perception–decision–action loops. 
Meanwhile, SemCom also creates new threats where adversaries manipulate meaning without violating syntactic integrity. Layered defenses identified in Section \ref{sec:4layered_defense} are essential to ensure semantic integrity, provenance, and trust. 
The tradeoff between security and efficiency could vary across different domain requirement, emphasizing that secure SemCom is context-dependent and requires joint optimization across robustness, semantic fidelity, and real-time constraints. 

\section{Open Research Directions} \label{sec:open}
While recent advances in SemCom and robust AI have established foundational principles, translating these developments into secure and deployable systems for real-world wireless environments remains a significant challenge. This section outlines key research directions needed to move from proof-of-concept demonstrations to trustworthy and integrated SemCom deployments.

\vspace{.5em}
\paragraph{Human-Centered Evaluation and Semantic Trust Metrics}
Traditional metrics such as bit error rate (BER), classification accuracy, or BLEU scores fail to capture the semantically grounded objectives of secure communication, including meaning preservation, intent alignment, and robustness to adversarial or contextual shifts. New evaluation metrics are needed that better reflect the goals of SemCom systems, namely preserving semantic correctness, enabling downstream task success, and adapting gracefully to uncertainty and context changes.

Effective metrics should be task-aware, model-agnostic, and interpretable. For example, semantic distortion may be quantified through changes in control accuracy, degradation in policy performance, or inconsistencies at the concept level. Besides, human-in-the-loop evaluation should be incorporated \cite{glikson2020human,kumar2024applications}. In safety-critical domains such as autonomous driving, healthcare, and mission planning, systems must expose meaningful indicators—including confidence scores, semantic anomaly signals, and fallback prompts—that allow human operators to identify and mitigate semantic degradation or ambiguity during runtime.

Future benchmarks should further include human-centered robustness measures, such as perceived utility loss, semantic disagreement rates, or erosion of user trust \cite{glikson2020human,lai2023towards,tocchetti2025ai}. These measures, collected through user studies or simulated evaluation agents, can complement task-level metrics and reveal failure modes that are not visible through standard performance curves. Recent work on semantic intent modeling and shared control interfaces suggests promising pathways for aligning semantic representations with human goals and expectations \cite{wang2025capsule,vaithilingam2025semantic}.

Finally, standardized evaluation suites and red-teaming benchmarks discussed in Section~\ref{sec:deployment} should explicitly support real-time alerts, abstention mechanisms, and interaction modalities between AI agents and human supervisors. Incorporating human feedback into evaluation protocols is essential for developing SemCom systems that remain robust under adversarial conditions and operational uncertainty.

\vspace{.5em}
\paragraph{Composable System-Wide Assurance}
Robustness guarantees derived in isolation often fail when systems are integrated \cite{hooker2021unrestricted}. Achieving system-wide assurance requires frameworks that can compose guarantees across semantic encoders, channel coders, semantic decoders, and application logic, even under runtime adaptation, partial failure, or adversarial interference.

Future research should develop composable assurance frameworks that (i) formalize how per-layer certificates can be composed and verified across protocol stacks, (ii) support runtime reconfiguration in response to envelope violations or degraded trust signals, and (iii) expose control hooks that enable graceful degradation and fallback behaviors. Such frameworks may leverage modular runtime governors, programmable semantics-aware control planes, or hybrid verification techniques to enforce multi-stage robustness guarantees.

\vspace{.5em}
\paragraph{Secure Semantic Knowledge Management}
SemCom pipelines increasingly rely on shared semantic assets, including pretrained models, task embeddings, and knowledge graphs. These assets are vulnerable to poisoning, exfiltration, and unauthorized adaptation, which can silently degrade system performance or introduce subtle backdoors \cite{carlini2021poisoning,sabir2021machine}.

Securing semantic assets requires protocols that ensure provenance, integrity, access control, and confidentiality, while maintaining low-latency inference and lightweight memory footprints. Promising research directions include cryptographic tagging of semantic units, differential privacy mechanisms for knowledge sharing \cite{zhao2022survey}, and federated update schemes with semantic-aware version control \cite{nguyen2024efficient}. In particular, the rise of knowledge-centric AI pipelines demands renewed attention to knowledge hygiene and integrity in SemCom, where shared models can become a vector for semantic compromise \cite{deveaux2021definition}. Maintaining semantic knowledge hygiene is critical for reliable inference and trustworthy decision-making in distributed SemCom settings.

\vspace{.5em}
\paragraph{Interoperability with Cryptographic Security}
Semantic security must coexist with traditional cryptographic guarantees, including confidentiality, authentication, and integrity \cite{golomb2005signal,kartalopoulos2006primer,sklavos2017wireless}. However, combining semantic processing with cryptographic primitives introduces nontrivial interactions. For instance, encryption may obscure semantic content from downstream inference modules, while semantic inspection may conflict with confidentiality requirements.

Addressing this challenge calls for joint designs that support encrypted semantic processing, secure feature extraction, and zero-knowledge verification of semantic properties. Research on semantically aware encryption schemes, homomorphic operations over semantic embeddings \cite{acar2018survey}, and cryptographic protocols for semantic provenance and policy enforcement can enable principled integration of security and semantics. Integrating privacy-preserving mechanisms, such as differential privacy \cite{zhao2022survey} or secure multiparty computation \cite{lindell2020secure}, could support semantic robustness in cross-device SemCom scenarios, including vehicular edge networks or federated smart sensors, without compromising trust or confidentiality.

\vspace{.5em}
\paragraph{Trustworthiness Dimensions in Semantic Communication}
Beyond technical security mechanisms, SemCom systems must align with broader AI assurance principles \cite{li2023trustworthy}, such as fairness, explainability, and accountability, to support safe and ethical deployment. While preceding sections address robustness, abstention, and cryptographic protection, these capabilities must be situated within frameworks that promote transparent and human-aligned communication \cite{doshi2021towards}.

For example, semantic encoders and inference modules should be evaluated for potential bias amplification, particularly in domains like healthcare or autonomous systems \cite{mehrabi2021survey,rajkomar2018ensuring}. Explainability remains underdeveloped: when semantic miscommunication occurs, systems should expose interpretable reasoning traces or semantic provenance to aid diagnosis and recovery \cite{lipton2018mythos,wang2025capsule}. Accountability is also essential, where SemCom pipelines should support traceable decision chains across multi-agent, multi-model systems to assign responsibility when semantic failures arise \cite{mittelstadt2016ethics}.

Future research should embed these trust dimensions into both system design and evaluation, such as fairness-aware encoder/decoder training, interpretable semantic reconstruction, contract-based traceability, and human-verifiable semantic intents. By explicitly aligning semantic security with trustworthiness goals, SemCom can evolve into an ethically grounded infrastructure for AI-native communication.

\section{Conclusion}
SemCom transforms wireless systems by shifting the focus from symbol reproduction to preserving task-relevant meaning, introducing AI-induced dependencies that render conventional communication security models insufficient. By embedding learned models, shared knowledge, and inference into the communication pipeline, SemCom exposes new semantic-level attack surfaces in which failures may arise even when lower-layer reliability and cryptographic protections remain intact, making semantic integrity a system-level security challenge.

This survey presented a system-level, AI-defense–oriented synthesis of security in SemCom, organized around an AI-centric threat model and a pipeline-spanning taxonomy of countermeasures. Our analysis reveals fundamental gaps in robustness transfer across layers and highlights shared semantic knowledge as a first-class vulnerability that demands explicit protection and governance rather than isolated model hardening.

Looking ahead, securing SemCom requires principled system design that integrates human-centered evaluation, composable assurance, and secure semantic knowledge management. Future SemCom systems should expose verifiable semantic interfaces and adopt evaluation practices that capture task performance, user trust, and robustness under adversarial conditions, within explicit security–utility operating envelopes. Interoperability with cryptographic mechanisms and alignment with AI assurance principles, such as fairness, explainability, and accountability, are essential. By unifying these elements, this survey aims to guide the development of robust, trustworthy, and deployable SemCom systems for real-world and safety-critical applications.

\bibliography{refs.bib}

\begin{thebibliography}{100}
\providecommand{\url}[1]{#1}
\csname url@samestyle\endcsname
\providecommand{\newblock}{\relax}
\providecommand{\bibinfo}[2]{#2}
\providecommand{\BIBentrySTDinterwordspacing}{\spaceskip=0pt\relax}
\providecommand{\BIBentryALTinterwordstretchfactor}{4}
\providecommand{\BIBentryALTinterwordspacing}{\spaceskip=\fontdimen2\font plus
\BIBentryALTinterwordstretchfactor\fontdimen3\font minus \fontdimen4\font\relax}
\providecommand{\BIBforeignlanguage}[2]{{%
\expandafter\ifx\csname l@#1\endcsname\relax
\typeout{** WARNING: IEEEtran.bst: No hyphenation pattern has been}%
\typeout{** loaded for the language `#1'. Using the pattern for}%
\typeout{** the default language instead.}%
\else
\language=\csname l@#1\endcsname
\fi
#2}}
\providecommand{\BIBdecl}{\relax}
\BIBdecl

\bibitem{weaver1949recent}
W.~Weaver, \emph{Recent Contributions to the Mathematical Theory of Communication}.\hskip 1em plus 0.5em minus 0.4em\relax University of Illinois Press, 1949.

\bibitem{xie2021deep}
H.~Xie, Z.~Qin, G.~Y. Li, and B.-H. Juang, ``Deep learning enabled semantic communication systems,'' \emph{IEEE Trans. Signal Process.}, vol.~69, pp. 2663--2675, 2021.

\bibitem{bourtsoulatze2019deep}
E.~Bourtsoulatze, D.~B. Kurka, and D.~G{\"u}nd{\"u}z, ``Deep joint source-channel coding for wireless image transmission,'' \emph{IEEE Trans. Cogn. Commun. Netw.}, vol.~5, no.~3, pp. 567--579, 2019.

\bibitem{yang2022semantic}
W.~Yang, H.~Du, Z.~Q. Liew, W.~Y.~B. Lim, Z.~Xiong, D.~Niyato, X.~Chi, X.~Shen, and C.~Miao, ``Semantic communications for future internet: Fundamentals, applications, and challenges,'' \emph{IEEE Commun. Surveys Tuts.}, vol.~25, no.~1, pp. 213--250, 2022.

\bibitem{liang2023vista}
C.~Liang, X.~Deng, Y.~Sun, R.~Cheng, L.~Xia, D.~Niyato, and M.~A. Imran, ``{VISTA: Video Transmission over A Semantic Communication Approach},'' in \emph{Proc. IEEE Int. Conf. Commun. Workshops (ICC Workshops)}.\hskip 1em plus 0.5em minus 0.4em\relax IEEE, 2023, pp. 1777--1782.

\bibitem{chaccour2024less}
C.~Chaccour, W.~Saad, M.~Debbah, Z.~Han, and H.~V. Poor, ``Less data, more knowledge: Building next-generation semantic communication networks,'' \emph{IEEE Commun. Surveys Tuts.}, vol.~27, no.~1, pp. 37--76, 2024.

\bibitem{xin2024entropy}
Y.~Xin, M.~Chen, and J.~Zhang, ``Semantic communication: A survey of its theoretical foundations and applications,'' \emph{Entropy}, vol.~26, no.~7, p. 547, 2024.

\bibitem{guo2024survey}
S.~Guo, Y.~Wang, N.~Zhang, Z.~Su, T.~H. Luan, Z.~Tian, and X.~Shen, ``A survey on semantic communication networks: Architecture, security, and privacy,'' \emph{IEEE Commun. Surveys Tuts.}, vol.~27, no.~5, pp. 2860--2894, 2025.

\bibitem{yang2024secure}
Z.~Yang, M.~Chen, G.~Li, Y.~Yang, and Z.~Zhang, ``Secure semantic communications: Fundamentals and challenges,'' \emph{IEEE Netw.}, vol.~38, no.~6, pp. 513--520, 2024.

\bibitem{sagduyu2023semantic}
Y.~E. Sagduyu and S.~Ulukus, ``Is semantic communication secure? a tale of multi-domain vulnerabilities,'' \emph{IEEE Commun. Mag.}, vol.~61, no.~11, pp. 40--46, 2023.

\bibitem{chen2025secure}
W.~Chen, Q.~Yang, Y.~Jia, J.~Pan, S.~Shao, J.~Dai, M.~Tao, and P.~Zhang, ``Secure digital semantic communications: Fundamentals, challenges, and opportunities,'' \emph{arXiv preprint arXiv:2512.24602}, 2025.

\bibitem{yuan2019adversarial}
X.~Yuan, P.~He, Q.~Zhu, and X.~Li, ``Adversarial examples: Attacks and defenses for deep learning,'' \emph{IEEE Trans. Neural Netw. Learn. Syst.}, vol.~30, no.~9, pp. 2805--2824, 2019.

\bibitem{li2023trustworthy}
B.~Li, P.~Qi, B.~Liu, S.~Di, J.~Liu, J.~Pei, J.~Yi, and B.~Zhou, ``Trustworthy {AI}: From principles to practices,'' \emph{ACM Comput. Surveys}, vol.~55, no.~9, pp. 1--46, 2023.

\bibitem{wei2025trustworthy}
W.~Wei and L.~Liu, ``Trustworthy distributed {AI} systems: Robustness, privacy, and governance,'' \emph{ACM Comput. Surveys}, vol.~57, no.~6, pp. 1--42, 2025.

\bibitem{wen2023survey}
J.~Wen, Z.~Zhang, Y.~Lan, Z.~Cui, J.~Cai, and W.~Zhang, ``A survey on federated learning: challenges and applications,'' \emph{Int. J. Mach. Learn. Cybern.}, vol.~14, no.~2, pp. 513--535, 2023.

\bibitem{won2024resource}
D.~Won, G.~Woraphonbenjakul, A.~B. Wondmagegn, A.-T. Tran, D.~Lee, D.~S. Lakew, and S.~Cho, ``Resource management, security, and privacy issues in semantic communications: A survey,'' \emph{IEEE Commun. Surveys Tuts.}, vol.~27, no.~3, pp. 1758--1797, 2025.

\bibitem{li2024robustsemcom}
X.~Zhang, \emph{Robust Semantic Communications and Privacy Protection}.\hskip 1em plus 0.5em minus 0.4em\relax Wiley, 2024, pp. 67--86.

\bibitem{meng2025securesemcom}
R.~Meng, S.~Gao, D.~Fan, H.~Gao, Y.~Wang, X.~Xu, B.~Wang, S.~Lv, Z.~Zhang, M.~Sun, S.~Han, C.~Dong, X.~Tao, and P.~Zhang, ``A survey of secure semantic communications,'' \emph{J. Netw. Comput. Appl.}, p. 104181, 2025.

\bibitem{shen2023secure}
M.~Shen, J.~Wang, H.~Du, D.~Niyato, X.~Tang, J.~Kang, Y.~Ding, and L.~Zhu, ``Secure semantic communications: Challenges, approaches, and opportunities,'' \emph{IEEE Netw.}, vol.~38, no.~4, pp. 197--206, 2023.

\bibitem{liang2025safeguarded}
C.~Liang, Y.~Sun, D.~Liu, D.~Yu, and M.~A. Imran, ``{Safeguarded AI-Driven Semantic Communication: Design Principles, Architecture, and Challenges},'' \emph{IEEE Commun. Standards Mag.}, 2025.

\bibitem{oshea2017intro}
T.~O'Shea and J.~Hoydis, ``An introduction to deep learning for the physical layer,'' \emph{IEEE Trans. Cogn. Commun. Netw.}, vol.~3, no.~4, pp. 563--575, 2017.

\bibitem{jiang2022semantic}
S.~Jiang, Y.~Liu, Y.~Zhang, P.~Luo, K.~Cao, J.~Xiong, H.~Zhao, and J.~Wei, ``Reliable semantic communication system enabled by knowledge graph,'' \emph{Entropy}, vol.~24, no.~6, p. 846, 2022.

\bibitem{wang2023knowledge}
B.~Wang, R.~Li, J.~Zhu, Z.~Zhao, and H.~Zhang, ``Knowledge enhanced semantic communication receiver,'' \emph{IEEE Commun. Lett.}, vol.~27, no.~7, pp. 1794--1798, 2023.

\bibitem{hello2024semantic}
N.~Hello, P.~Di~Lorenzo, and E.~C. Strinati, ``Semantic communication enhanced by knowledge graph representation learning,'' in \emph{Proc. IEEE Int. Workshop Signal Process. Adv. Wireless Commun. (SPAWC)}.\hskip 1em plus 0.5em minus 0.4em\relax IEEE, 2024, pp. 876--880.

\bibitem{liang2025knowledge}
C.~Liang, Y.~Sun, D.~Niyato, and M.~A. Imran, ``{Knowledge Graph Fusion Based Semantic Communication Framework},'' \emph{IEEE Trans. Mobile Comput.}, vol.~24, no.~11, pp. 11\,416--11\,429, 2025.

\bibitem{liu2023transformer}
S.~Liu, Z.~Gao, G.~Chen, Y.~Su, and L.~Peng, ``Transformer-based joint source channel coding for textual semantic communication,'' in \emph{Proc. IEEE/CIC Int. Conf. Commun. China (ICCC)}, 2023, pp. 1--6.

\bibitem{xu2023deep}
J.~Xu, T.-Y. Tung, B.~Ai, W.~Chen, Y.~Sun, and D.~G{\"u}nd{\"u}z, ``Deep joint source-channel coding for semantic communications,'' \emph{IEEE Commun. Mag.}, vol.~61, no.~11, pp. 42--48, 2023.

\bibitem{zhang2023prompt}
L.~Zhang, M.~Das, Y.~Sun, D.~Niyato, and X.~Yuan, ``Prompt-based transceiver cooperation for semantic communications with domain-incremental background knowledge,'' in \emph{Proc. IEEE Global Commun. Conf. (GLOBECOM)}.\hskip 1em plus 0.5em minus 0.4em\relax IEEE, 2023, pp. 2087--2092.

\bibitem{wang2025llmsc}
Z.~Wang, L.~Zou, S.~Wei, K.~Li, F.~Liao, H.~Mi, and R.~Lai, ``Llm-sc: Large language model--enabled text semantic communication systems,'' \emph{Applied Sciences}, vol.~15, no.~13, p. 7227, 2025.

\bibitem{salehi2025llm}
S.~Salehi, M.~Erol-Kantarci, and D.~Niyato, ``Llm-enabled data transmission in end-to-end semantic communication,'' \emph{arXiv preprint arXiv:2504.07431}, 2025.

\bibitem{cheng2024wireless}
R.~Cheng, Y.~Sun, D.~Niyato, L.~Zhang, L.~Zhang, and M.~A. Imran, ``{A Wireless AI-generated Content (AIGC) Provisioning Framework Empowered by Semantic Communication},'' \emph{IEEE Trans. Mobile Comput.}, vol.~24, no.~3, pp. 2137--2150, 2024.

\bibitem{xia2025generative}
L.~Xia, Y.~Sun, C.~Liang, L.~Zhang, M.~A. Imran, and D.~Niyato, ``{Generative AI for Semantic Communication: Architecture, Challenges, and Outlook},'' \emph{IEEE Wireless Commun.}, vol.~32, no.~1, pp. 132--140, 2025.

\bibitem{liang2024generative}
C.~Liang, H.~Du, Y.~Sun, D.~Niyato, J.~Kang, D.~Zhao, and M.~A. Imran, ``{Generative AI-driven Semantic Communication Networks: Architecture, Technologies and Applications},'' \emph{IEEE Trans. Cogn. Commun. Netw.}, 2024.

\bibitem{zhao2025multi}
H.~Zhao, H.~Li, D.~Xu, S.~Song, and K.~B. Letaief, ``Multi-modal self-supervised semantic communication,'' in \emph{Proc. IEEE Int. Mediterranean Conf. Commun. Netw. (MeditCom)}, 2025, pp. 1--6.

\bibitem{tang2024contrastive}
S.~Tang, Q.~Yang, L.~Fan, X.~Lei, A.~Nallanathan, and G.~K. Karagiannidis, ``Contrastive learning-based semantic communications,'' \emph{IEEE Trans. Commun.}, vol.~72, no.~10, pp. 6328--6343, 2024.

\bibitem{zou2025self}
J.~Zou, Z.~Wan, F.~Wang, S.~Ye, and S.~Liu, ``The self supervised multimodal semantic transmission mechanism for complex network environments,'' \emph{Scientific Reports}, vol.~15, no.~1, p. 29899, 2025.

\bibitem{yan2025review}
X.~Yan, F.~Xiumei, K.-L.~A. Yau, X.~Zhixin, M.~Rui, and Y.~Gang, ``A review of reinforcement learning for semantic communications,'' \emph{J. Netw. Syst. Manage.}, vol.~33, no.~3, p.~52, 2025.

\bibitem{lu2021rl_semcom}
K.~Lu, R.~Li, X.~Chen, Z.~Zhao, and H.~Zhang, ``Reinforcement learning-powered semantic communication via semantic similarity,'' \emph{arXiv preprint arXiv:2108.12121}, 2021.

\bibitem{zhao2023joint}
F.~Zhao, G.~Bagwe, E.~Mohammed, L.~Feng, L.~Zhang, and Y.~Sun, ``Joint computing resource and bandwidth allocation for semantic communication networks,'' in \emph{Proc. IEEE Veh. Technol. Conf. (VTC)}, 2023, pp. 1--5.

\bibitem{xu2024federated}
J.~Xu, H.~Yao, R.~Zhang, T.~Mai, S.~Huang, and S.~Guo, ``Federated learning powered semantic communication for uav swarm cooperation,'' \emph{IEEE Wireless Commun.}, vol.~31, no.~4, pp. 140--146, 2024.

\bibitem{nguyen2024efficient}
L.~X. Nguyen, H.~Q. Le, Y.~L. Tun, P.~S. Aung, Y.~K. Tun, Z.~Han, and C.~S. Hong, ``An efficient federated learning framework for training semantic communication systems,'' \emph{IEEE Trans. Veh. Technol.}, vol.~73, no.~10, pp. 15\,872--15\,877, 2024.

\bibitem{si2024post}
P.~Si, R.~Liu, L.~Qian, J.~Zhao, and K.-Y. Lam, ``Post-deployment fine-tunable semantic communication,'' \emph{IEEE Trans. Wireless Commun.}, vol.~24, no.~1, pp. 35--50, 2024.

\bibitem{zhang2025o2sc}
G.~Zhang, K.~Kang, Y.~Cai, Q.~Hu, Y.~C. Eldar, and A.~L. Swindlehurst, ``{O2SC}: Realizing channel-adaptive semantic communication with one-shot online-learning,'' \emph{IEEE Trans. Commun.}, vol.~73, no.~5, pp. 3268--3282, 2025.

\bibitem{liu2024ofdm}
C.~Liu, C.~Guo, Y.~Yang, W.~Ni, and T.~Q. Quek, ``{OFDM}-based digital semantic communication with importance awareness,'' \emph{IEEE Trans. Commun.}, vol.~72, no.~10, pp. 6301--6315, 2024.

\bibitem{liu2025knowledge}
X.~Liu, Y.~Sun, R.~Cheng, L.~Xia, H.~Abumarshoud, L.~Zhang, and M.~A. Imran, ``Knowledge-assisted privacy preserving in semantic communication,'' \emph{IEEE Wireless Commun.}, vol.~32, no.~2, pp. 76--83, 2025.

\bibitem{goodfellow2014explaining}
I.~J. Goodfellow, J.~Shlens, and C.~Szegedy, ``Explaining and harnessing adversarial examples,'' \emph{arXiv preprint arXiv:1412.6572}, 2014.

\bibitem{biggio2018wild}
B.~Biggio and F.~Roli, ``Wild patterns: Ten years after the rise of adversarial machine learning,'' in \emph{Proc. ACM SIGSAC Conf. Comput. Commun. Secur. (CCS)}, 2018, pp. 2154--2156.

\bibitem{madry2018towards}
A.~Madry, A.~Makelov, L.~Schmidt, D.~Tsipras, and A.~Vladu, ``Towards deep learning models resistant to adversarial attacks,'' in \emph{Proc. Int. Conf. Learn. Representations (ICLR)}, 2017.

\bibitem{zhang2019theoretically}
H.~Zhang, Y.~Yu, J.~Jiao, E.~Xing, L.~El~Ghaoui, and M.~Jordan, ``Theoretically principled trade-off between robustness and accuracy,'' in \emph{Proc. Int. Conf. Mach. Learn. (ICML)}.\hskip 1em plus 0.5em minus 0.4em\relax PMLR, 2019, pp. 7472--7482.

\bibitem{sinha2018certifying}
A.~Sinha, H.~Namkoong, and J.~Duchi, ``Certifying some distributional robustness with principled adversarial training,'' \emph{Proc. Int. Conf. Learn. Representations (ICLR)}, 2018.

\bibitem{cohen2019certified}
J.~Cohen, E.~Rosenfeld, and Z.~Kolter, ``Certified adversarial robustness via randomized smoothing,'' in \emph{Proc. Int. Conf. Mach. Learn. (ICML)}.\hskip 1em plus 0.5em minus 0.4em\relax PMLR, 2019, pp. 1310--1320.

\bibitem{zou2016survey}
Y.~Zou, J.~Zhu, X.~Wang, and L.~Hanzo, ``A survey on wireless security: Technical challenges, recent advances, and future trends,'' \emph{Proc. IEEE}, vol. 104, no.~9, pp. 1727--1765, 2016.

\bibitem{nichols2001wireless}
R.~K. Nichols, P.~Lekkas, and P.~C. Lekkas, \emph{Wireless security}.\hskip 1em plus 0.5em minus 0.4em\relax McGraw-Hill Professional Publishing, 2001.

\bibitem{biggio2012poisoning}
B.~Biggio, B.~Nelson, and P.~Laskov, ``Poisoning attacks against support vector machines,'' \emph{arXiv preprint arXiv:1206.6389}, 2012.

\bibitem{gu2017badnets}
T.~Gu, B.~Dolan-Gavitt, and S.~Garg, ``Badnets: Identifying vulnerabilities in the machine learning model supply chain,'' \emph{arXiv preprint arXiv:1708.06733}, 2017.

\bibitem{bhagoji2019analyzing}
A.~N. Bhagoji, S.~Chakraborty, P.~Mittal, and S.~Calo, ``Analyzing federated learning through an adversarial lens,'' in \emph{Proc. Int. Conf. Mach. Learn. (ICML)}.\hskip 1em plus 0.5em minus 0.4em\relax PMLR, 2019, pp. 634--643.

\bibitem{wang2020attack}
H.~Wang, K.~Sreenivasan, S.~Rajput, H.~Vishwakarma, S.~Agarwal, J.-y. Sohn, K.~Lee, and D.~Papailiopoulos, ``Attack of the tails: Yes, you really can backdoor federated learning,'' \emph{Proc. Conf. Neural Inf. Process. Syst. (NeurIPS)}, vol.~33, pp. 16\,070--16\,084, 2020.

\bibitem{ilyas2019adversarial}
A.~Ilyas, S.~Santurkar, D.~Tsipras, L.~Engstrom, B.~Tran, and A.~Madry, ``Adversarial examples are not bugs, they are features,'' \emph{Proc. Conf. Neural Inf. Process. Syst. (NeurIPS)}, vol.~32, 2019.

\bibitem{shokri2017membership}
R.~Shokri, M.~Stronati, C.~Song, and V.~Shmatikov, ``Membership inference attacks against machine learning models,'' in \emph{Proc. IEEE Symp. Secur. Priv. (SP)}.\hskip 1em plus 0.5em minus 0.4em\relax IEEE, 2017, pp. 3--18.

\bibitem{fredrikson2015model}
M.~Fredrikson, S.~Jha, and T.~Ristenpart, ``Model inversion attacks that exploit confidence information and basic countermeasures,'' in \emph{Proc. ACM SIGSAC Conf. Comput. Commun. Secur. (CCS)}, 2015, pp. 1322--1333.

\bibitem{tramer2016stealing}
F.~Tram{\`e}r, F.~Zhang, A.~Juels, M.~K. Reiter, and T.~Ristenpart, ``Stealing machine learning models via prediction $\{$APIs$\}$,'' in \emph{Proc. USENIX Secur. Symp.}, 2016, pp. 601--618.

\bibitem{nist80012}
\BIBentryALTinterwordspacing
{National Institute of Standards and Technology}, ``An introduction to information security,'' NIST, Tech. Rep. Special Publication 800-12 Rev. 1, 2017. [Online]. Available: \url{https://nvlpubs.nist.gov/nistpubs/SpecialPublications/NIST.SP.800-12r1.pdf}
\BIBentrySTDinterwordspacing

\bibitem{huang2011adversarial}
L.~Huang, A.~D. Joseph, B.~Nelson, B.~I. Rubinstein, and J.~D. Tygar, ``Adversarial machine learning,'' in \emph{Proc. ACM Workshop Secur. Artif. Intell. (AISec)}, 2011, pp. 43--58.

\bibitem{barreno2010security}
M.~Barreno, B.~Nelson, A.~D. Joseph, and J.~D. Tygar, ``The security of machine learning,'' \emph{Machine learning}, vol.~81, no.~2, pp. 121--148, 2010.

\bibitem{papernot2016towards}
N.~Papernot, P.~McDaniel, A.~Sinha, and M.~Wellman, ``Towards the science of security and privacy in machine learning,'' \emph{arXiv preprint arXiv:1611.03814}, 2016.

\bibitem{huang2017adversarial}
S.~Huang, N.~Papernot, I.~Goodfellow, Y.~Duan, and P.~Abbeel, ``Adversarial attacks on neural network policies,'' \emph{arXiv preprint arXiv:1702.02284}, 2017.

\bibitem{papernot2017practical}
N.~Papernot, P.~McDaniel, I.~Goodfellow, S.~Jha, Z.~B. Celik, and A.~Swami, ``Practical black-box attacks against machine learning,'' in \emph{Proc. ACM Asia Conf. Comput. Commun. Secur. (ASIACCS)}, 2017, pp. 506--519.

\bibitem{hendrycks2020augmix}
D.~Hendrycks, N.~Mu, E.~D. Cubuk, B.~Zoph, J.~Gilmer, and B.~Lakshminarayanan, ``Augmix: A simple data processing method to improve robustness and uncertainty,'' in \emph{Proc. Int. Conf. Learn. Representations (ICLR)}, 2020.

\bibitem{zhang2018mixup}
H.~Zhang, M.~Cisse, Y.~N. Dauphin, and D.~Lopez-Paz, ``mixup: Beyond empirical risk minimization,'' in \emph{Proc. Int. Conf. Learn. Representations (ICLR)}, 2018.

\bibitem{yun2019cutmix}
S.~Yun, D.~Han, S.~J. Oh, S.~Chun, J.~Choe, and Y.~Yoo, ``Cutmix: Regularization strategy to train strong classifiers with localizable features,'' in \emph{Proc. IEEE/CVF Int. Conf. Comput. Vision (ICCV)}, 2019.

\bibitem{cubuk2019autoaugment}
E.~D. Cubuk, B.~Zoph, D.~Mane, V.~Vasudevan, and Q.~V. Le, ``Autoaugment: Learning augmentation policies from data,'' in \emph{Proc. IEEE/CVF Conf. Comput. Vision Pattern Recognit. (CVPR)}, 2019.

\bibitem{rifai2011contractive}
S.~Rifai, P.~Vincent, X.~Muller, X.~Glorot, and Y.~Bengio, ``Contractive auto-encoders: Explicit invariance during feature extraction,'' in \emph{Proc. Int. Conf. Mach. Learn. (ICML)}, 2011.

\bibitem{ross2018improving}
A.~Ross and F.~Doshi-Velez, ``Improving the adversarial robustness and interpretability of deep neural networks by regularizing their input gradients,'' in \emph{Proc. AAAI Conf. Artif. Intell. (AAAI)}, vol.~32, no.~1, 2018.

\bibitem{yoshida2017spectral}
Y.~Yoshida and T.~Miyato, ``Spectral norm regularization for improving the generalizability of deep learning,'' \emph{arXiv preprint arXiv:1705.10941}, 2017.

\bibitem{cisse2017parseval}
M.~Cisse, P.~Bojanowski, E.~Grave, Y.~Dauphin, and N.~Usunier, ``Parseval networks: Improving robustness to adversarial examples,'' in \emph{Proc. Int. Conf. Mach. Learn. (ICML)}, 2017.

\bibitem{tsuzuku2018lmt}
Y.~Tsuzuku, I.~Sato, and M.~Sugiyama, ``Lipschitz-margin training: Scalable certification of perturbation invariance,'' in \emph{Proc. Adv. Neural Inf. Process. Syst. (NeurIPS)}, 2018.

\bibitem{ma2018lid}
X.~Ma, B.~Li, Y.~Wang, S.~M. Erfani, S.~Wijewickrema, G.~Schoenebeck, D.~Song, M.~E. Houle, and J.~Bailey, ``Characterizing adversarial subspaces using local intrinsic dimensionality,'' in \emph{Proc. Int. Conf. Learn. Representations (ICLR)}, 2018.

\bibitem{meng2017magnet}
D.~Meng and H.~Chen, ``Magnet: A two-pronged defense against adversarial examples,'' in \emph{Proc. ACM SIGSAC Conf. Comput. Commun. Secur. (CCS)}, 2017.

\bibitem{xu2018featuresqueezing}
W.~Xu, D.~Evans, and Y.~Qi, ``Feature squeezing: Detecting adversarial examples in deep neural networks,'' in \emph{Proc. Netw. Distrib. Syst. Secur. Symp. (NDSS)}, 2018.

\bibitem{guo2018countering}
C.~Guo, M.~Rana, M.~Cisse, and L.~Van Der~Maaten, ``Countering adversarial images using input transformations,'' in \emph{Proc. Int. Conf. Learn. Representations (ICLR) Workshop}, 2018.

\bibitem{xie2018randomization}
C.~Xie, J.~Wang, Z.~Zhang, Z.~Ren, and A.~Yuille, ``Mitigating adversarial effects through randomization,'' in \emph{Proc. Int. Conf. Learn. Representations (ICLR)}, 2018.

\bibitem{carlini2017detecting}
N.~Carlini and D.~Wagner, ``Adversarial examples are not easily detected: Bypassing ten detection methods,'' in \emph{Proc. ACM Workshop Artif. Intell. Secur. (AISec)}, 2017.

\bibitem{athalye2018obfuscated}
A.~Athalye, N.~Carlini, and D.~Wagner, ``Obfuscated gradients give a false sense of security: Circumventing defenses to adversarial examples,'' in \emph{Proc. Int. Conf. Mach. Learn. (ICML)}, 2018.

\bibitem{salman2020denoised}
H.~Salman, M.~Sun, G.~Yang, A.~Kapoor, and J.~Z. Kolter, ``Denoised smoothing: A provable defense for pretrained classifiers,'' in \emph{Proc. Adv. Neural Inf. Process. Syst. (NeurIPS)}, 2020.

\bibitem{gowal2018ibp}
S.~Gowal, K.~Dvijotham, R.~Stanforth, R.~Bunel, C.~Qin, J.~Uesato, R.~Arandjelovic, T.~Mann, and P.~Kohli, ``On the effectiveness of interval bound propagation for training verifiably robust models,'' \emph{arXiv preprint arXiv:1810.12715}, 2018.

\bibitem{wong2018provable}
E.~Wong and J.~Z. Kolter, ``Provable defenses against adversarial examples via the convex outer adversarial polytope,'' in \emph{Proc. Int. Conf. Mach. Learn. (ICML)}, 2018.

\bibitem{zhang2019crownibp}
H.~Zhang, H.~Chen, C.~Xiao, S.~Gowal, R.~Stanforth, B.~Li, D.~Boning, and C.-J. Hsieh, ``Towards stable and efficient training of verifiably robust neural networks,'' in \emph{Proc. Adv. Neural Inf. Process. Syst. (NeurIPS)}, 2019.

\bibitem{li2025synchronizing}
L.~Li, Y.~He, R.~Xu, B.~Chen, B.~Han, Y.~Zhao, and J.~Li, ``Synchronizing llm-based semantic knowledge bases via secure federated fine-tuning in semantic communication,'' \emph{Front. Artif. Intell.}, vol.~8, p. 1690950, 2025.

\bibitem{hoang2025adversarial}
V.-T. Hoang, V.-L. Nguyen, R.-G. Chang, P.-C. Lin, R.-H. Hwang, and T.~Q. Duong, ``Adversarial attacks against shared knowledge interpretation in semantic communications,'' \emph{IEEE Trans. Cogn. Commun. Netw.}, vol.~11, no.~2, pp. 1024--1040, 2025.

\bibitem{dreossi2018semantic}
T.~Dreossi, S.~Jha, and S.~A. Seshia, ``Semantic adversarial deep learning,'' in \emph{Proc. Int. Conf. Comput. Aided Verification (CAV)}.\hskip 1em plus 0.5em minus 0.4em\relax Springer, 2018, pp. 3--26.

\bibitem{chen2023model}
Y.~Chen, Q.~Yang, Z.~Shi, and J.~Chen, ``The model inversion eavesdropping attack in semantic communication systems,'' in \emph{Proc. IEEE Global Commun. Conf. (GLOBECOM)}.\hskip 1em plus 0.5em minus 0.4em\relax IEEE, 2023, pp. 5171--5177.

\bibitem{peng2024adversarial}
J.~Peng, H.~Xing, L.~Xu, S.~Luo, P.~Dai, L.~Feng, J.~Song, B.~Zhao, and Z.~Xiao, ``Adversarial reinforcement learning based data poisoning attacks defense for task-oriented multi-user semantic communication,'' \emph{IEEE Trans. Mobile Comput.}, vol.~23, no.~12, pp. 14\,834--14\,851, 2024.

\bibitem{sagduyu2023vulnerabilities}
Y.~E. Sagduyu, T.~Erpek, S.~Ulukus, and A.~Yener, ``Vulnerabilities of deep learning-driven semantic communications to backdoor (trojan) attacks,'' in \emph{Proc. Annu. Conf. Inf. Sci. Syst. (CISS)}.\hskip 1em plus 0.5em minus 0.4em\relax IEEE, 2023, pp. 1--6.

\bibitem{guo2025persistent}
Z.~Guo, A.~Kumar, and R.~Tourani, ``Persistent backdoor attacks in continual learning,'' in \emph{Proc. USENIX Secur. Symp.}, 2025, pp. 6379--6397.

\bibitem{zhou2024backdoor}
Y.~Zhou, R.~Q. Hu, and Y.~Qian, ``Backdoor attacks and defenses on semantic-symbol reconstruction in semantic communications,'' in \emph{Proc. IEEE Int. Conf. Commun. (ICC)}.\hskip 1em plus 0.5em minus 0.4em\relax IEEE, 2024, pp. 734--739.

\bibitem{tang2025towards}
S.~Tang, Y.~Chen, Q.~Yang, R.~Zhang, D.~Niyato, and Z.~Shi, ``Towards secure semantic communications in the presence of intelligent eavesdroppers,'' \emph{arXiv preprint arXiv:2503.23103}, 2025.

\bibitem{yuan2022membership}
X.~Yuan and L.~Zhang, ``Membership inference attacks and defenses in neural network pruning,'' in \emph{Proc. USENIX Secur. Symp.}\hskip 1em plus 0.5em minus 0.4em\relax USENIX Association, 2022, pp. 4561--4578.

\bibitem{liu2024manipulating}
J.~Liu, Y.~He, W.~Xu, Y.~Xie, and J.~Han, ``Manipulating semantic communication by adding adversarial perturbations to wireless channel,'' in \emph{Proc. IEEE/ACM Int. Symp. Quality of Service (IWQoS)}.\hskip 1em plus 0.5em minus 0.4em\relax IEEE, 2024, pp. 1--10.

\bibitem{rong2025semantic}
Y.~Rong, G.~Nan, M.~Zhang, S.~Chen, S.~Wang, X.~Zhang, N.~Ma, S.~Gong, Z.~Yang, Q.~Cui, X.~Tao, and T.~Q.~S. Quek, ``Semantic entropy can simultaneously benefit transmission efficiency and channel security of wireless semantic communications,'' \emph{IEEE Trans. Inf. Forensics Secur.}, vol.~20, pp. 2067--2082, 2025.

\bibitem{fallahreyhani2024countering}
M.~Fallahreyhani, P.~Azmi, and N.~Mokari, ``Countering physical adversarial attacks in semantic communication networks: Innovations and strategies,'' in \emph{Proc. Int. Symp. Telecommun. (IST)}.\hskip 1em plus 0.5em minus 0.4em\relax IEEE, 2024, pp. 637--642.

\bibitem{tang2023gan_jamming}
R.~Tang, D.~Gao, M.~Yang, T.~Guo, H.~Wu, and G.~Shi, ``{GAN}-inspired intelligent jamming and anti-jamming strategy for semantic communication systems,'' in \emph{Proc. IEEE Int. Conf. Commun. (ICC)}, 2023, pp. 1623--1628.

\bibitem{zhou2025rome}
K.~Zhou, G.~Zhang, Y.~Cai, Q.~Hu, and G.~Yu, ``{ROME}: Robust model ensembling for semantic communication against semantic jamming attacks,'' \emph{arXiv preprint arXiv:2501.01172}, 2025.

\bibitem{chen2025coding_jamming}
W.~Chen, Q.~Yang, S.~Shao, Z.~Shi, J.~Chen, and X.~Shen, ``A coding-enhanced jamming approach for secure semantic communication over wiretap channels,'' \emph{arXiv preprint arXiv:2504.16960}, 2025.

\bibitem{zhou2022adaptive}
Q.~Zhou, R.~Li, Z.~Zhao, Y.~Xiao, and H.~Zhang, ``Adaptive bit rate control in semantic communication with incremental knowledge-based {HARQ},'' \emph{IEEE Open J. Commun. Soc.}, vol.~3, pp. 1076--1089, 2022.

\bibitem{gong2023adaptive}
W.~Gong, H.~Tong, S.~Wang, Z.~Yang, X.~He, and C.~Yin, ``Adaptive bitrate video semantic communication over wireless networks,'' in \emph{Proc. Int. Conf. Wireless Commun. Signal Process. (WCSP)}.\hskip 1em plus 0.5em minus 0.4em\relax IEEE, 2023, pp. 122--127.

\bibitem{zhou2024stealthy}
Y.~Zhou, R.~Q. Hu, and Y.~Qian, ``Stealthy backdoor attacks on semantic symbols in semantic communications,'' in \emph{Proc. IEEE Global Commun. Conf. (GLOBECOM)}.\hskip 1em plus 0.5em minus 0.4em\relax IEEE, 2024, pp. 4975--4981.

\bibitem{ren2024knowledge}
J.~Ren, Z.~Zhang, J.~Xu, G.~Chen, Y.~Sun, P.~Zhang, and S.~Cui, ``Knowledge base enabled semantic communication: A generative perspective,'' \emph{IEEE Wireless Commun.}, vol.~31, no.~4, pp. 14--22, 2024.

\bibitem{han2025manipulating}
B.~Han, Y.~He, R.~Xu, D.~Xiao, N.~Ruan, and J.~Li, ``Manipulating digital twin networks by poisoning semantic knowledge base,'' in \emph{Proc. IEEE Int. Conf. Commun. (ICC)}.\hskip 1em plus 0.5em minus 0.4em\relax IEEE, 2025, pp. 720--725.

\bibitem{he2025invisible}
Y.~He, X.~Yang, G.~Li, and J.~Li, ``On the invisible backdoor attacks on peer-to-peer semantic vehicular networks,'' \emph{Electronics Letters}, vol.~61, no.~1, p. e70431, 2025.

\bibitem{yang2024inviins}
X.~Yang, G.~Li, M.~Dong, K.~Ota, J.~Wu, and J.~Li, ``Inviins: Invisible instruction backdoor attacks on peer-to-peer semantic networks,'' in \emph{Proc. IEEE Int. Symp. Parallel Distrib. Process. Appl. (ISPA)}.\hskip 1em plus 0.5em minus 0.4em\relax IEEE, 2024, pp. 956--964.

\bibitem{li2023catfl}
G.~Li, Y.~Zhao, and Y.~Li, ``Catfl: Certificateless authentication-based trustworthy federated learning for 6g semantic communications,'' in \emph{Proc. IEEE Wireless Commun. Netw. Conf. (WCNC)}.\hskip 1em plus 0.5em minus 0.4em\relax IEEE, 2023, pp. 1--6.

\bibitem{shi2021semantic}
G.~Shi, Y.~Xiao, Y.~Li, and X.~Xie, ``From semantic communication to semantic-aware networking: Model, architecture, and open problems,'' \emph{IEEE Commun. Mag.}, vol.~59, no.~8, pp. 44--50, 2021.

\bibitem{uysal2022semantic}
E.~Uysal, O.~Kaya, A.~Ephremides, J.~Gross, M.~Codreanu, P.~Popovski, M.~Assaad, G.~Liva, A.~Munari, B.~Soret \emph{et~al.}, ``Semantic communications in networked systems: A data significance perspective,'' \emph{IEEE Network}, vol.~36, no.~4, pp. 233--240, 2022.

\bibitem{he2019model}
Z.~He, T.~Zhang, and R.~B. Lee, ``Model inversion attacks against collaborative inference,'' in \emph{Proc. Annu. Comput. Secur. Appl. Conf. (ACSAC)}, 2019, pp. 148--162.

\bibitem{ding2024patrol}
S.~Ding, L.~Zhang, M.~Pan, and X.~Yuan, ``Patrol: Privacy-oriented pruning for collaborative inference against model inversion attacks,'' in \emph{Proc. IEEE/CVF Winter Conf. Appl. Comput. Vis. (WACV)}, 2024, pp. 4716--4725.

\bibitem{shen2021coordinated}
W.~Shen, H.~Li, and Z.~Zheng, ``Coordinated attacks against federated learning: A multi-agent reinforcement learning approach,'' in \emph{Proc. Int. Conf. Learn. Representations Workshops (ICLR Workshops)}.\hskip 1em plus 0.5em minus 0.4em\relax ICLR 2021 Workshop on Security and Safety in Machine Learning Systems (SecML), 2021.

\bibitem{zhao2020shielding}
L.~Zhao, S.~Hu, Q.~Wang, J.~Jiang, C.~Shen, X.~Luo, and P.~Hu, ``Shielding collaborative learning: Mitigating poisoning attacks through client-side detection,'' \emph{IEEE Trans. Dependable Secure Comput.}, vol.~18, no.~5, pp. 2029--2041, 2020.

\bibitem{hu2022robust}
Q.~Hu, G.~Zhang, Z.~Qin, Y.~Cai, G.~Yu, and G.~Y. Li, ``Robust semantic communications against semantic noise,'' in \emph{Proc. IEEE Veh. Technol. Conf. (VTC)}.\hskip 1em plus 0.5em minus 0.4em\relax IEEE, 2022, pp. 1--6.

\bibitem{wei2025robust}
K.~Wei, R.~Xie, W.~Xu, Z.~Lu, and H.~Xiao, ``Robust semantic communication via adversarial training,'' \emph{IEEE Trans. Veh. Technol.}, vol.~74, no.~12, pp. 19\,849--19\,853, 2025.

\bibitem{hu2023robust}
Q.~Hu, G.~Zhang, Z.~Qin, Y.~Cai, G.~Yu, and G.~Y. Li, ``Robust semantic communications with masked vq-vae enabled codebook,'' \emph{IEEE Trans. Wireless Commun.}, vol.~22, no.~12, pp. 8707--8722, 2023.

\bibitem{chen2024lightweight}
G.~Chen, G.~Nan, Z.~Jiang, H.~Du, R.~Shi, Q.~Cui, and X.~Tao, ``Lightweight and robust wireless semantic communications,'' \emph{IEEE Commun. Lett.}, vol.~28, no.~11, pp. 2633--2637, 2024.

\bibitem{peng2022robust}
X.~Peng, Z.~Qin, D.~Huang, X.~Tao, J.~Lu, G.~Liu, and C.~Pan, ``A robust deep learning enabled semantic communication system for text,'' in \emph{Proc. IEEE Global Commun. Conf. (GLOBECOM)}.\hskip 1em plus 0.5em minus 0.4em\relax IEEE, 2022, pp. 2704--2709.

\bibitem{peng2024robust_text}
X.~Peng, Z.~Qin, X.~Tao, J.~Lu, and L.~Hanzo, ``A robust semantic text communication system,'' \emph{IEEE Trans. Wireless Commun.}, vol.~23, no.~9, pp. 11\,372--11\,385, 2024.

\bibitem{song2018pixeldefend}
Y.~Song, T.~Kim, S.~Nowozin, S.~Ermon, and N.~Kushman, ``Pixeldefend: Leveraging generative models to understand and defend against adversarial examples,'' in \emph{Proc. Int. Conf. Learn. Representations (ICLR)}, 2018.

\bibitem{weng2025generative}
Z.~Weng, Z.~Wang, Z.~Qin, and X.~Tao, ``Generative semantic communications for robust speech-to-text translation,'' \emph{IEEE Trans. Wireless Commun.}, p. early access, 2025.

\bibitem{cai2025robust}
Y.~Cai, ``Robust and adaptive semantic noise for complex secure communication networks,'' \emph{Physical Communication}, vol.~72, p. 102763, 2025.

\bibitem{tung2023deepjscec}
T.-Y. Tung and D.~G{\"u}nd{\"u}z, ``Deep joint source-channel and encryption coding: Secure semantic communications,'' in \emph{Proc. IEEE Int. Conf. Commun. (ICC)}, 2023, pp. 5620--5625.

\bibitem{kalkhoran2023securedeepjscc}
S.~A. Ameli~Kalkhoran, M.~Letafati, E.~Erdemir, B.~H. Khalaj, H.~Behroozi, and D.~Gündüz, ``Secure deep-jscc against multiple eavesdroppers,'' in \emph{Proc. IEEE Global Commun. Conf. (GLOBECOM)}, 2023, pp. 3433--3438.

\bibitem{nan2023physical}
G.~Nan, Z.~Li, J.~Zhai, Q.~Cui, G.~Chen, X.~Du, X.~Zhang, X.~Tao, Z.~Han, and T.~Q. Quek, ``Physical-layer adversarial robustness for deep learning-based semantic communications,'' \emph{IEEE J. Sel. Areas Commun.}, vol.~41, no.~8, pp. 2592--2608, 2023.

\bibitem{zhao2025secdiff}
C.~Zhao, J.~Wang, R.~Zhang, D.~Niyato, H.~Du, Z.~Xiong, D.~I. Kim, and P.~Zhang, ``Secdiff: Diffusion-aided secure deep joint source-channel coding against adversarial attacks,'' \emph{arXiv preprint arXiv:2511.01466}, 2025.

\bibitem{peng2024robust_image}
X.~Peng, Z.~Qin, X.~Tao, J.~Lu, and K.~B. Letaief, ``A robust semantic communication system for image transmission,'' in \emph{Proc. IEEE Global Commun. Conf. (GLOBECOM)}.\hskip 1em plus 0.5em minus 0.4em\relax IEEE, 2024, pp. 2154--2159.

\bibitem{lu2024efficient}
X.~Lu, K.~Zhu, J.~Li, and Y.~Zhang, ``Efficient knowledge base synchronization in semantic communication network: A federated distillation approach,'' in \emph{Proc. IEEE Wireless Commun. Netw. Conf. (WCNC)}.\hskip 1em plus 0.5em minus 0.4em\relax IEEE, 2024, pp. 1--6.

\bibitem{hu2022knowledge}
L.~Hu, Y.~Li, H.~Zhang, L.~Yuan, F.~Zhou, and Q.~Wu, ``Robust semantic communication driven by knowledge graph,'' in \emph{Proc. Int. Conf. Internet Things: Syst. Manage. Secur. (IOTSMS)}.\hskip 1em plus 0.5em minus 0.4em\relax IEEE, 2022, pp. 1--5.

\bibitem{liu2024semantic}
X.~Liu, H.~Liang, C.~Dong, and X.~Xu, ``Semantic synchronization for enhanced reliability in communication systems,'' in \emph{Proc. IEEE Wireless Commun. Netw. Conf. (WCNC)}.\hskip 1em plus 0.5em minus 0.4em\relax IEEE, 2024, pp. 1--6.

\bibitem{erdemir2023generative}
E.~Erdemir, T.-Y. Tung, P.~L. Dragotti, and D.~G{\"u}nd{\"u}z, ``Generative joint source-channel coding for semantic image transmission,'' \emph{IEEE J. Sel. Areas Commun.}, vol.~41, no.~8, pp. 2645--2657, 2023.

\bibitem{chen2023commin}
J.~Chen, D.~You, D.~G{\"u}nd{\"u}z, and P.~L. Dragotti, ``{CommIN}: Semantic image communications as an inverse problem with {INN}-guided diffusion models,'' in \emph{Proc. IEEE Int. Conf. Acoust. Speech Signal Process. (ICASSP)}, 2024, pp. 6675--6679.

\bibitem{luo2022relay}
X.~Luo, B.~Yin, Z.~Chen, and J.~Wang, ``Autoencoder-based semantic communication systems with relay channels,'' in \emph{Proc. IEEE Int. Conf. Commun. (ICC)}, 2022, pp. 711--716.

\bibitem{ma2023intelligentrelay}
S.~Ma, W.~Liang, B.~Zhang, and D.~Wang, ``An investigation on intelligent relay assisted semantic communication networks,'' in \emph{Proc. IEEE Wireless Commun. Netw. Conf. (WCNC)}, 2023, pp. 1--6.

\bibitem{gao2024esanet}
N.~Gao, Q.~Huang, C.~Li, S.~Jin, and M.~Matthaiou, ``{EsaNet}: Environment semantics enabled physical layer authentication,'' \emph{IEEE Wireless Commun. Lett.}, vol.~13, no.~1, pp. 178--182, 2024.

\bibitem{tan2024sae}
H.~Tan, N.~Xie, and A.~X. Liu, ``An optimization framework for active physical-layer authentication,'' \emph{IEEE Trans. Mobile Comput.}, vol.~23, no.~1, pp. 164--179, 2024.

\bibitem{shokrnezhad2025arc}
M.~Shokrnezhad and T.~Taleb, ``An autonomous network orchestration framework integrating large language models with continual reinforcement learning,'' \emph{arXiv preprint arXiv:2502.16198}, 2025.

\bibitem{weng2024ross}
Z.~Weng, Z.~Qin, and G.~Y. Li, ``Robust semantic communications for speech transmission,'' in \emph{Proc. IEEE Int. Conf. Acoust. Speech Signal Process. (ICASSP)}, 2025.

\bibitem{papineni2002bleu}
K.~Papineni, S.~Roukos, T.~Ward, and W.-J. Zhu, ``Bleu: a method for automatic evaluation of machine translation,'' in \emph{Proc. Annu. Meeting Assoc. Comput. Linguist. (ACL)}, 2002, pp. 311--318.

\bibitem{cer2017semeval}
D.~Cer, M.~Diab, E.~Agirre, I.~Lopez-Gazpio, and L.~Specia, ``Semeval-2017 task 1: Semantic textual similarity-multilingual and cross-lingual focused evaluation,'' \emph{arXiv preprint arXiv:1708.00055}, 2017.

\bibitem{wijesinghe2025taco}
A.~Wijesinghe, W.~Wang, S.~Wanninayaka, S.~Zhang, and Z.~Ding, ``Taco: Rethinking semantic communications with task adaptation and context embedding,'' \emph{arXiv preprint arXiv:2505.10834}, 2025.

\bibitem{zhan2025sparse}
X.~Zhan, J.~Cao, X.~Zhu, Y.~Zhang, Z.~Dong, and C.~Fan, ``Sparse vector coding based robust semantic communication for dynamic environment,'' in \emph{Proc. IEEE Int. Workshop Radio Freq. Antenna Technol. (iWRF\&AT)}.\hskip 1em plus 0.5em minus 0.4em\relax IEEE, 2025, pp. 425--430.

\bibitem{peng2025deepscri}
X.~Peng, Z.~Qin, X.~Tao, J.~Lu, and K.~B. Letaief, ``A robust image semantic communication system with multi-scale vision transformer,'' \emph{IEEE J. Sel. Areas Commun.}, vol.~43, no.~1, pp. 53--68, 2025.

\bibitem{tian2025model}
Z.~Tian, W.~Wang, C.~Zhang, and S.~Yu, ``Model-enabled task-oriented semantic communications through knowledge synchronization,'' \emph{IEEE Trans. Cogn. Commun. Netw.}, 2025.

\bibitem{gao2025agenticsemcom}
H.~Gao, M.~Sun, R.~Zhang, Y.~Wang, X.~Xu, N.~Ma, D.~Niyato, and P.~Zhang, ``Agentic ai-enhanced semantic communications: Foundations, architecture, and applications,'' \emph{arXiv preprint arXiv:2512.23294}, 2025.

\bibitem{wang2022measure}
X.~Wang, H.~Wang, and D.~Yang, ``Measure and improve robustness in nlp models: A survey,'' in \emph{Proc. Conf. North Amer. Chapter Assoc. Comput. Linguist.: Human Lang. Technol. (NAACL-HLT)}, 2022, pp. 4569--4586.

\bibitem{goyal2023survey}
S.~Goyal, S.~Doddapaneni, M.~M. Khapra, and B.~Ravindran, ``A survey of adversarial defenses and robustness in nlp,'' \emph{ACM Comput. Surveys}, vol.~55, no. 14s, pp. 1--39, 2023.

\bibitem{drenkow2021systematic}
N.~Drenkow, N.~Sani, I.~Shpitser, and M.~Unberath, ``A systematic review of robustness in deep learning for computer vision: Mind the gap?'' \emph{arXiv preprint arXiv:2112.00639}, 2021.

\bibitem{li2023trade}
Y.~Li and C.~Xu, ``Trade-off between robustness and accuracy of vision transformers,'' in \emph{Proc. IEEE/CVF Conf. Comput. Vision Pattern Recognit. (CVPR)}, 2023, pp. 7558--7568.

\bibitem{hendrycks2019benchmarking}
D.~Hendrycks and T.~Dietterich, ``Benchmarking neural network robustness to common corruptions and perturbations,'' in \emph{Proc. Int. Conf. Learn. Representations (ICLR)}, 2019.

\bibitem{nie2020adversarial}
Y.~Nie, A.~Williams, E.~Dinan, M.~Bansal, J.~Weston, and D.~Kiela, ``Adversarial nli: A new benchmark for natural language understanding,'' in \emph{Proc. Annu. Meeting Assoc. Comput. Linguist. (ACL)}, 2020, pp. 4885--4901.

\bibitem{aastrom2021feedback}
K.~J. {\AA}str{\"o}m and R.~Murray, \emph{Feedback systems: an introduction for scientists and engineers}.\hskip 1em plus 0.5em minus 0.4em\relax Princeton university press, 2021.

\bibitem{liu2000real}
J.~W.~S. Liu, \emph{Real-Time Systems}.\hskip 1em plus 0.5em minus 0.4em\relax Prentice Hall, 2000.

\bibitem{zappone2019model}
A.~Zappone, M.~Di~Renzo, M.~Debbah, T.~T. Lam, and X.~Qian, ``Model-aided wireless artificial intelligence: Embedding expert knowledge in deep neural networks for wireless system optimization,'' \emph{IEEE Veh. Technol. Mag.}, vol.~14, no.~3, pp. 60--69, 2019.

\bibitem{li2023sok}
L.~Li, T.~Xie, and B.~Li, ``Sok: Certified robustness for deep neural networks,'' in \emph{Proc. IEEE Symp. Secur. Priv. (SP)}.\hskip 1em plus 0.5em minus 0.4em\relax IEEE, 2023, pp. 1289--1310.

\bibitem{zhang2022rethinking}
B.~Zhang, D.~Jiang, D.~He, and L.~Wang, ``Rethinking lipschitz neural networks and certified robustness: A boolean function perspective,'' \emph{Proc. Adv. Neural Inf. Process. Syst. (NeurIPS)}, vol.~35, pp. 19\,398--19\,413, 2022.

\bibitem{carlini2017towards}
N.~Carlini and D.~Wagner, ``Towards evaluating the robustness of neural networks,'' in \emph{Proc. IEEE Symp. Secur. Priv. (SP)}.\hskip 1em plus 0.5em minus 0.4em\relax IEEE, 2017, pp. 39--57.

\bibitem{nasr2019comprehensive}
M.~Nasr, R.~Shokri, and A.~Houmansadr, ``Comprehensive privacy analysis of deep learning: Passive and active white-box inference attacks against centralized and federated learning,'' in \emph{Proc. IEEE Symp. Secur. Priv. (SP)}.\hskip 1em plus 0.5em minus 0.4em\relax IEEE, 2019, pp. 739--753.

\bibitem{liu2021machine}
B.~Liu, M.~Ding, S.~Shaham, W.~Rahayu, F.~Farokhi, and Z.~Lin, ``When machine learning meets privacy: A survey and outlook,'' \emph{ACM Comput. Surveys}, vol.~54, no.~2, pp. 1--36, 2021.

\bibitem{elliott2022ai}
D.~Elliott and E.~Soifer, ``Ai technologies, privacy, and security,'' \emph{Front. Artif. Intell.}, vol.~5, p. 826737, 2022.

\bibitem{saad2019vision}
W.~Saad, M.~Bennis, and M.~Chen, ``A vision of 6g wireless systems: Applications, trends, technologies, and open research problems,'' \emph{IEEE Netw.}, vol.~34, no.~3, pp. 134--142, 2019.

\bibitem{zhu2020toward}
G.~Zhu, D.~Liu, Y.~Du, C.~You, J.~Zhang, and K.~Huang, ``Toward an intelligent edge: Wireless communication meets machine learning,'' \emph{IEEE Commun. Mag.}, vol.~58, no.~1, pp. 19--25, 2020.

\bibitem{zhang2025resource}
C.~Zhang, L.~Huang, and Q.~Ning, ``Resource allocation in wireless semantic communications: A comprehensive survey,'' \emph{IEEE Commun. Surveys Tuts.}, vol.~28, pp. 2965--3001, 2025.

\bibitem{hua2025bandwidth}
S.~Hua, Y.~Sun, K.~Ma, L.~Feng, M.~Chen, Z.~Yang, and M.~A. Imran, ``{Bandwidth Management in Semantic Communications: A Tradeoff Between Data Sensing and Transmission},'' \emph{IEEE Trans. Veh. Technol.}, 2025.

\bibitem{ma2025power}
K.~Ma, H.~Abumarshoud, S.~Hua, M.~Imran, and Y.~Sun, ``{Power Allocation for Throughput Maximization in NOMA-Based Semantic Communication System},'' in \emph{Proc. IEEE Int. Conf. Commun. (ICC)}.\hskip 1em plus 0.5em minus 0.4em\relax IEEE, 2025, pp. 4288--4293.

\bibitem{liu2025joint}
X.~Liu, Y.~Liu, H.~Tang, F.~Zhao, L.~Xia, and Y.~Sun, ``{Joint Knowledge and Power Management for Secure Semantic Communication Networks},'' \emph{arXiv preprint arXiv:2504.15260}, 2025.

\bibitem{getu2025semantic}
T.~M. Getu, G.~Kaddoum, and M.~Bennis, ``Semantic communication: A survey on research landscape, challenges, and future directions,'' \emph{Proc. IEEE}, vol. 112, no.~11, pp. 1649--1685, 2024.

\bibitem{morris2020textattack}
J.~Morris, E.~Lifland, J.~Y. Yoo, J.~Grigsby, D.~Jin, and Y.~Qi, ``Textattack: A framework for adversarial attacks, data augmentation, and adversarial training in nlp,'' in \emph{Proc. Conf. Empir. Methods Nat. Lang. Process. (EMNLP) Syst. Demonstrations}, 2020, pp. 119--126.

\bibitem{zhao2023evaluating}
Y.~Zhao, T.~Pang, C.~Du, X.~Yang, C.~Li, N.-M.~M. Cheung, and M.~Lin, ``On evaluating adversarial robustness of large vision-language models,'' \emph{Proc. Adv. Neural Inf. Process. Syst. (NeurIPS)}, vol.~36, pp. 54\,111--54\,138, 2023.

\bibitem{brundage2020toward}
M.~Brundage, S.~Avin, J.~Wang, H.~Belfield, G.~Krueger, G.~Hadfield, H.~Khlaaf, J.~Yang, H.~Toner, R.~Fong \emph{et~al.}, ``Toward trustworthy ai development: mechanisms for supporting verifiable claims,'' \emph{arXiv preprint arXiv:2004.07213}, 2020.

\bibitem{bonati2021intelligence}
L.~Bonati, S.~D'Oro, M.~Polese, S.~Basagni, and T.~Melodia, ``Intelligence and learning in o-ran for data-driven nextg cellular networks,'' \emph{IEEE Commun. Mag.}, vol.~59, no.~10, pp. 21--27, 2021.

\bibitem{polese2023understanding}
M.~Polese, L.~Bonati, S.~D'oro, S.~Basagni, and T.~Melodia, ``Understanding o-ran: Architecture, interfaces, algorithms, security, and research challenges,'' \emph{IEEE Commun. Surveys Tuts.}, vol.~25, no.~2, pp. 1376--1411, 2023.

\bibitem{ma2024implementation}
J.~Ma, D.~Xu, T.~Zhang, and K.~Yu, ``Implementation and evaluation of semantic communication on sdr based lora platform,'' in \emph{Proc. IEEE Wireless Commun. Netw. Conf. (WCNC)}.\hskip 1em plus 0.5em minus 0.4em\relax IEEE, 2024, pp. 1--6.

\bibitem{qiu2018can}
T.~Qiu, N.~Chen, K.~Li, M.~Atiquzzaman, and W.~Zhao, ``How can heterogeneous internet of things build our future: A survey,'' \emph{IEEE Commun. Surveys Tuts.}, vol.~20, no.~3, pp. 2011--2027, 2018.

\bibitem{powderrenew}
``{POWDER-RENEW Wireless Research Platform},'' \url{https://www.powderwireless.net}, accessed: 2025-12-17.

\bibitem{b5gplayground}
``{6G Flagship B5G Playground},'' \url{https://www.6gflagship.com/b5g-playground/}, university of Oulu, Finland. Accessed: 2025-12-17.

\bibitem{oai2023}
``{OpenAirInterface},'' \url{https://www.openairinterface.org}, 2023, accessed: 2025-12-17.

\bibitem{ara2020testbed}
``{ARA: Advanced Wireless Research Testbed},'' \url{https://arawireless.org}, 2020, accessed: 2025-12-17.

\bibitem{rise6g2023}
``{RISE-6G Project},'' \url{https://rise-6g.eu/}, 2023, accessed: 2025-12-17.

\bibitem{chen2020performance}
X.~Chen, Z.~Feng, Z.~Wei, F.~Gao, and X.~Yuan, ``Performance of joint sensing-communication cooperative sensing uav network,'' \emph{IEEE Trans. Veh. Technol.}, vol.~69, no.~12, pp. 15\,545--15\,556, 2020.

\bibitem{yang2023spatio}
K.~Yang, D.~Yang, J.~Zhang, M.~Li, Y.~Liu, J.~Liu, H.~Wang, P.~Sun, and L.~Song, ``Spatio-temporal domain awareness for multi-agent collaborative perception,'' in \emph{Proc. IEEE/CVF Int. Conf. Comput. Vis.}, 2023, pp. 23\,383--23\,392.

\bibitem{dao2024practical}
M.-Q. Dao, J.~S. Berrio, V.~Fr{\'e}mont, M.~Shan, E.~H{\'e}ry, and S.~Worrall, ``Practical collaborative perception: A framework for asynchronous and multi-agent 3d object detection,'' \emph{IEEE Trans. Intell. Transp. Syst.}, vol.~25, no.~9, pp. 12\,163--12\,175, 2024.

\bibitem{sheng2024semantic}
Y.~Sheng, H.~Ye, L.~Liang, S.~Jin, and G.~Y. Li, ``Semantic communication for cooperative perception based on importance map,'' \emph{J. Franklin Inst.}, vol. 361, no.~6, p. 106739, 2024.

\bibitem{lu2025cross}
M.~Lu, G.~Liu, L.~Liang, C.~Guo, H.~Ye, and S.~Jin, ``Cross-modal semantic communication for heterogeneous collaborative perception,'' \emph{arXiv preprint arXiv:2511.20000}, 2025.

\bibitem{Sheridan1992Telerobotics}
T.~B. Sheridan, \emph{Telerobotics, Automation, and Human Supervisory Control}.\hskip 1em plus 0.5em minus 0.4em\relax MIT Press, 1992.

\bibitem{Niemeyer2008telerobotics}
G.~Niemeyer, C.~Preusche, G.~Hirzinger, and M.~Buss, ``Telerobotics,'' \emph{Springer Handbook of Robotics}, pp. 741--757, 2008.

\bibitem{Talli2023SemanticEffective}
P.~Talli, F.~Pase, F.~Chiariotti, A.~Zanella, and M.~Zorzi, ``Semantic and effective communication for remote control tasks with dynamic feature compression,'' in \emph{Proc. IEEE INFOCOM Workshop}.\hskip 1em plus 0.5em minus 0.4em\relax IEEE, 2023, pp. 1--6.

\bibitem{zeng2024knowledge}
Q.~Zeng, Z.~Wang, Y.~Zhou, H.~Wu, L.~Yang, and K.~Huang, ``Knowledge-based ultra-low-latency semantic communications for robotic edge intelligence,'' \emph{IEEE Trans. Commun.}, 2024.

\bibitem{acharya2025agentic}
D.~B. Acharya, K.~Kuppan, and B.~Divya, ``Agentic ai: Autonomous intelligence for complex goals--a comprehensive survey,'' \emph{IEEE Access}, 2025.

\bibitem{sapkota2025ai}
R.~Sapkota, K.~I. Roumeliotis, and M.~Karkee, ``Ai agents vs. agentic ai: A conceptual taxonomy, applications and challenges,'' \emph{arXiv preprint arXiv:2505.10468}, 2025.

\bibitem{li2024semantic}
P.~Li, Z.~Liu, W.~Pang, and J.~Cao, ``Semantic collaboration: A collaborative approach for multi-agent systems based on semantic communication,'' in \emph{Proc. Int. Conf. Comput., Netw. Internet Things (CNIOT)}, 2024, pp. 123--132.

\bibitem{jiang2025large}
F.~Jiang, C.~Pan, L.~Dong, K.~Wang, O.~A. Dobre, and M.~Debbah, ``From large ai models to agentic ai: A tutorial on future intelligent communications,'' \emph{arXiv preprint arXiv:2505.22311}, 2025.

\bibitem{tang2026rethinking}
S.~Tang, Y.~Jia, Z.~Yang, Q.~Yang, R.~Zhang, J.~Du, J.~Park, Z.~Shi, and K.~B. Letaief, ``Rethinking secure semantic communications in the age of generative and agentic ai: Threats and opportunities,'' \emph{arXiv preprint arXiv:2601.01791}, 2026.

\bibitem{zuehlke2010smartfactory}
D.~Zuehlke, ``Smartfactory—towards a factory-of-things,'' \emph{Annu. Rev. Control}, vol.~34, no.~1, pp. 129--138, 2010.

\bibitem{lee2015cyber}
J.~Lee, B.~Bagheri, and H.-A. Kao, ``A cyber-physical systems architecture for industry 4.0-based manufacturing systems,'' \emph{Manuf. Lett.}, vol.~3, pp. 18--23, 2015.

\bibitem{luo2022semantic}
X.~Luo, H.-H. Chen, and Q.~Guo, ``Semantic communications: Overview, open issues, and future research directions,'' \emph{IEEE Wireless Commun.}, vol.~29, no.~1, pp. 210--219, 2022.

\bibitem{pokhrel2022learning}
S.~R. Pokhrel, ``Learning from data streams for automation and orchestration of 6g industrial iot: toward a semantic communication framework,'' \emph{Neural Comput. Appl.}, vol.~34, no.~18, pp. 15\,197--15\,206, 2022.

\bibitem{jagatheesaperumal2023semantic}
S.~K. Jagatheesaperumal, Z.~Yang, Q.~Yang, C.~Huang, W.~Xu, M.~Shikh-Bahaei, and Z.~Zhang, ``Semantic-aware digital twin for metaverse: A comprehensive review,'' \emph{IEEE Wireless Commun.}, vol.~30, no.~4, pp. 38--46, 2023.

\bibitem{thomas2023causal}
C.~K. Thomas, W.~Saad, and Y.~Xiao, ``Causal semantic communication for digital twins: A generalizable imitation learning approach,'' \emph{IEEE J. Sel. Areas Inf. Theory}, vol.~4, pp. 698--717, 2023.

\bibitem{glikson2020human}
E.~Glikson and A.~W. Woolley, ``Human trust in artificial intelligence: Review of empirical research,'' \emph{Academy of management annals}, vol.~14, no.~2, pp. 627--660, 2020.

\bibitem{kumar2024applications}
S.~Kumar, S.~Datta, V.~Singh, D.~Datta, S.~K. Singh, and R.~Sharma, ``Applications, challenges, and future directions of human-in-the-loop learning,'' \emph{IEEE Access}, vol.~12, pp. 75\,735--75\,760, 2024.

\bibitem{lai2023towards}
V.~Lai, C.~Chen, A.~Smith-Renner, Q.~V. Liao, and C.~Tan, ``Towards a science of human-ai decision making: An overview of design space in empirical human-subject studies,'' in \emph{Proc. ACM Conf. Fairness, Accountability, Transparency (FAccT)}, 2023, pp. 1369--1385.

\bibitem{tocchetti2025ai}
A.~Tocchetti, L.~Corti, A.~Balayn, M.~Yurrita, P.~Lippmann, M.~Brambilla, and J.~Yang, ``Ai robustness: a human-centered perspective on technological challenges and opportunities,'' \emph{ACM Comput. Surveys}, vol.~57, no.~6, pp. 1--38, 2025.

\bibitem{wang2025capsule}
S.~Wang, Y.~Zhuang, R.~Zhang, and Z.~Song, ``Capsule network-based semantic intent modeling for human-computer interaction,'' \emph{arXiv preprint arXiv:2507.00540}, 2025.

\bibitem{vaithilingam2025semantic}
P.~Vaithilingam, M.~Kim, F.-C. Acosta-Parenteau, D.~Lee, A.~Mhedhbi, E.~L. Glassman, and I.~Arawjo, ``Semantic commit: Helping users update intent specifications for ai memory at scale,'' in \emph{Proc. Annu. ACM Symp. User Interface Softw. Technol. (UIST)}, 2025, pp. 1--18.

\bibitem{hooker2021unrestricted}
G.~Hooker, L.~Mentch, and S.~Zhou, ``Unrestricted permutation forces extrapolation: variable importance requires at least one more model, or there is no free variable importance,'' \emph{Stat. Comput.}, vol.~31, no.~6, p.~82, 2021.

\bibitem{carlini2021poisoning}
N.~Carlini and A.~Terzis, ``Poisoning and backdooring contrastive learning,'' in \emph{Proc. Int. Conf. Learn. Representations (ICLR)}, 2021.

\bibitem{sabir2021machine}
B.~Sabir, F.~Ullah, M.~A. Babar, and R.~Gaire, ``Machine learning for detecting data exfiltration: A review,'' \emph{ACM Comput. Surveys}, vol.~54, no.~3, pp. 1--47, 2021.

\bibitem{zhao2022survey}
Y.~Zhao and J.~Chen, ``A survey on differential privacy for unstructured data content,'' \emph{ACM Comput. Surveys}, vol.~54, no. 10s, pp. 1--28, 2022.

\bibitem{deveaux2021definition}
D.~Deveaux, T.~Higuchi, S.~U{\c{c}}ar, J.~H{\"a}rri, and O.~Altintas, ``A definition and framework for vehicular knowledge networking: An application of knowledge-centric networking,'' \emph{IEEE Veh. Technol. Mag.}, vol.~16, no.~2, pp. 57--67, 2021.

\bibitem{golomb2005signal}
S.~W. Golomb and G.~Gong, \emph{Signal design for good correlation: for wireless communication, cryptography, and radar}.\hskip 1em plus 0.5em minus 0.4em\relax Cambridge University Press, 2005.

\bibitem{kartalopoulos2006primer}
S.~V. Kartalopoulos, ``A primer on cryptography in communications,'' \emph{IEEE Commun. Mag.}, vol.~44, no.~4, pp. 146--151, 2006.

\bibitem{sklavos2017wireless}
N.~Sklavos, M.~Manninger, X.~Zhang, O.~Koufopavlou, V.~Hassler, P.~Kitsos, M.~Mcloone, P.~Hamalainen, A.~P. Fournaris, V.~Rijmen \emph{et~al.}, \emph{Wireless security and cryptography: specifications and implementations}.\hskip 1em plus 0.5em minus 0.4em\relax CRC Press, 2017.

\bibitem{acar2018survey}
A.~Acar, H.~Aksu, A.~S. Uluagac, and M.~Conti, ``A survey on homomorphic encryption schemes: Theory and implementation,'' \emph{ACM Comput. Surveys}, vol.~51, no.~4, pp. 1--35, 2018.

\bibitem{lindell2020secure}
Y.~Lindell, ``Secure multiparty computation,'' \emph{Commun. ACM}, vol.~64, no.~1, pp. 86--96, 2020.

\bibitem{doshi2021towards}
F.~Doshi-Velez and B.~Kim, ``Towards a rigorous science of interpretable machine learning,'' \emph{Nat. Mach. Intell.}, vol.~3, no.~6, pp. 422--431, 2021.

\bibitem{mehrabi2021survey}
N.~Mehrabi, F.~Morstatter, N.~Saxena, K.~Lerman, and A.~Galstyan, ``A survey on bias and fairness in machine learning,'' \emph{ACM Comput. Surveys}, vol.~54, no.~6, pp. 1--35, 2021.

\bibitem{rajkomar2018ensuring}
A.~Rajkomar, M.~Hardt, and M.~Howell, ``Ensuring fairness in machine learning to advance health equity,'' \emph{Ann. Intern. Med.}, vol. 169, no.~12, pp. 866--872, 2018.

\bibitem{lipton2018mythos}
Z.~C. Lipton, ``The mythos of model interpretability,'' \emph{Commun. ACM}, vol.~61, no.~10, pp. 36--43, 2018.

\bibitem{mittelstadt2016ethics}
B.~D. Mittelstadt, P.~Allo, M.~Taddeo, S.~Wachter, and L.~Floridi, ``The ethics of algorithms: Mapping the debate,'' \emph{Big Data Soc.}, vol.~3, no.~2, p. 2053951716679679, 2016.

\end{thebibliography}
\bibliographystyle{IEEEtran}

\vfill

\end{document}